\documentclass[12pt]{article}
\usepackage{amsmath}
\usepackage{graphicx}
\usepackage{enumerate}
\usepackage[authoryear]{natbib}
\usepackage{url} 

\newcommand{\blind}{1}

\usepackage{subcaption}
\usepackage{amsthm}
\usepackage{verbatim}
\usepackage{multirow}
\usepackage{color}
\usepackage{longtable}
\usepackage{tabularx}
\usepackage{threeparttable}
\usepackage{booktabs}

\newtheorem{thm}{Theorem}[section] 
\newtheorem{lemma}{Lemma}[section] 
\newtheorem{cor}{Corollary}[section]

\newtheorem{definition}{Definition}[section]

\newcommand{\bed}{\begin{definition}}
\newcommand{\eed}{\end{definition}}
\newcommand{\beq}{\begin{equation}}
\newcommand{\eeq}{\end{equation}}

\newcommand{\goto}{\rightarrow}

\newcommand{\cF}{{\cal F}}

\newcommand{\beqn}{\begin{equation}}
\newcommand{\eeqn}{\end{equation}}
\newcommand{\balign}{\begin{align}}
\newcommand{\ealign}{\end{align}}

\newcommand{\cShat}{\hat{\cS}}

\newcommand{\cG}{{\cal G}}

\newcommand{\cC}{{\cal C}}
\newcommand{\cS}{{\cal S}}

\newcommand{\reals}{{\mathcal{R}}}

\newcommand{\bA}{\boldsymbol{A}}
\newcommand{\bB}{\boldsymbol{B}}
\newcommand{\bD}{\boldsymbol{D}}
\newcommand{\bI}{\boldsymbol{I}}
\newcommand{\bL}{\boldsymbol{L}}
\newcommand{\bP}{\boldsymbol{P}}
\newcommand{\bU}{\boldsymbol{U}}
\newcommand{\bV}{\boldsymbol{V}}
\newcommand{\bZ}{\boldsymbol{Z}}
\newcommand{\bX}{\boldsymbol{X}}
\newcommand{\bY}{\boldsymbol{Y}}
\newcommand{\bM}{\boldsymbol{M}}
\newcommand{\ba}{\boldsymbol{a}}
\newcommand{\bx}{\boldsymbol{x}}
\newcommand{\by}{\boldsymbol{y}}
\newcommand{\bLambda}{\boldsymbol{\Lambda}}

\newcommand{\bSigma}{\boldsymbol{\Sigma}}
\newcommand{\bTheta}{\boldsymbol{\Theta}}
\newcommand{\bXi}{\mathbf{\Xi}}
\newcommand{\balpha}{\boldsymbol{\alpha}}
\newcommand{\bxi}{\boldsymbol{\xi}}
\newcommand{\bbeta}{\boldsymbol{\beta}}
\newcommand{\bgamma}{\boldsymbol{\gamma}}
\newcommand{\bell}{\boldsymbol{\ell}}
\newcommand{\bfeta}{\boldsymbol{\eta}}
\newcommand{\bmu}{\boldsymbol{\mu}}
\newcommand{\bz}{\boldsymbol{z}}
\newcommand{\bveps}{\boldsymbol{\varepsilon}}
\newcommand{\btheta}{\boldsymbol{\theta}}

\newcommand{\cN}{{\cal N}}

\newcommand{\bXtilde}{\tilde{\bX}}
\newcommand{\bYtilde}{\tilde{\bY}}

\newcommand{\cshat}{\hat{s}}

\newcommand{\E}{\mathbb{E}}

\newcommand{\cB}{\mathcal{B}}

\newcommand{\Khat}{{\hat{K}}}

\newcommand{\bu}{\bar{u}}

\def\limsup{\mathop{\overline{\rm lim}}}

\newcommand{\tmax}{\theta_0}

\newcommand{\gammahat}{\hat{\bgamma}}

\newcommand{\Prob}{\mathbf{P}}

\newcommand{\be}{\begin{equation}}
\newcommand{\ee}{\end{equation}}

\newcommand{\bO}{\boldsymbol{O}}

\newcommand{\bfe}{\boldsymbol{e}}
\newcommand{\bE}{\boldsymbol{E}}
\newcommand{\bW}{\boldsymbol{W}}
\newcommand{\bw}{\boldsymbol{w}}
\newcommand{\bm}{\boldsymbol{m}}
\newcommand{\beps}{\boldsymbol{\epsilon}}
\newcommand{\bdelta}{\boldsymbol{\delta}}
\newcommand{\RR}{\mathcal{R}}

\newcommand{\signal}{\tau}

\def\ben{\begin{equation*}}
\def\een{\end{equation*}}
\def\bea{\begin{eqnarray}}
\def\eea{\end{eqnarray}}
\usepackage{amssymb}

\newtheorem{aspt}{Assumption}
\newtheorem{prop}[thm]{Proposition}

\newtheorem{example}[thm]{Example}

\begin{document}

\def\spacingset#1{\renewcommand{\baselinestretch}%
{#1}\small\normalsize} \spacingset{1}


\if1\blind
{
  \title{\bf Optimal Network-Guided Covariate Selection for High-Dimensional Data Integration}
  \author{Tao Shen\hspace{.2cm}\\
    Department of Statistics and Data Science, National University of Singapore\\
    and \\
    Wanjie Wang\thanks{
    The authors gratefully acknowledge Singapore MOE grant Tier-1-A-8001451-00-00 and NUS Research Scholarship (IRP).}\\
    Department of Statistics and Data Science, National University of Singapore}
  \maketitle
} \fi

\if0\blind
{
  \bigskip
  \bigskip
  \bigskip
  \begin{center}
    {\LARGE\bf Optimal Network-Guided Covariate Selection for High-Dimensional Data Integration}
\end{center}
  \medskip
} \fi

\bigskip
\begin{abstract}
Modern data often arise with multiple modalities. For example, covariates and a network are observed on the same subjects, and both contain useful information. Effectively integrating these modalities is important and challenging, especially when the response is unavailable. We study the fundamental covariate selection problem for high-dimensional data by leveraging network information.

We propose the Network-Guided Covariate Selection (NGCS) algorithm. NGCS exploits the spectral structure of the network to construct a network-guided screening statistic, and employs data-driven Higher Criticism Thresholding for covariate recovery. We establish consistency guarantees for NGCS under general networks. In particular, under two commonly used network models, we relate the projected signal strength to the individual signal strength, and demonstrate that NGCS is optimal for covariate selection. It could achieve the same rate as supervised learning.

We further consider a two-study setting for downstream applications, where the network is observed only in Study 1. For clustering and regression, we propose NG-clu and NG-reg algorithms. 
NG-clu accurately clusters all subjects, while NG-reg improves prediction by using the post-selection covariate matrix.
Experiments on synthetic and real datasets demonstrate the robustness and superior performance of our algorithms across various network models, noise distributions, and signal strengths.
\end{abstract}

\noindent%
\vfill

\newpage
\spacingset{1.9} 

\section{Introduction}
\subsection{Background and Motivation}\label{subsec:background}

Modern datasets have become increasingly complex, and this complexity may come from two dimensions. First, multiple types of information may be observed on the same set of subjects, giving rise to multimodal data \citep{baltruvsaitis2018multimodal, feng2018angle}. Second, in multi-study settings, data are collected across different studies on different subjects, but toward a common scientific goal \citep{patil2018training, curran2009integrative}. Methods that integrate multiple modalities or multiple studies are therefore of high practical value.

Consider a study with $n$ subjects, where both high-dimensional covariates $\bX_i \in \reals^p$ and subject-subject relational data $\bA \in \reals^{n \times n}$ are collected. Since both $\bA$ and $\bX$ describe the same subjects, methods that integrate the two should offer improved performance. The intuition has motivated developments in community detection, dimension reduction and auto-regression with fruitful results \citep{zhu2019portal, zhao2022dimension, hu2024network, SpecMultiNet}. In high-dimensional settings where $p \gg n$, covariate selection is essential in both theory and practice. However, there are very limited work when both $\bA$ and $\bX$ are observed.

One example that illustrates the necessity of covariate selection is the multi-study setting. 
Suppose there is a second dataset with $N - n$ subjects. In this study, only covariates $\bX_i \in \reals^p$ are observed, due to the cost of collecting network data. Let $\bX_{(2)} \in \reals^{(N-n)\times p}$ be the covariate matrix for this dataset and $\bXtilde \in \reals^{N \times p}$ be the stacked matrix for all $N$ subjects. Since the network $\bA_{(2)}$ is not observed, standard methods that require both network data and covariates cannot be applied. Even if two adjacency matrices were available, stacking them does not yield a valid full network because there is no relational information between subjects across the two studies. This challenge makes covariate selection particularly crucial in the multi-study setting.

These complexities motivate the two goals of this paper. Our primary goal is to identify the informative covariates using both $\bA$ and $\bX$. Our second goal is to use the selected covariates to improve downstream performance in the second study.

Our first goal is to perform network-guided covariate selection using both $\bA$ and $\bX$. For high-dimensional data where $p \gg n$, the curse of dimensionality poses challenges to classical methods \citep{donoho2000}. 
A commonly adopted assumption is that only a small fraction of covariates, denoted as $\mathcal{S}$, are related to the latent structure. 
Existing methods for recovery $\cS$ differ depending on whether auxiliary information (like response variables) is available. 
When only $\bX$ is available, informative covariates can be recovered using model-based clustering \citep{EM1} or covariate-wise weighting schemes \citep{cosa, sparsekmeans}. When responses are available, i.e., in supervised learning, methods range from penalized regression methods \citep{tibshirani1996regression, zhang2010nearly} to nonparametric approaches \citep{janitza2016random}. The Higher Criticism (HC) framework \citep{DJ08, DJ14} provides a powerful tool for selecting weak and sparse signals, and can be adapted to both supervised and unsupervised settings \citep{IFPCA, jin2017phase, DJ14}.

The substantial gap between supervised and unsupervised performance naturally raises the question of whether the network $\bA$ can be used as a supervising response for covariate selection. Because $\bA$ and $\bX$ share the same latent structure, this is feasible; however, the diversity of network models makes it challenging to design a method that is powerful, computationally efficient, and robust.

Although the literature on network-guided covariate selection is limited, there is growing interest in how network information can enhance covariate-based latent structure recovery. In microarray analysis, \cite{li2008network} employs covariate network Laplacian regularization with $\ell_1$ penalties to enhance biomarker selection. \cite{wu2018network} introduces a Markov chain approach that combines ranking statistics with network structure, which is later generalized by \cite{wang2021network} for survival traits. \cite{zhu2019portal} studies a network autoregression model and identifies a small set of ``portal nodes". 
These works imposes sparsity assumptions on subjects rather than on covariates. Other methods, such as \cite{ACMfs1} and \cite{zhao2022dimension}, focus on covariate dimension reduction, but yield reduced interpretability compared with covariate selection. \cite{Wang2025refine} proposes a dual-stage U-statistic for network-enhanced covariate screening, although optimality guarantees are not established. 




Our second goal is to improve clustering and regression performance using the selected covariates. These downstream tasks highlight the broader value of covariate selection. In the two-study setting, our aim for clustering is to group all $N$ subjects from two studies into several clusters. 
For regression, we assume that a response variable $z_i$ is observed only in the second study, and we are interested in estimating $z_i$ for subjects in the first study as well as for new subjects. 
A straightforward approach is to apply high-dimensional clustering and regression methods directly to the stacked covariate matrix $\bXtilde$. However, this strategy ignores the information in the network $\bA$. 
We instead propose to first select covariates using both $\bA$ and $\bX$, and then apply classical approaches on the post-selection data. By leveraging the rich information in the network, this approach effectively transforms a high-dimensional problem into a low-dimensional one, thereby improving accuracy.

\subsection{Our Contributions}\label{subsec:contri}

Our work studies multimodal data in which ultra-high-dimensional covariates are paired with a network. 
Rather than estimating the latent structure that is often sensitive to model specification and tuning, we propose a Network-Guided Covariate Selection (NGCS) method that directly recovers covariates associated with the latent structure. In the multi-study setting, we further show that these selected covariates enhance downstream performance, even without recovering the latent structure itself. Our main contributions are summarized as follows.

\begin{itemize}
    \item We propose a novel network-guided screening statistic, that is robust across a wide range of network models. Under sub-Gaussian covariates and model-free assumptions on the network, we establish screening consistency and introduce a data-driven threshold with theoretical guarantees. 
    \item We establish the lower bound of the signal strength for covariate selection when both $\bA$ and $\bX$ are available. It is not addressed in the existing literature. Under two widely used network models, the upper bound of NGCS matches this lower bound, demonstrating the optimality of our approach.
    \item We consider two downstream applications: clustering and regression. For these two tasks, we propose NG-clu and NG-reg algorithms, respectively. Both methods rely on the selected covariates and come with theoretical guarantees, illustrating the power of covariate selection in multi-study data integration. 
    \item Through extensive simulations, we demonstrate that NGCS outperforms existing methods in covariate selection, clustering, and prediction, across a variety of network models and covariate distributions. A real dataset analysis further illustrates the robustness and effectiveness of our method.
\end{itemize}
Finally, we highlight the semi-supervised nature of our setting: network data provides high-quality but not exact information. Through covariate selection, our spectral approach effectively transfers this information from network-augmented settings to covariate-only environments through interpretable steps. This flexibility, together with our theoretical guarantees and empirical performance, suggests broad potential for NGCS in high-dimensional and network-based data integration.

\subsection{Organization and Notations}
The remainder of this paper is organized as follows. Section \ref{sec:method} focuses on the covariate selection problem, where we introduce the network and covariate models, develop the NGCS approach, and establish its consistency and optimality. Section \ref{sec:application} formulates the two-study setting, and presents the downstream clustering and regression algorithms, together with their theoretical guarantees. Sections \ref{sec:simulation} and \ref{sec:data} provide numerical results on simulated and real data, respectively. Technical proofs and additional numerical results are in the supplementary materials \citep{supp}.

Throughout this paper, we use lower case for subject-wise vectors, such as $\bx_i$ and $\by_i$, and upper case for covariate-wise vectors, such as $\bX_j$ and $\bM_j$. 
We call $X\sim \mathcal{SG}(\sigma^2_{sg}, \bI)$ if it follows sub-Gaussian distribution with covariance matrix $\bI$ and variance proxy $\sigma^2_{sg}$. 
For a vector $\ba$, $\|\ba\|$, $\|\ba\|_1$, and $\|\ba\|_{\infty}$ gives the $\ell_2$ norm, $\ell_{1}$ norm, and $\ell_{\infty}$ norm of $\ba$, respectively. 
For a matrix $\bA$, $\lambda_k(\bA)$ denotes the $k$-th largest singular value of $\bA$, and $\|\bA\| = \lambda_1(\bA)$. 
For two series $a_n$ and $b_n$, we say $a_n \asymp b_n$ if there is a constant $C$, such that $a_n \leq C b_n$ and $b_n \leq C a_n$ when $n$ is large enough. We say $a_n \lesssim b_n$ if $\limsup_{n \goto \infty} a_n/b_n \leq 1$, and $a_n \gtrsim b_n$ similarly. Finally, we use the notation $[N]:=\{1,\ldots,N\}$ for any integer $N$. 

\section{Network-Guided Covariate Selection}
\label{sec:method}

\subsection{Covariate and Network Models}\label{subsec:model}

In this section, we focus on the covariate selection problem with $\bX \in \RR^{n \times p}$ and $\bA \in \RR^{n \times n}$. Suppose each subject $i\in[n]$ has a $K$-dimensional latent vector $\by_i\in\RR^K$ that governs both the covariates and the network, where $K\ll\min\{p,n\}$. Let $\bY = (\by_1, \cdots, \by_n)^\top \in \reals^{n \times K}$ be the latent information matrix. We introduce the following assumptions to formalize the relationship among $\bA$, $\bX$ and $\bY$.
\begin{aspt}\label{aspt:indep}
$\bA$ and $\bX$ are conditionally independent given $\{\by_i\}_{i=1}^n$.
\end{aspt}
Many studies have examined both shared and individual latent structures in multimodal data \citep{feng2018angle, wu2018network, wang2021network}. In this paper, we adopt this standard ``shared latent cause" condition \citep{casc, hu2024network}, which assumes that once the common latent vectors $\by_i$'s are given, neither the network $\bA$ nor the covariates $\bX$ carry additional information about each other. Practical examples include gene interaction networks paired with omics data \citep{hofree2013network, jiang2019microbiome}, and social relational networks coupled with profile-based covariates \citep{lazega2001collegial, o2011longitudinal}.

Given the latent vectors, we link them to the covariates via a linear structure, that 
$\E\big[X_{ij}|\by_i\big] = \by_i^\top \bM_j$ with a loading vector $\bM_j\in\RR^K$, $j \in [p]$. 
In matrix form, there is $\E[\bX_j|\bY] = \bY \bM_j$. Hence $\bM_j$ captures the dependence between covariate $j$ and the latent structure. A covariate is informative when $\|\bM_j\| \neq 0$. 
Let $\bM=[\bM_1,\dots,\bM_p]\in\RR^{K\times p}$ be the loading matrix. 
We formalize this assumption using sub-Gaussian noise.
\begin{aspt} \label{aspt:cov}
Given the latent vector $\by_i$, the covariate $\bx_i$ follows  
\begin{equation}\label{eq:cov}
\bx_i = \bM^\top \by + \bveps_i, \qquad i\in[n],
\end{equation}
where $\bm \varepsilon_i = [\epsilon_{i1},\dots,\epsilon_{ip}]^\top$ is sub-Gaussian with
$\E[\bm \varepsilon_i\bm \varepsilon_i^\top| \by_i]=\bI_p$ and the sub-Gaussian parameter $\sigma^2_{sg}$.
A covariate is informative if $\|\bM_j\| \neq 0$, and the set of informative covariates is defined by
\begin{equation}\label{eqn:signals}
    \cS \;=\; \{\, j\in[p]: \|\bM_j\|\neq 0 \,\}.
\end{equation}
\end{aspt}
The linear latent structure in Assumption \ref{aspt:cov} is commonly adopted in clustering, high-dimensional factor models, and data integration \citep{fan2013large, casc, jin2017phase, feng2018angle}.
We define the signal strength and sparsity parameter as
\begin{equation}\label{eqn:kappa}
\kappa = \min_{j \in \cS} \|\bM_j\|, \qquad 
\epsilon = |\cS|/p,
\end{equation}
where $\kappa$ measures the minimum signal strength among informative covariates and $\epsilon$ represents the proportion of informative covariates. Motivated by applications such as genomics, we assume that informative covariates are sparse and individually weak.
\begin{aspt}\label{aspt:rw}
When $p \to \infty$, both $\kappa \to 0$ and $\epsilon \to 0.$
\end{aspt}

Next, we link the network $\bA$ to latent factors $\bY$. Considering an undirected network $\bA$, we assume that edges follow an independent Bernoulli distributions that 
\begin{equation}
    \E[A_{ij}\mid \by_i,\by_j] \;=\; g\big(\by_i, \by_j; \btheta), \qquad i<j,
\end{equation}
where $g$ is a symmetric link function possibly involving parameters $\btheta$. 
Thus, the probability of an edge depends only on their latent vectors. 
To better understand possible link functions, we present two popular network models as examples, including the degree-corrected stochastic block model (DCSBM) \citep{DCBickel, DCzhuji, SCORE} with the degree-corrected mixed-membership model (DCMM) as a variant \citep{mixedscore, mao2021estimating},  and the random dot product graph (RDPG) \citep{RDPGfixadj, RDPGtutorial, LatentSpec}.
\begin{example}[DCSBM \& DCMM]\label{exam:dcsbm}
The network $\bA$ follows DCSBM if
\begin{equation}
    A_{ij}|[\by_i, \by_j] \sim Bernoulli(\theta_i \theta_j \by_i^\top \bB \by_j), \qquad i < j,
\end{equation}
where $\by_i(k) = 1$ if subject $i$ belongs to group $k$ and 0 otherwise, $\theta_i$ denotes the degree heterogeneity and $\bB \in \reals^{K \times K}$ is a constant matrix. The link function $g(\by_i, \by_j; \btheta, \bB) = \theta_i \theta_j \by_i^\top \bB \by_j$, which means the connection between subjects $i$ and $j$ depends on their group labels and the network parameter $\btheta$ and $\bB$. Allowing $\by_i$ to take soft memberships where $0 \leq \by_i(k) \leq 1$ for all $k \in [K]$ and $\sum_{k\in[K]} \by_i(k) = 1$, then it becomes the DCMM model. 
\end{example}
\begin{example}[RDPG]\label{exam:rdpg}
The network $\bA$ follows RDPG if
\begin{equation}
    A_{ij}|[\by_i, \by_j] \sim Bernoulli(\rho_n \by_i^\top \by_j),
\end{equation}
where $\by_i$ and $\by_j$ are the latent positions of subjects $i$ and $j$, respectively, and $\rho_n$ is the overall connection density. Hence, the link function $g(\by_i, \by_j; \rho_n) = \rho_n \by_i^\top \by_j$. It can be generalized to the latent position model, where $\E[A_{ij}|\by_i, \by_j] = g(\by_i, \by_j)$; see   \cite{hoff2002latent}.
\end{example}

\subsection{Network-Guided Screening Statistic}\label{sec:cov}

We now construct the network-guided screening statistic for covariates. Let's begin with the oracle case where $\bY$ is known. 
For covariate $j$, Assumption \ref{aspt:cov} gives the linear model $\E[\bX_j|\bY] = \bY \bM_j$. Hence, testing whether covariate $j$ is informative amounts to testing the hypothesis $\bM_j = 0$, i.e., the overall significance of this linear model. 
In the special case that $\bveps \sim \mathcal N({\bf0}, \bI)$ and the latent dimension $K$ is small, the likelihood ratio statistic is $t_j^o = \bX_j^\top \bY (\bY^\top \bY)^{-1}\bY^\top \bX_j$.  
Let $\bXi(\bY)=[\bxi_1^o,\ldots,\bxi_K^o]$ be the top $K$ left singular vectors of $\bY$. Then some algebra shows that $t_j^o =\|\bXi(\bY)^\top \bX_j\|^2 = \sum_{k = 1}^K \{(\bxi_k^o)^\top \bX_j\}^2$. If covariate $j \in \cS^c$, then $t_j^o \sim \chi^2_{K}$; if $j \in \cS$, then $\E[t_j^o|\bY]\gtrsim n\|\bM_j\|^2$ provided that $\bY$ is non-degenerate. 

This motivates us to build a projection-based screening statistic, $t_j(\bU) = \|\bU^\top \bX_j\|^2$, where $\bU$ is any projection matrix satisfying $\bU^\top\bU=\bI$. 
If $\bU$ suitably captures the information in $\bXi(\bY)$, then the empirical statistic $t_j(\bU)$ will have the same power as the oracle case $t_j^o$. 
A natural way is to construct $\bU$ is to use the network $\bA$, which provides high-quality information about the latent structure. Motivated by this observation, we propose the network-guided screening statistic:
\begin{equation}\label{eqn:stat}
t_j = \sum\nolimits_{k = 1}^{\Khat} (\bxi_k^\top \bX_j)^2,
\end{equation}
where $\bxi_k$ is the $k$-th eigenvector of $\bA$ and $\Khat$ is a tuning parameter. As an illustration, consider the RDPG model. Since $\E[\bA] = \rho_n \bY \bY^T$, the eigenvectors of $\bA$ are close to those of $\bY$ by random matrix theory, so using $\bxi_k$ is well-motivated.

The choice of $\bxi_k$ is not limited to the adjacency matrix. In network analysis, the regularized Laplacian $\bL=\bD^{-1/2}\bA\bD^{-1/2}$ is often used to alleviate degree heterogeneity, where $\bD_{ii}=\sum_{j}\bA_{ij}$ \citep{joseph2016impact}. Letting $\bxi_k$ be the top $k$-th largest eigenvector of $\bL$ gives a variant of $t_j$. In Section~\ref{sec:rw}, we establish theoretical guarantees of $t_j$ using eigenvectors of either $\bA$ or $\bL$, showing that $t_j$ is flexible and robust.

The robustness of the statistic also appears in the choice of $\Khat$. 
In practice, the true latent dimension $K$ is unknown, but can be reasonably bounded using domain knowledge. 
Our theory in Sections~\ref{sec:rw}--\ref{sec:lowerbound} shows an interesting phenomenon: when $\Khat$ is overestimates $K$ by an $O(1)$ factor, $t_j$ still achieves rate-optimal recovery; when $\Khat < K$, $t_j$ retains power but may become suboptimal. This tolerance to overspecification allows us to $t_j$ without specifying the exact network model or latent dimension. 

Based on the screening statistic, we compute $p$-values by quantifying the deviation of $t_j$ from the null distribution that $\|\bM_j\| = 0$. 
Under Gaussian noise, $t_j \sim \chi^2_{\Khat}$. Under sub-Gaussian noise, the exact $p$-values are not available. Hence, we construct an upper bound of its $p$-value $p_j$, where for a constant $c > 0$, 
\begin{equation}\label{eqn:HW}
p_{j} \leq \pi_j := \min\{\exp(-c\min\{(t_j-\Khat)^2/\Khat^2\sigma^4_{sg},(t_j-\Khat)/\sigma^2_{sg}\}), 1\}.
\end{equation}
We refer to $\pi_j$ as the HW-$\chi^2$ $p$-value, derived from the Hanson-Wright inequality for sub-Gaussian random variables \citep{hw}. This bound is  accurate for large $t_j$ (mostly when $j \in \cS$), and guarantees screening consistency.

In practice, Gaussian noise often have a heavier tail than sub-Gaussian noise, meaning the $\chi^2$ distribution can serve as a practical upper bound. Thus, instead of using a fixed constant $c$, we may take $\pi_j = F^{-1}_{\Khat}(t_j)$, where $F_{\Khat}$ is the cumulative density function of $\chi^2_{\Khat}$ distribution. 
Numerical results in Section~\ref{sec:simulation} show that it works well.

To conclude, although the network $\bA$ is used to guide the screening of $\bX$, our goal is not to recover $\bY$ itself, but rather the information in $\bXi(\bY)$. Leveraging the spectral information from $\bA$ yields a statistic that is robust to the choice of $\Khat$, flexible across different network models and matrices ($\bA$, $\bL$, or other variants), and achieves an optimal rate. This demonstrates an important phenomenon: exact latent structure recovery is not necessary for achieving optimal performance in data integration. 

\subsection{Network-Guided Covariate Selection (NGCS) Approach}\label{sec:algo}

The HW-$\chi^2$ $p$-values $\pi_j$ provide an estimate $\cShat = \{j \in [p]: \pi_j \leq T_{thre}\}$. Now the question is how to decide $T_{thre}$. Equivalently, if we order the $p$-values from smallest to largest, $\pi_{(1)}\leq\pi_{(2)}\leq\cdots\pi_{(p)}$, then the question is to decide how many of the $p$-values corresponding to informative covariates.  
The Higher Criticism (HC) framework provides a data-driven solution that is particularly effective when the signals are rare and weak \citep{DJ08, DJ14, IFPCA, jin2017phase}. We adopt this framework in combination with the HW-$\chi^2$ $p$-values.

The HC framework is built on the HC score, which is a function of the ordered $p$-value $\pi_{(j)}$. In our setting, the HC score is defined as 
$$
HC(j) = \sqrt{p} \cdot \frac{j/p - \pi_{(j)}}{\sqrt{\pi_{(j)}(1 - \pi_{(j)})}}, \qquad j \in [p].
$$ 
Here, $\pi_{(j)}$ is the $j$-th smallest $p$-value and $j/p$ being the corresponding uniform quantile. The HC score measures the deviation of empirical $p$-values from the uniform quantiles.  
When $\max_{1\leq j \leq p/2} HC(j) \leq \sqrt{2\log \log p}$, the network-guided screening statistic cannot recover any informative covariates \citep{DJ14}. 
This may occur when the network carries little information about the latent structure, or when the signals are extremely sparse and weak.  Otherwise, a data-driven cutoff is selected at the index maximizing the HC score, and all covariates with $p$-values below this threshold are retained:
$$
T_{thre} = \pi_{(\hat{s})}, \quad \cShat = \{1 \leq j \leq p: \pi_j \leq T_{thre}\}, \quad \mbox{where }\hat{s} = \arg\max_{1 \leq j \leq p/2} HC(j).
$$
\indent Combining Sections~\ref{sec:cov} and \ref{sec:algo}, we propose NGCS, a network-guided covariate selection method. Given the screening statistics $t_j$ and the HW-$\chi^2$ $p$-values $\pi_j$, NGCS ranks the covariates and selects $\cShat$ via HC thresholding. The full procedure is provided in Table~\ref{tab:alg1}.

\begin{table}[htbp!]
\caption{Network-Guided Covariate Selection Algorithm} 
\begin{tabular}{ll} \hline
& \underline{Input}: adjacency matrix $\bA$, covariate matrix $\bX$, tuning parameter $\Khat$. \\
\hline
Step 1  & Calculate the top $\hat{K}$ eigenvectors of $\bA$ (or $\bL$), denoted as $\bxi_1, \cdots, \bxi_{\hat{K}}$.\\
Step 2 & Calculate the Network-Guided statistic: 
$t_j = \sum_{k = 1}^{\hat{K}} (\xi_k^\top \bX_j)^2$. 
   Find the HW-$\chi^2$ \\
   &$p$-values: for covariate $j$, $\pi_j = \exp(-c_0\min\{(t_j-\Khat)^2/\Khat^2,t_j-\Khat\})$, $j \in [p]$.\\
Step 3  & Order the $p$-values so that $\pi_{(1)} \leq \pi_{(2)} \leq \cdots \leq \pi_{(p)}$. \\
  & Calculate HC score: $HC(j) = \sqrt{p}\frac{j/p - \pi_{(j)}}{\sqrt{\pi_{(j)}(1 - \pi_{(j)})}}$.\\
Step 4 & Let $S = \max_{1 \leq j \leq p/2} HC(j)$. If $S \leq \sqrt{2\log\log p}$, return $\cShat = \emptyset$.\\
 & Otherwise, define the threshold  $T_{thre} = \pi_{(\cshat)}$, where $\cshat = \arg\max_{1 \leq j \leq p/2} HC(j)$.\\
Output & Recovered set $\cShat = \{j: \pi_j < T_{thre}\}$. \\
 \hline
\end{tabular}
\label{tab:alg1} 
\end{table}

\subsection{Consistency of NGCS}\label{sec:rw}
In this section, we establish the theoretical guarantee of our NGCS approach. We begin with the consistency of $t_j(\bU)$ for any projection matrix $\bU$. Then we specialize $\bU$ to be the spectral matrix of $\bA$ (or the Laplacian $\bL$) to characterize the information required from the network. Finally, we examine the DCSBM and the RDPG as two special cases to illustrate the conditions on the network $\bA$. 

We first define the signal strength of $t_j(\bU)$. Given the latent factors $\bY$, define the aggregated signal strength of projection $\bU$ to be
\be\label{eqn:signal}
\signal(\bU) = \signal(\bU; \bM, \bY)={\min_{j\in \cS}\|\bU^\top \bY \bM_j\|}.
\ee
Theorem \ref{thm:main} provides a condition on $\signal(\bU)$, under which there is a clear separation between $\cS$ and $\cS^c$ using $\bU$. Moreover, under this condition, the HC procedure selects the correct threshold with high probability. 
\begin{thm}\label{thm:main}
Suppose Assumptions \ref{aspt:indep}--\ref{aspt:rw} hold and the sparsity parameter $\epsilon = \epsilon_p = p^{-\beta}$ for a constant $0 < \beta < 1$. 
Let $\pi_i$ be the HW-$\chi^2$ $p$-value of $i$-th covariate using a projection matrix $\bU \in \reals^{n \times \Khat}$ with $\bU^\top \bU = \bI$. 
Suppose $\signal(\bU)^2\geq 16\max\{(1-\beta)/\sigma^2_{sg}, 7\}\log p$, then there is exists $p_0$ so that if $p>p_0$, with probability at least $1-O(1/p)$, 
\[
\max_{i\in \cS}\pi_{i}<\min_{i\in \cS^c}\pi_{i}.
\]
Further, let $\cShat$ be the set of selected covariates by HCT in Algorithm \ref{tab:alg1}. 
There is a constant $C > 0$, so that with probability at least $1 - O(1/p)$, 
\[
\cS \subset \cShat, \mbox{ and } |\cShat\backslash\cS| \leq C\log^2p \ll |\cS|.
\]
\end{thm}

Theorem \ref{thm:main} guarantees that the HC procedure applied to $t_j(\bU)$ will almost exactly recover $\cS$, provided the aggregated signal strength $\signal(\bU) \geq \sqrt{c\log p}$. This condition requires that the individual signal strength $\kappa$ is not too weak and that the projection matrix $\bU$ captures the relevant latent information. The constant $c$ depends on the sparsity level of informative covariates. 

We next connect the projection matrix $\bU$ to the network $\bA$.
In the NGCS approach, we take $\bU = \bXi(\bA; \Khat)$ or $\bU = \bXi(\bL; \Khat)$, where $\bXi$ denotes the matrix formed by the top $\Khat$ eigenvectors of $\bA$ or $\bL$. This leads to the following corollary. 
\begin{cor}\label{cor:main}
Under the assumptions of Theorem \ref{thm:main}, let $\cShat$ be the set of selected covariates by Algorithm \ref{tab:alg1}, where $\bU = \bXi(\bA; \Khat)$ or $\bU = \bXi(\bL; \Khat)$. If the aggregated signal strength $\signal(\bU)^2 \geq 16\max\{(1-\beta)/\sigma^2_{sg}, 7\}\log p$, then there is a constant $C > 0$, so that with probability at least $1 - O(1/p)$, 
$\cS \subset \cShat$ and $|\cShat\backslash\cS| \leq C\log^2p \ll |\cS|.$
\end{cor}

To interpret the condition $\tau(\bU)^2 \geq 16\max\{(1-\beta)/\sigma^2_{sg}, 7\}\log p$ for $\bU = \bXi(\bA, \Khat)$ or $\bU = \bXi(\bL, \Khat)$, we examine two typical network models, the DCSBM in Example~\ref{exam:dcsbm} and the RDPG in Example~\ref{exam:rdpg}, and specify the assumptions needed for consistency.
\begin{aspt}\label{aspt:dcsbm}
Suppose the network $\bA$ follows DCSBM with parameters $\{\theta_i\}_{i \in [n]}$ and block matrix $\bB$. Assume the following conditions hold. (i)(Non-degenerate network) there is a constant $c_B > 0$ so that $\lambda_K(\bB)\geq c_B$ and $\max_{k\in[K]} B_{jk} \geq c_B$ for any $j \in [K]$; (ii) (Network density) when $n \to \infty$, there is a constant $c_{\theta}>0$ such that $\tmax/c_{\theta} \leq \theta_i \leq \tmax$ for all $i \in [n]$, where $\tmax = \max_i\theta_i$. Furthermore, $n\tmax^2\geq c_{\theta}\log n$. (iii) (Community size) there is a constant $\rho > 0$ so that $\sum_{i\in[n]} \by_i(k)/n \geq \rho$ for any $k \in [K]$.
\end{aspt}
These are standard regularity conditions for DCSBM in community detection. Assumption 4(i) ensures that inter-community connection probabilities are non-zero and the connection matrix has full rank. Assumption 4(ii) prevents the network from being too sparse, and Assumption 4(iii) rules out vanishing communities.
\begin{thm}\label{thm:dcsbm}
Suppose network $\bA$ is generated from a DCSBM satisfying Assumption \ref{aspt:dcsbm}. 
Then there is a constant $c_{D} = c_D(c_{\theta}, c_B, \rho, K)$, such that with probability $1 - O(1/n)$, 
\[
\signal(\bU)^2 \geq nc_D\kappa^2,
\]
where $\Khat \geq K$ and $\bU = \bXi(\bA; \Khat)$ or $\bU = \bXi(\bL; \Khat)$. 

Furthermore, under Assumptions~\ref{aspt:indep}--\ref{aspt:rw}, if $\kappa \geq c\sqrt{\log p/n}$, then NGCS almost exactly recovers $\cShat$ with probability $1 - O(1/p+1/n)$.
\end{thm}

Under regular conditions on DCSBM, Theorem \ref{thm:dcsbm} connects the aggregated signal strength $\tau(\bXi_{\Khat})$ to the individual signal strength  $\kappa$. When $\kappa \gtrsim \sqrt{\log p/n}$ and $\Khat \geq K$, NGCS guarantees an exact recovery of $\cS$. This rate is significantly better than the unsupervised setting, which requires $\kappa \gtrsim n^{-1/4}$ \citep{jin2017phase}.
When $\Khat < K$, deriving a direct relationship between $\tau(\bU)$ and $\kappa$ requires an eigen-gap between the $\Khat$-th eigenvalue and $(\Khat+1)$-th eigenvalue in $\bA$, where the conditions are difficult to verify. Therefore, we do not include it in the discussion, though empirically it still works well; see the supplementary materials \citep{supp}.
\begin{aspt}\label{aspt:latentrandomnew}
Consider a RDPG $\bA\sim RDPG(F)$ where the latent positions $\by_i \sim F$, $i \in [n]$. Let $D$ be the support  of $F$. Suppose the following conditions hold. (i) (Regularized support) $\max_{\by_i \in D}\|\by_i\| = 1$ and $0 \leq \by_i^\top \by_j \leq 1$ for any $\by_i, \by_j\in D$. (ii) (Non-degeneration) The covariance matrix $cov(\by_i)$ has full rank $K$. (iii) (Network density) There exist constants $c, c_d >0$, so that $\by_i^\top \E[F] \geq c$ for all $\by_i \in D$ and $n\rho_n \geq c_d\log n$. 
\end{aspt}
\begin{thm}
\label{thm:lp}
Suppose Assumptions \ref{aspt:indep}--\ref{aspt:rw} and \ref{aspt:latentrandomnew} hold. 
Consider NGCS with $\Khat \geq K$. 
Then there exists a constant $c_{lp} = c_{lp}(c,c_d, K)$, such that with probability $1 - O(1/n)$, $\tau(\bU)^2 \geq n c_{lp}\kappa^2,$  
where $\bU = \bXi(\bA; \Khat)$ or $\bU = \bXi(\bL; \Khat)$. Therefore, when $\kappa \geq c\sqrt{\log p/n}$,  NGCS exactly recovers $\cS$ with high probability. 
\end{thm}
Under RDPG, the required rate $\kappa \gtrsim \sqrt{\log p/n}$ matches the result under DCSBM and does not require knowing $K$ exactly. This further demonstrates the robustness of NGCS.

\subsection{Statistical Lower Bound}\label{sec:lowerbound}
We investigate the statistical lower bound for the covariate selection problem when both $\bA$ and $\bX$ are available. Consider a simplified model for $\bX$, where each $\bM_j$ is nonzero with probability $\epsilon$, independent across $j \in [p]$. In other words,  
\be\label{eqn:lbmodel}
I\{\|\bM_j\| \neq 0\} \sim Bernoulli(\epsilon) \qquad j \in [p]. 
\ee
We treat $\epsilon$ as a constant parameter in the Bernoulli distribution. Since $p\epsilon \to \infty$ as $p \to \infty$, the sparsity level satisfies $|\cS|/p = \epsilon(1+o(1))$ with high probability. Hence, the role of $\epsilon$ is coherent with the notion of sparsity.

\begin{thm} 
\label{thm:lowerbound}
Suppose Assumptions \ref{aspt:indep}--\ref{aspt:rw} and \eqref{eqn:lbmodel} hold. 
For any constants $0 < q_1, q_2 < 1$, suppose the individual signal strength satisfies 
$\max_{j \in \cS}\|\bM_j\| \leq {\sqrt{-2\log (q_1/\epsilon+q_2)}}/{\sqrt{n}}.$
Then any statistical estimator $\cShat(\bX, \bA)$ of $\cS$ fails to control both types of error. In other words, the following inequality cannot both hold for any $\cShat(\bX, \bA)$: 
\begin{equation}
\label{eqn:FDR}
P(j \in \cShat(\bX, \bA)|j \notin \cS)\leq  q_1, \quad
P(j \notin \cShat(\bX, \bA)|j \in \cS)\leq q_2.
\end{equation}
In particular, if $\epsilon=p^{-\beta}$ and we further set $q_1=p^{-1}$ and $q_2=0$ which indicates an exact recovery, then no statistical method will succeed when
$\max_{j \in \cS}\|\bM_j\|\leq {\sqrt{2(1-\beta)\log p/n}}. $
\end{thm}
Theorem \ref{thm:lowerbound} provides the first lower bound for covariate selection when both $\bA$ and $\bX$ are available.
With network information $\bA$, exact recovery of $\cS$ requires that the signal strength in $\bX$ satisfy $\max_{j \in \cS}\|\bM_j\| \geq \sqrt{2(1-\beta)\log p / n}$. 
It matches the supervised learning rate \citep{FanLv} because of the additional information contained in $\bA$. 
This significantly improves upon the unsupervised learning requirement $\kappa \gtrsim n^{-1/4}$, when only $\bX$ is available \citep{clusteringbound, jin2017phase}.

Under the DCSBM and RDPG models, we have shown that NGCS recovers $\cS$ when the signal strength $\kappa \geq c\sqrt{\log p/n}$. This rate $n^{-1/2}$ matches the lower bound in Theorem \ref{thm:lowerbound}. It further demonstrates the optimality of the NGCS approach. 
Although our setting may be viewed as ``semi-supervised", NGCS achieves the supervised learning rate.

\section{Network-Guided Clustering and Regression}\label{sec:application}

\subsection{Two-study Setting}
\label{sec:setting}
We now turn to downstream applications in a multi-study setting. Consider a two-study scenario, in which two independent studies are conducted on a common $p$-dimensional covariate space.
Study~1 contains $n$ subjects, indexed by $i\in [n]$, with covariate matrix $\bX\in\RR^{n\times p}$ and a subject-subject network $\bA\in\RR^{n\times n}$.
Study~2 contains $N-n$ additional subjects, indexed by $i\in[N] \backslash[n]$, with covariate matrix $\bX_{(2)}\in\RR^{(N-n)\times p}$. For regression analysis, we further assume that an
additional response $z_i$ is observed in Study~2, denoted collectively as $\bz\in\RR^{N-n}$. The stacked covariate matrix over both studies is $\tilde{\bX} = \bigl[\bX^\top,  \bX_{(2)}^\top\bigr]^\top \in \RR^{N\times p}$.
Similarly, we denote $\bY$, $\bY_{(2)}$ and $\bYtilde$  the latent factor matrices for Study 1, Study 2 and all $N$ data points, respectively. In summary, given $\bXtilde$ and $\bA$, our goal is to recover the underlying structure for all $N$ subjects. We study both  clustering and regression under this two-study framework.

In the clustering setting, each subject $i \in [N]$ has a latent class label $\ell(i) \in [K]$, where $K$ is known. 
The latent vector $\by_i \in \reals^K$ encodes this label, where $\by_i(k) = 1$ if $\ell(i) = k$ and 0 otherwise. 
Thus, recovering $\bell$ is equivalent to recovering $\bYtilde$. 
High-dimensional clustering approaches applied to $\bXtilde$ \citep{sparsekmeans, IFPCA, arias2017simple} cannot incorporate the informative network $\bA$, while community detection methods \citep{casc, hu2024network} cannot be applied to subjects in Study~2 because $\bA$ is only observed in Study~1.
Our goal is therefore to develop a network-guided clustering approach that utilizes both $\bA$ and $\bXtilde$. 

In the regression setting, only the $N-n$ subjects in Study~2 have an observed response $z_i$. We aim to predict $z_{new}$ for a new subject with covariate data $\bx_{new}$. Suppose $z_i$ follows a linear model with respect to the latent factor $\by_i$, where  
\begin{equation} \label{eqn:regmodel}
    z_i = \bbeta^\top \by_i + \delta_i, \qquad \delta_i \sim \mathcal{N}(0,\sigma^2_{\delta}), \quad i \in [N]/[n].
\end{equation}
Since only $\bx_i$ is observed, this corresponds to a factor regression model \citep{agarwal2009regression, basilevsky2009statistical, fan2024latent}.
Prediction in Study~2 utilizes $\bX_{(2)}$ and $\bz$, but cannot leverage the network $\bA$ or covariates $\bX$ in Study~1. Our aim is to improve prediction of $z_{new}$ by leveraging both $\bX$ and $\bA$.

This two-study configuration reflects common data integration scenarios in statistical learning, where the data may come from heterogeneous sources (e.g. labeled vs. unlabeled cohorts), differ in format, and present varying levels of data quality. Joint analysis improves statistical efficiency and robustness \citep{wang2023distributionally, niu2024covariate}. 
The proposed framework naturally extends to more general multi-source and multi-modal settings.

\subsection{Algorithm: Network-Guided Clustering and Regression}\label{sec:apps}

To leverage both $\bA$ and the full covariate matrix $\bXtilde$ for clustering and regression, a natural strategy is to use the NGCS approach in Section \ref{sec:algo} to recover the informative covariates $\cShat$. 
We propose the Network-Guided clustering (NG-clu) and Network-Guided regression (NG-reg) methods. Both methods the latent structural information encoded in the network into the covariates through a two-stage procedure. In stage (i), we apply NGCS to $(\bA, \bX)$ from Study 1 and obtain the informative set $\cShat$. In stage (ii), we conduct a singular value decomposition (SVD) on the post-selection matrix $\bXtilde^{\cShat} \in \RR^{N \times |\cShat|}$ to extract leading factors for downstream tasks. This framework leverages both $\bA$ and $\bXtilde$ through interpretable covariates, and the additional SVD step further extracts the latent factors. Below we introduce the two approaches in detail.

{\bf Clustering}. 
We first obtain $\cShat$ by applying NGCS to $\bA$ and $\bX$ in Study~1. Given the post-selection data matrix $\bXtilde^{\cShat}$, we embed the data points using its leading singular space \citep{opca}. Denote the SVD as $\bXtilde^{\cShat} = \bU \bLambda \bV^\top$. Let $\bLambda_{\Khat} \in \reals^{\Khat \times \Khat}$ be the diagonal matrix of the largest $\Khat$ singular values, and let $\bU_{\Khat} \in \reals^{N \times \Khat}$ contain the corresponding left singular vectors. Each subject is embedded into a \(\Khat\)-dimensional space using the weighted representation \(\bU_{\Khat}\bLambda_{\Khat} \in \reals^{N \times \Khat}\). Finally, we apply \(k\)-means to \(\bU_{\Khat}\bLambda_{\Khat}\) and obtain the estimated labels $\hat{\bell}$ for all \(N\) subjects. This procedure is summarized in Table~\ref{tab:cluster}.

There is a tuning parameter $\Khat$ in NG-clu. Since the number of clusters is $K$, it is natural to assume the latent factors lie in a $K$-dimensional space and to 
set $\Khat = K$. However, this assumption may be overly optimistic. In our framework, the choice of $\Khat$ used in NGCS and SVD is allowed to differ from $K$ used in the final $k$-means step. For this reason, including the singular values $\bLambda_{\Khat}$ is crucial, as they provide weights reflecting the relative importance of the singular vectors and ensure clustering consistency when $\Khat \geq K$; see Theorem \ref{thm:cluster}. This novel property makes our method notably robust to overestimating the number of clusters, a challenge frequently encountered in applications.

\begin{table}[ht!]
\caption{Network-Guided Clustering (NG-clu) Algorithm} 
\begin{tabular}{ll}
\hline
& \underline{Input}: $\bA$, $\bXtilde$, $\bX$, number of classes $K$, tuning parameter $\Khat$. \\  
\hline
Step 1. & Covariate Selection: Apply NGCS Algorithm \ref{tab:alg1} with $(\bA, \bX)$ to get $\cShat$.\\
& Let $\bXtilde^{\cShat}$ be the sub-matrix of $\bXtilde$ restricted on columns in $\cShat$ only. \\
Step 2. & SVD: Let $\bLambda_{\Khat}$ be the diagonal matrix consisting of the largest $\Khat$ singular \\ 
&values of $\bXtilde^{\cShat}$ and $\bU_{\Khat}$ contain the corresponding left singular vectors.\\
Step 3. & Clustering: Apply $k$-means to $\bU_{\Khat}\bLambda_{\Khat}$ with each row being a data point.\\
Output. & Return the estimated label $
\hat{\bell}$. \\
 \hline
\end{tabular}
\label{tab:cluster} 
\end{table}

{\bf Regression.} In the regression setting, the response $z_i$ can be predicted through $\bx_i$, due to the relationship between $\bx_i$ and $\by_i$ in Assumption \ref{aspt:cov}.  

Recall that $\E[\bx_i|\by_i] = \bM^\top \by_i$. 
Thus, the response satisfies  $\E[z_i|\by_i] = \bgamma^\top \bM^\top \by_i = \bgamma^\top \E[\bx_i|\by_i]$, where $\bgamma$ is any solution to the constraint $\bM \bgamma = \bbeta$. This representation provides the possibility of using $\bx_i$ to estimate $z_i$, and a critical insight: covariates not in $\cS$ contribute negligible information to the response prediction.

Motivated by these insights, we propose the NG-reg algorithm. We begin by applying NGCS to $(\bA, \bX)$ and obtain $\cShat$. Then we perform SVD on the post-selection matrix $\bX_{(2)}^{\cShat} \in \reals^{(N-n) \times |\cShat|}$ in Study~2. The coefficient vector $\bgamma$ is estimated by the spectral projection: 
\begin{equation}
\label{eqn:gammaestimate}
    \hat{\bgamma} = \bV_{\Khat} \bLambda_{\Khat} \bU_{\Khat}^\top \bz.
\end{equation} 
For a new subject with covariate vector $\bx_{new}$, we first restrict it to $\cShat$ and obtain $\bx_{new}^{\cShat}$. Then we predict its response as $\hat{z}_{new} = \hat{\bgamma}^\top\bx_{new}^{\cShat}$. This approach naturally handles the fact that new subjects lack network data, reduces dimensionality through covariate selection and spectral projection, and remains effective even when the observed responses are very few, i.e. $N-n \ll p$. 
Details are in Table \ref{tab:regression}.

\begin{table}[ht!]
\caption{Network-Guided Regression (NG-reg) Algorithm} 
\begin{tabular}{ll}
\hline
& \underline{Input}: $\bA, \bX$, $\bX_{(2)}, \bz$, tuning parameter $\hat{K}$, new sample $\bx_{new}$ \\  
\hline
Step 1. & Covariate Selection: Apply NGCS Algorithm \ref{tab:alg1} on $(\bA,\bX)$ to get $\cShat$.\\
& Let $\bX^{\cShat}_{(2)}$ be the sub-matrix of $\bX_{(2)}$ restricted on columns in $\cShat$ only. \\
Step 2. & SVD: Let $\bLambda_{\Khat}$ consist of the largest $\Khat$ singular values of $\bX_{(2)}^{\cShat}$. Let $\bU_\Khat$ and \\
&$\bV_\Khat$ contain the corresponding left and right singular vectors, respectively.\\
Step 3. & Regression: Let $\hat{\bgamma} = \bV_{\Khat} \bLambda_{\Khat} \bU_{\Khat}^\top \bz$. \\
Output: & Return the estimate $
\hat{z}_{i} = \hat{\bgamma}^\top \bx_{i}^{\cShat}$ for $i \in [n]$ and $
\hat{z}_{new} = \hat{\bgamma}^\top \bx_{new}^{\cShat}$ for new points. \\
 \hline
\end{tabular}
\label{tab:regression} 
\end{table}

\subsection{Clustering Consistency under Degree-Corrected SBM}\label{sec:sbm}

In this section, we establish the theoretical guarantee for NG-clu under DCSBM. Because DCSBM naturally exhibits a clustering structure, it serves as a convenient model for illustrating our results. We emphasize, however, that the derivation is not restricted to DCSBM; for other network models under which NGCS succeeds, clustering consistency can be established using similar arguments.

Recall the two-study setting in Section~\ref{sec:setting}. With a known number of clusters $K$, our goal is to recover the label vector $\bell$. 
Let $\bM_j$ denote the $j$-th column of the loading matrix $\bM$, and let $\bmu_k$ denote the $k$-th row of $\bM$. Recall that the individual signal strength is $\kappa = \min_{j \in \cS} \|\bM_j\|$. Let $s = |\cS|$ and $\cshat = |\cShat|$. 
Since $\by_i$ is fully decided by $\ell(i)$, we update Assumption~\ref{aspt:cov} to Assumption~\ref{aspt:covcluster}. 

\begin{aspt}\label{aspt:covcluster}
    Let $\bM \in \reals^{K \times p}$ be the covariate loading matrix and $\bmu_k$ be the $k$-th row of $\bM$, then the covariates follow 
    $\bx_i|[\ell(i) = k] \sim \mathcal{N}(\bmu_k, \bI), \quad i \in [N], j \in [p].$
\end{aspt}
Unlike covariate selection, clustering relies on the overall signal strength. We therefore impose Assumption~\ref{aspt:cluster} on $\bM$ about the signal strength of $\|\bmu_k\|$. Since the individual signal strength is bounded by $\kappa$, we reasonably take $\sqrt{s}\kappa$ as the bound for $\|\bmu_k\|$ and $\|\bmu_k - \bmu_j\|$.
\begin{aspt}\label{aspt:cluster}
Suppose $rank(\bM)=r\leq K$, and there is a constant $c_M > 0$ such that  
 \[
\lambda_r(\bM)\geq c_M \|\bM\|, \quad 
\|\bmu_k\|\geq  c_M\sqrt{s}\kappa ,\quad \|\bmu_k-\bmu_j\|\geq c_M\sqrt{s}\kappa, \quad 1 \leq k \neq j \leq K.
\]
\end{aspt}
Assumption~\ref{aspt:cluster} allows the loading matrix $\bM$ to have a rank of $r < K$, i.e., $\bM$ is rank-deficient. This flexibility substantially widens the applicability of the model. This flexibility is enabled by the fact that NGCS allows an inflated choice of $\Khat$. Under these assumptions, we establish the clustering consistency of NG-clu in Theorem \ref{thm:cluster}.
\begin{thm}\label{thm:cluster}
Suppose Assumptions \ref{aspt:indep}, \ref{aspt:rw}--\ref{aspt:dcsbm}, and \ref{aspt:covcluster}--\ref{aspt:cluster} hold, and $\epsilon = p^{-\beta}$ for $0 < \beta < 1$. If the signal strength $\kappa > \sqrt{\max\{16-16\beta, 14\}\log p/(c_Dn)}$, and 
$\kappa>\frac{6\sqrt{K}}{\rho^{3/2} c^2_M} \frac{\sqrt{\cshat} + \sqrt{N\log^2p}}{\sqrt{Ns}}$, 
then with high probability, the clustering error by Algorithm \ref{tab:cluster} satisfies
 \[
 Err= \frac{1}{N}\min_{permutation\; \pi: [K]\to[K]}I\{\pi(\hat{\ell}(i)) \neq \ell(i)\}\leq \frac{(\cshat+N\log^2p)}{2Ns\kappa^2}.
 \]
 In particular, if $\kappa^2>(\cshat+N\log^2p)/(Ns)$, then there are no misclassified nodes. 
\end{thm}

According to Corollary \ref{cor:main} and Theorem \ref{thm:dcsbm}, when $\kappa \geq \sqrt{(c_{\beta}/c_D) \log p/n}$, the number of selected covariates satisfies $\cshat \leq s(1 + C\log^2 p/s)$ with high probability. Combining this with Theorem \ref{thm:cluster}, NG-clu achieves strong clustering consistency when $\kappa \geq C\sqrt{\log p}\max\{1/\sqrt{n}, 1/\sqrt{s}\}$ for a sufficiently large $p$. 
The term $1/\sqrt{n}$ corresponds to the requirement for successful covariate selection, and $1/\sqrt{s}$ corresponds to the requirement for successful clustering. This matches the statistical bounds in \cite{jin2017phase}, and it bridges the gap between the statistical bound and the computational tractable bound there. 

\subsection{Prediction Error under the RDPG model}\label{sec:latent}

In this section, we establish the prediction accuracy of NG-reg under an RDPG network, where each node has a low-dimensional latent position governing network connections \citep{RDPGfixadj, RDPGtutorial, LatentSpec}. Although we focus on RDPG here for concreteness, the arguments extends to other network models under which NGCS succeeds.

Recall the two-study setting in Section~\ref{sec:setting}. Our goal is to predict $\E[z|\bx]$. To analyze the prediction error, we impose that the loading matrix $\bM$ has full rank with sufficiently large eigenvalues. We first establish the error bound for the estimator $\gammahat$ in Theorem \ref{thm:regression},  assuming that the informative covariates are exactly recovered.

\begin{aspt}\label{aspt:regression}
Suppose $rank(\bM) = K$, and $\lambda_K(\bM)\geq c_M \|\bM\|$ for a constant $c_M > 0$.
\end{aspt}
\begin{thm}\label{thm:regression}
Consider the post-selection covariate matrix $\bX_{(2)}^{\cShat} \in \reals^{(N-n) \times |\cShat|}$ for Study~2 with $\cShat \supset  \cS$. Let $s = |\cS|$ and $\cshat = |\cShat|$. 
Suppose Assumption \ref{aspt:regression} holds and $\kappa >3(\sqrt{N-n}+\sqrt{\cshat})/\sqrt{(N-n)s}$. 
Then for any new data point $\bx_0$ with latent factor $\by_0$, with probability $1 - O(1/n)$, for a constant $C>0$, $\gammahat$ defined in 
\eqref{eqn:gammaestimate} follows that
\[
|\gammahat^\top \bx^{\cShat}_{0}-\E[z|\by_0]|
\leq C\|\bbeta\|\frac{\sqrt{N-n}+\sqrt{\cshat}}{\kappa\sqrt{(N-n)s}}
+ \frac{C\sigma_{\delta}}{\sqrt{N-n}}.
\]
\end{thm}
This error bound consists of two components. The first term is determined by the underlying regression structure, i.e., the coefficients characterizing $\E[z|\by_0]$. The second term relects the noise $\delta_i$ in the response vector. 

Our NG-reg approach suggests to use NGCS to estimate $\cShat$. Combining Corollary \ref{cor:main} and Theorem \ref{thm:regression} yields the prediction error bound for NG-clu in Corollary \ref{cor:NGreg} below. Even without responses, Study~1 significantly improves prediction by reducing it into a low-dimensional regression problem. 
\begin{cor}\label{cor:NGreg}
    Consider the model \eqref{eqn:regmodel} for $\bz$ and RDPG($F$) for the network, where Assumptions \ref{aspt:indep}--\ref{aspt:rw}, \ref{aspt:latentrandomnew}, and \ref{aspt:regression} hold. When $\kappa \gg C(1/\sqrt{N-n} + 1/\sqrt{s})$, the prediction error of the NG-reg approach goes to $\sigma_{\delta}/\sqrt{N-n}$, at the same order as the ordinary linear regression on $N-n$ data points. 
\end{cor}

\section{Simulation}\label{sec:simulation}

We compare our proposed NGCS, NG-clu, and NG-reg approaches with other methods on synthetic datasets. We cross three network models with three noise families to obtain nine simulation scenarios.
For all settings, we take $N=1000$ subjects in total ($n=800$, $N-n=200$), $p=1200$ covariates, and $|\cS|=50$ informative coordinates.

Here are the three network models and their parameter setups: (i) DCSBM ($K=3$): class labels $\ell(i)\sim\mathrm{Unif}([K])$, one-hot $\by_i$, and $A_{ij}\sim\mathrm{Bernoulli}\!\left(\theta_i\theta_j\,\by_i^\top\bB\by_j\right)$ with $\theta_i\sim\text{Exp}(5)+0.06$ and $\bB$ having $1/2$ on the diagonal and $1/4$ off-diagonal; (ii) DCMM ($K=3$): class labels $\ell(i)\sim\mathrm{Unif}([K])$, $Y_{ik} \sim I\{\ell(i)=k\}+{\rm Unif}(0,0.3)$ first and then normalize each row as $\by_i = \by_i/\|\by_i\|_1$, with edges as in DCSBM; (iii) RDPG ($K=10$): $\by_i\sim{\cal N}(0.2\mathbf 1,\bSigma)$ where $\bSigma$ is block-diagonal (unit diagonals, within-block correlations $\mathrm{Unif}(0,1)$), and $A_{ij}\sim\mathrm{Bernoulli}\!\left(0.01\,\by_i^\top \by_j\right)$.

Covariates follow $\bX_j=\bY\bM_j+\bm\epsilon$. The latent factors $\bM$ are set as follows: for $j\notin\cS$, $\bM_j=\mathbf 0$; for $j\in\cS$, in DCSBM/DCMM take $M_{ij}\sim \tfrac12{\cal N}(\mu,0.05^2)+\tfrac12{\cal N}(-\mu,0.05^2)$, while in RDPG take $M_{ij}\sim \tfrac12\,{\rm Unif}(0.05,\mu)+\tfrac12\,{\rm Unif}(-\mu,-0.05)$. For the noise $\bveps$, three settings are considered:  (a) Gaussian ${\cal N}(0,1)$; (b) Wilson--Hilferty transformation of $\chi^2_5$ (centered and scaled); (c) mixing sub-Gaussian: first, for each covariate, choose the distribution from \{Rademacher$(-1, 1)$, Unif($-\sqrt{3}, \sqrt{3}$), Bernoulli(0.5, centered and scaled), $0.02\delta_{-5} + 0.02\delta_{5} + 0.96\delta_{0}$\} with $\delta_a$ being the delta function, then generate $\bx_i$ independently, with $X_{ij}$ following the selected distribution of covariate $j$.  In addition, to compare NG-reg with classical regression, we additionally set $z_i=\balpha^\top \by_i+\delta_i$ for 200 samples in Study~2 for the RDPG network, where $\balpha\sim{\cal N}(\mathbf 0,\bI)$ and $\delta_i\sim{\cal N}(0,0.5)$.

The three network models and three noise distributions give us 9 scenarios. For each scenario, we measure the reliability of covariate selection methods using the false discovery rate (FDR). A lower FDR indicates a more reliable selection of covariates. We consider the following methods: 1) NGCS and its variants, including NGCS with $\bXi(\bA, K)$ and HCT (NGCS-HCT-A), NGCS with $\bXi(\bL, K)$ (NGCS-HCT-L), NGCS with a hard threshold that keeps covariates with the smallest 50 $p$-values (NGCS-50), and NGCS with HW-$\chi^2$ $p$-values (NGCS-HW); 
2) Ranking methods based on $\bX$, including the marginal chi-square statistics (Chi) in \cite{IFPCA} and sparse $k$-means (SKmeans) in \cite{sparsekmeans}. We keep 50 covariates with top rankings.

\begin{figure}
    \centering
    \includegraphics[width=0.85\linewidth]{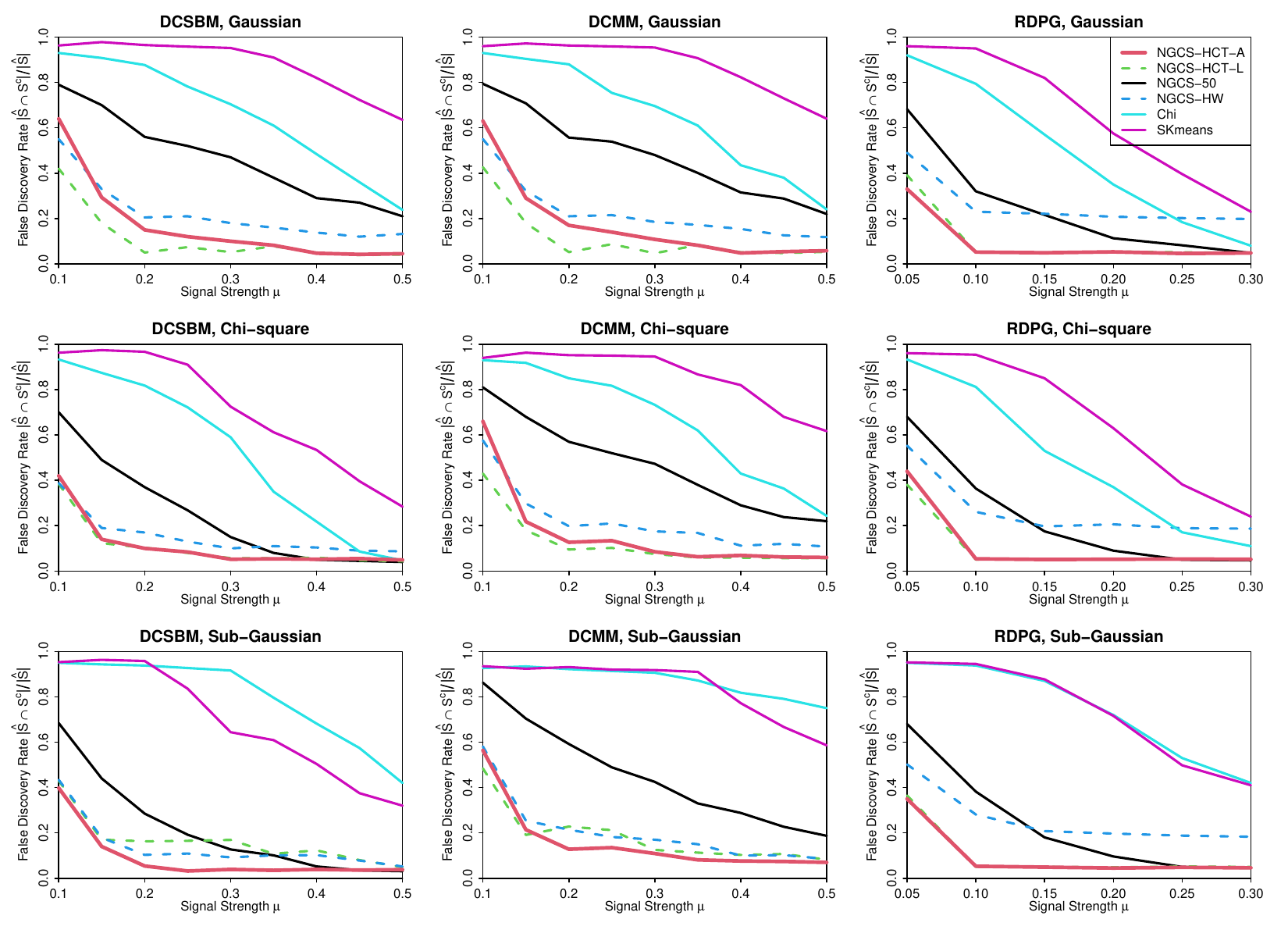}
    \caption{Summary plots for FDRs under different network and noise settings.}
    \label{fig:fdr}
\end{figure}

Figure~\ref{fig:fdr} reports FDR versus the signal strength parameter $\mu$ ($[0.1,0.5]$ for DCSBM/DCMM; $[0.05,0.3]$ for RDPG), each with 50 repetitions. There are several findings. First, our NGCS algorithms consistently outperform other approaches, especially when $\mu$ is small. It proves the importance of integrating data from other sources. Furthermore, NGCS-HCT algorithms, where the threshold is data-driven, outperform NGCS-50 when signals are weak. For strong signals, all NGCS algorithms perform very well. It demonstrates the optimality of HC threshold in numerical analysis. Second, our NGCS algorithms are robust to combinations of network models and covariate distributions. Third, despite the conservativeness of HW-$\chi^2$ $p$-values, NGCS-HW remains comparable to the best performers under Gaussian noise.

We then explore the clustering task, where our goal is to identify $\bell\in\reals^{1000}$ under DCSBM-leading settings. We consider two sets of clustering methods: 1) NG-clu algorithm with $\bA$ and $K$ (NG-clu), and with $\bA$ and $2K$ (NG-clu($2K$)); 2) $\bX$-based high-dimensional clustering methods, including spectral clustering (Spec), influential covariates PCA (IF-PCA) in \cite{IFPCA}, Sparse K-means (SKmeans) in \cite{sparsekmeans}, and Sparse Alternate Sum clustering (SAS) in \cite{arias2017simple}. The network-based methods cannot be applied here, because $\bA$ is not available for Study 2. Table~\ref{tab:combined-results} summarizes the average clustering error rates over 50 repetitions, where our proposed NG-clu algorithm consistently outperforms other methods, especially when signal is weak. Meanwhile, our method is robust to slight inflation of $\hat K$, as verified by great performance for NG-clu($2K$).

The regression task is explored under the RDPG network and three covariate distributions, where our goal is to predict the responses $\bz$. We compare our proposed NG-reg and NG-reg($2K$) algorithm with popular high-dimensional regression approaches, including: penalized high-dimensional regression with Lasso (Lasso), Minimax Concave Penalty (MCP), and Smoothly Clipped Absolute Deviation (SCAD) penalty, Principal Component Regression (PCR in \cite{liu2003principal}), and high-dimensional regression using correlation-adjusted marginal correlation (CAR in \cite{zuber2011high}).  The mean squared error between the estimation $\hat{\bz}$ and $\bY\balpha$ is used to assess the performance.  
The average errors among 50 repetitions are presented in Table~\ref{tab:combined-results}. It shows that our NG-reg algorithm outperforms all the other methods, with larger improvements for a smaller $\mu$. This highlights that enhancing data quality by additional network information is crucial for complex data.

\begin{table}[htbp]
\centering
\resizebox{0.9\textwidth}{!}{
\setlength{\tabcolsep}{6pt}
\renewcommand{\arraystretch}{0.6}
\setlength{\aboverulesep}{0pt}
\setlength{\belowrulesep}{0pt}
\setlength{\cmidrulesep}{0pt}
\begin{tabular}{llccc|ccc|ccc}
\toprule
& & \multicolumn{9}{c}{Noise Family} \\
\cmidrule(lr){3-11}
& & \multicolumn{3}{c}{(a)} & \multicolumn{3}{c}{(b)} & \multicolumn{3}{c}{(c)} \\
\midrule
\multicolumn{2}{l}{\textbf{Clustering}} & $\mu = 0.1$ & $\mu = 0.3$ & $\mu = 0.5$ & $\mu = 0.1$ & $\mu = 0.3$ & $\mu = 0.5$ & $\mu = 0.1$ & $\mu = 0.3$ & $\mu = 0.5$\\
\midrule
Spec        &       & 0.5670 & 0.1808 & 0.0170 & 0.5700 & 0.2274 & 0.0284 & 0.5886 & 0.2122 & 0.0318  \\
SKmeans     &       & 0.6500 & 0.5286 & 0.2040 & 0.6264 & 0.5266 & 0.3710 & 0.6492 & 0.5150 & 0.2170 \\
SAS         &       & 0.6354 & 0.0606 & 0.0068 & 0.6224 & 0.3948 & 0.0100 & 0.6440 & 0.6314 & 0.0144 \\
IF-PCA      &       & 0.4776 & 0.1144 & 0.0058 & 0.4926 & 0.1394 & 0.0126 & 0.5888 & 0.1758 & 0.0194  \\
NG-clu ($2K$) &       & 0.3138 & 0.0654 & 0.0070 & 0.2848 & 0.0836 & 0.0108 & 0.2642 & 0.0852 & 0.0140 \\
NG-clu      &       & 0.2656 & 0.0662 & 0.0058 & 0.2676 & 0.0834 & 0.0102 & 0.2610 & 0.0836 & 0.0150  \\
\midrule
\multicolumn{2}{l}{\textbf{Regression}} & $\mu = 0.5$ & $\mu = 1.0$ & $\mu = 2.0$ & $\mu = 0.5$ & $\mu = 1.0$ & $\mu = 2.0$ & $\mu = 0.5$ & $\mu = 1.0$ & $\mu = 2.0$\\
\midrule
Lasso       &       &  1.1014 & 0.8069 & 0.6761 & 1.4315 & 0.9609 & 0.7590 & 1.3013 & 0.8780 & 0.6989 \\
MCP         &       &   1.2733 & 0.8926 & 0.6604 & 1.5439 & 0.9978 & 0.7980 & 1.5300 & 1.0210 & 0.7429 \\
SCAD        &       &      1.3009 & 0.8990 & 0.6829 & 1.4814 & 1.0482 & 0.8029 & 1.5168 & 1.0080 & 0.7309 \\
CAR         &       &     1.0459 & 0.7553 & 0.6345 & 1.4449 & 0.9023 & 0.6783 & 1.3322 & 0.8603 & 0.6615 \\
PCR         &       &    1.2210 & 0.6836 & 0.5459 & 1.9125 & 0.9925 & 0.6253 & 1.8024 & 0.8753 & 0.5823 \\
NG-reg ($2K$) &       &    0.8162 & 0.6030 & 0.5374 & 1.0356 & 0.6889 & 0.5733 & 1.0175 & 0.6717 & 0.5601 \\
NG-reg      &       &     0.8205 & 0.6027 & 0.5376 & 1.0355 & 0.6882 & 0.5728 & 1.0160 & 0.6723 & 0.5601 \\
\bottomrule
\end{tabular}
}
\caption{Top: Clustering error rate across noise setups and signal strengths. Bottom: Mean squared error (MSE) across the same settings.}
\label{tab:combined-results}
\end{table}

\section{Sina$^{\text{TM}}$ Dataset}\label{sec:adddata} 

Sina-microblog website is the largest social platform in China, where each user can follow a list of users to read their new microblogs in a timely manner. 
\cite{jia2017node} extracted the follower--followee network with covariates for thousands of users from this website. In the network, each node represents a user, and each directed edge $(i,j)$ represents user $i$ following user $j$. 
We amplified the covariate vector to be $\bx_i \in \reals^{3000}$, where 10 out of 3000 are informative covariates and the others are noise. The 10 informative covariates give the user's interests on $K = 10$ topics. Code details can be found at \url{https://tinyurl.com/NGCSandRelated}.

\begin{figure}[!htbp]
    \centering
    \includegraphics[width=1.0\textwidth]{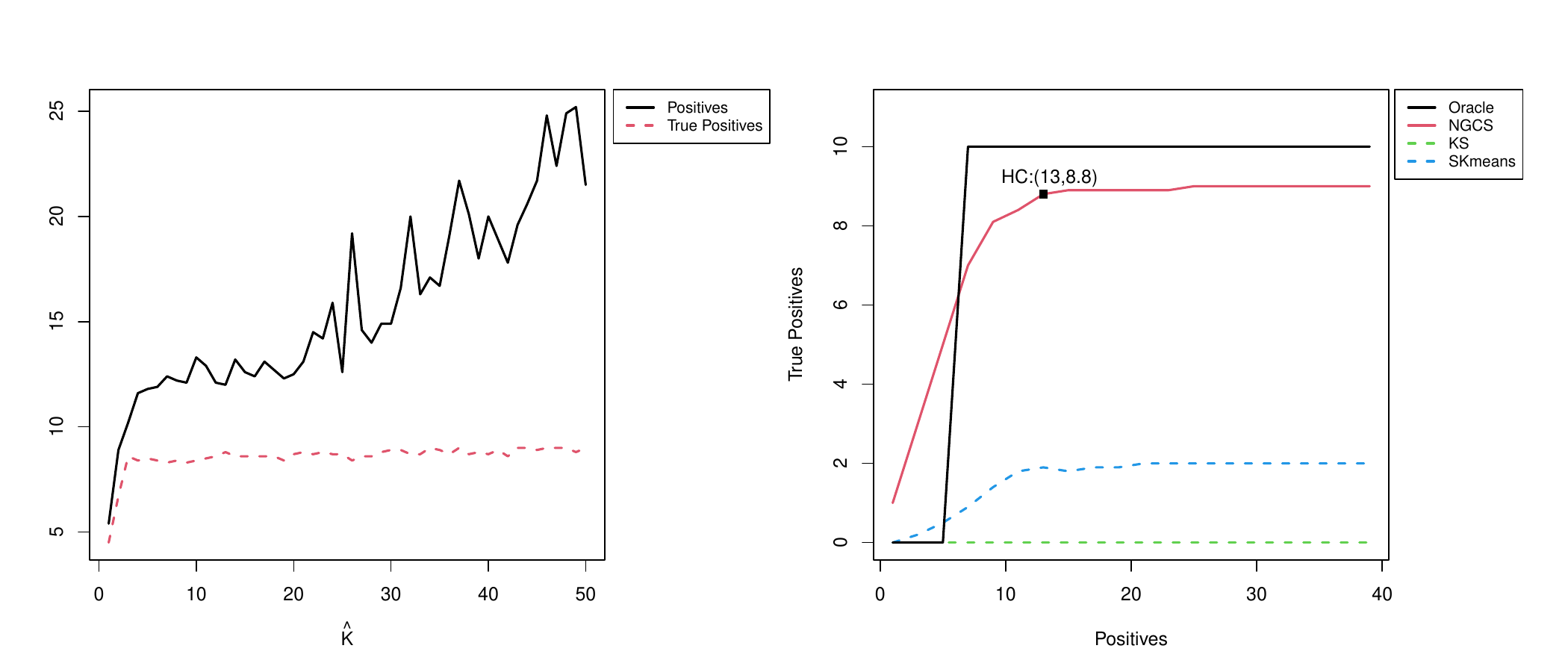}
    \caption{Covariate selection results.}
    \label{fig:sinase}
\end{figure}

Our first goal is to recover the 10 informative covariates. 
We first investigate NGCS with tuning parameter $\Khat \in [1, 50]$. 
For each possible $\Khat$, we randomly select $n = 2000$ users and apply NGCS with $\bA$ and $\Khat$. The average over 50 repetitions is summarized in Figure \ref{fig:sinase}. 
When $5 \leq \Khat \leq 20$, NGCS stably recovers 8 or 9 out of 10 informative covariates, with the number of all selected covariates $|\cShat| \leq 13$. Even with a large $\Khat = 50$, the recovered set $\cShat$ has a relatively small size of $\leq 25$. This demonstrates the robustness of NGCS with respect to the choice of $\Khat$. 
Fixing $\Khat = 10$, we compare NGCS with the oracle case where $Y$ is known, and with $\bX$-based methods including Sparse K-means (SKmeans) and the marginal Kolmogorov-Smirnov statistics (KS) used in \cite{IFPCA}. 
Our NGCS algorithm outperforms other methods, and HC chooses almost the best threshold.

We then investigate the regression problem. 
Given an informative covariate set $\cS$ with 9 covariates, 
the response is given by $z_i = 1 - \sum_{j \in \cS}X_{ij} + \cN(0, 0.5^2)$.
We take $n_1 = 2000$ users with $\bA$ and $n_2 \in \{100, 150, \cdots, 500\}$ users with $\bz$. 
The covariates $\bx_i \in \reals^{3000}$ are available for all $n_1 + n_2$ users. 
The goal is to estimate $\E[z_i|\bx_i]$ for $i \in [n_1]$. We compare our NG-reg algorithm with the penalized high-dimensional regression methods on the $n_2$ users, including Lasso, MCP, and SCAD. 
The root mean squared error (RMSE = $\sqrt{\sum_{i=1}^{n_1}(\hat z_i -  X_{ik})^2/n_1}$) on the $n_1$ users is used to evaluate the regression result, as summarized in Table~\ref{sinatable}.
 
In Table~\ref{sinatable}, the supervised high-dimensional methods on the $n_2$ users can hardly identify any informative covariates and the RMSE is always around 0.2 due to the large noise. Our method successfully identifies 8 informative covariates and gives a more accurate prediction of $\bz$, especially for large $n_2$. In the supplementary materials, we discuss the combination of NGCS and high-dimensional regression methods. 

\begin{table}[htbp]
\centering
\resizebox{\textwidth}{!}{
\begin{tabular}{l|lll|lll|lll|lll}
\toprule
\textbf{}   &\multicolumn{3}{c}{\underline{\textbf{NG-reg}}} & \multicolumn{3}{c}{\underline{\textbf{Lasso}}} & \multicolumn{3}{c}{\underline{\textbf{MCP}}}  & \multicolumn{3}{c}{\underline{\textbf{SCAD}}}    \\
 $n_2$ & RMSE & $|\cShat|$  & $|\cShat \cap \cS|$  & RMSE & $|\cShat|$  & $|\cShat \cap \cS|$ & RMSE & $|\cShat|$  & $|\cShat \cap \cS|$ & RMSE & $|\cShat|$  & $|\cShat \cap \cS|$ \\
\midrule
$100$ & 0.25 & 11.9 (1.79) & 8 (0.00) & 0.23 & 12.7 (23.30) & 0.2 (0.42) & 0.27 & 4.5 (7.21) & 0.2 (0.42) & 0.23 & 10.5 (16.39) & 0.2 (0.42)\\
$150$ & 0.21 & 12.5 (4.01) & 8 (0.00) & 0.20 & 14.0 (21.75) & 0.3 (0.67) &0.21 & 5.9 (6.37)& 0.3 (0.67) & 0.20 & 9.3 (16.91) & 0.5 (0.85)\\
$200$ & 0.17 & 12.2 (2.25) & 8 (0.00) & 0.20 & 2.4 (5.15) & 0.1 (0.32) & 0.20 & 1.4 (2.17)& 0.1 (0.32) & 0.19 & 3.7 (4.99) & 0.3 (0.48)\\
$250$ & 0.14 & 10.3 (1.16) & 8 (0.00) & 0.22 & 43.2 (51.47) & 0.2 (0.42) & 0.19 & 3.2 (5.79) & 0.0 (0.00) & 0.20 & 20.7 (28.23) & 0.0 (0.00)\\ 
$300$ & 0.13 & 11.0 (2.10) & 8 (0.00) & 0.19 & 11.5 (22.18) & 0.6 (0.70) & 0.19 & 5.4 (8.42)& 0.5 (0.53) & 0.20 & 13.9 (21.89) & 0.5 (0.53)\\
$350$ & 0.13 & 11.3 (1.64) & 8 (0.00) & 0.19 & 3.0 (7.15) & 0.2 (0.42) & 0.19 & 2.6 (4.86)& 0.2 (0.42) & 0.19 & 9.5 (16.66) & 0.4 (0.52)\\
$400$ & 0.11 & 12.5 (2.84) & 8 (0.00) & 0.19 & 12.9 (19.02) & 0.5 (0.97) & 0.19 & 4.6 (7.49)& 0.4 (0.70) & 0.19 & 5.1 (8.71) & 0.4 (0.70)\\
$450$ & 0.11 & 11.3 (1.77) & 8 (0.00) & 0.19 & 6.6 (7.82) &  0.7 (0.95) & 0.19 & 3.1 (4.63) & 0.5 (0.71) & 0.19 & 5.6 (6.69) & 0.6 (0.84)\\
$500$ & 0.11 & 12.2 (2.53) & 8 (0.00) & 0.19 & 10.5 (10.83) & 0.8 (0.79) & 0.19 & 6.7 (6.88) & 0.7 (0.82) & 0.19 & 13.5 (13.00) & 1.0 (1.05)\\
\bottomrule
\end{tabular}
}
\caption{Summary table for regression.}
\label{sinatable}
\end{table}

\section{Discussion}\label{sec:disc}
This work establishes a new framework for integrating heterogeneous data sources by leveraging partial information as guidance for covariate selection. Unlike conventional supervised approaches that require fully observed responses, our method exploits the network structure as an auxiliary signal, enabling effective selection of informative covariates even when exact responses are unavailable. This semi-supervised perspective highlights that high-quality but incomplete information can serve as a powerful surrogate for supervision. The principle is broadly adaptable, and future research may extend it to other multimodal datasets, such as combinations of manifold data with covariates.

In this framework, we focus on the linear assumption to demonstrate theoretical optimality. However, the underlying idea is not restricted to linearity. A natural extension is to allow arbitrary dependence between $\bX$ and $\bY$, such that $\E[\bX_j] = f(\bY)$. Along this line, researchers may start with non-parametric regression $\bX_j \sim f(\bY) + \bveps$ to find the test statistic $t_j(\bX_j, \bY)$ for the hypothesis $f(\cdot) = 0$. The next step is to extract partial information about $\bY$ from auxiliary structures such as networks, so that $t_j(\bX_j, \bY)$ enjoys similar performance to $t_j(\bX_j, \bA)$. An interesting discovery is that we only need  partial information from $\bA$, not the exact information.

\bibliographystyle{chicago}
\bibliography{network}

\begin{thebibliography}{}

\bibitem[\protect\citeauthoryear{Agarwal and Chen}{Agarwal and
  Chen}{2009}]{agarwal2009regression}
Agarwal, D. and B.-C. Chen (2009).
\newblock Regression-based latent factor models.
\newblock In {\em Proceedings of the 15th ACM SIGKDD international conference
  on Knowledge discovery and data mining}, pp.\  19--28.

\bibitem[\protect\citeauthoryear{Alex, Erd{\H{o}}s, Knowles, Yau, and Yin}{Alex
  et~al.}{2014}]{alex2014isotropic}
Alex, B., L.~Erd{\H{o}}s, A.~Knowles, H.-T. Yau, and J.~Yin (2014).
\newblock Isotropic local laws for sample covariance and generalized wigner
  matrices.
\newblock {\em Electronic Journal of Probability\/}~{\em 19}, 1--53.

\bibitem[\protect\citeauthoryear{Arias-Castro and Pu}{Arias-Castro and
  Pu}{2017}]{arias2017simple}
Arias-Castro, E. and X.~Pu (2017).
\newblock A simple approach to sparse clustering.
\newblock {\em Computational Statistics \& Data Analysis\/}~{\em 105},
  217--228.

\bibitem[\protect\citeauthoryear{Athreya, Fishkind, Tang, Priebe, Park,
  Vogelstein, Levin, Lyzinski, and Qin}{Athreya et~al.}{2017}]{RDPGtutorial}
Athreya, A., D.~E. Fishkind, M.~Tang, C.~E. Priebe, Y.~Park, J.~T. Vogelstein,
  K.~Levin, V.~Lyzinski, and Y.~Qin (2017).
\newblock Statistical inference on random dot product graphs: a survey.
\newblock {\em The Journal of Machine Learning Research\/}~{\em 18\/}(1),
  8393--8484.

\bibitem[\protect\citeauthoryear{Azizyan, Singh, and Wasserman}{Azizyan
  et~al.}{2013}]{clusteringbound}
Azizyan, M., A.~Singh, and L.~Wasserman (2013).
\newblock Minimax theory for high-dimensional gaussian mixtures with sparse
  mean separation.
\newblock {\em Advances in Neural Information Processing Systems\/}~{\em 26},
  2139--2147.

\bibitem[\protect\citeauthoryear{Baltru{\v{s}}aitis, Ahuja, and
  Morency}{Baltru{\v{s}}aitis et~al.}{2018}]{baltruvsaitis2018multimodal}
Baltru{\v{s}}aitis, T., C.~Ahuja, and L.-P. Morency (2018).
\newblock Multimodal machine learning: A survey and taxonomy.
\newblock {\em IEEE transactions on pattern analysis and machine
  intelligence\/}~{\em 41\/}(2), 423--443.

\bibitem[\protect\citeauthoryear{Bandeira and Van~Handel}{Bandeira and
  Van~Handel}{2016}]{bandeira2016sharp}
Bandeira, A.~S. and R.~Van~Handel (2016).
\newblock Sharp nonasymptotic bounds on the norm of random matrices with
  independent entries.
\newblock {\em The Annals of Probability\/}~{\em 44\/}(4), 2479--2506.

\bibitem[\protect\citeauthoryear{Basilevsky}{Basilevsky}{2009}]{basilevsky2009statistical}
Basilevsky, A.~T. (2009).
\newblock {\em Statistical factor analysis and related methods: theory and
  applications}.
\newblock John Wiley \& Sons.

\bibitem[\protect\citeauthoryear{Bickel and Chen}{Bickel and
  Chen}{2009}]{DCBickel}
Bickel, P.~J. and A.~Chen (2009).
\newblock A nonparametric view of network models and newman--girvan and other
  modularities.
\newblock {\em Proceedings of the National Academy of Sciences\/}~{\em
  106\/}(50), 21068--21073.

\bibitem[\protect\citeauthoryear{Binkiewicz, Vogelstein, and Rohe}{Binkiewicz
  et~al.}{2017}]{casc}
Binkiewicz, N., J.~T. Vogelstein, and K.~Rohe (2017).
\newblock Covariate-assisted spectral clustering.
\newblock {\em Biometrika\/}~{\em 104\/}(2), 361--377.

\bibitem[\protect\citeauthoryear{Chaudhuri, Chung, and Tsiatas}{Chaudhuri
  et~al.}{2012}]{opca}
Chaudhuri, K., F.~Chung, and A.~Tsiatas (2012).
\newblock Spectral clustering of graphs with general degrees in the extended
  planted partition model.
\newblock In {\em Conference on Learning Theory}, pp.\  35--1. JMLR Workshop
  and Conference Proceedings.

\bibitem[\protect\citeauthoryear{Chernoff}{Chernoff}{1952}]{chernoff}
Chernoff, H. (1952).
\newblock A measure of asymptotic efficiency for tests of a hypothesis based on
  the sum of observations.
\newblock {\em The Annals of Mathematical Statistics\/}, 493--507.

\bibitem[\protect\citeauthoryear{Curran and Hussong}{Curran and
  Hussong}{2009}]{curran2009integrative}
Curran, P.~J. and A.~M. Hussong (2009).
\newblock Integrative data analysis: the simultaneous analysis of multiple data
  sets.
\newblock {\em Psychological methods\/}~{\em 14\/}(2), 81.

\bibitem[\protect\citeauthoryear{Davis and Kahan}{Davis and
  Kahan}{1970}]{daviskahan}
Davis, C. and W.~M. Kahan (1970).
\newblock The rotation of eigenvectors by a perturbation. iii.
\newblock {\em SIAM Journal on Numerical Analysis\/}~{\em 7\/}(1), 1--46.

\bibitem[\protect\citeauthoryear{Donoho and Jin}{Donoho and Jin}{2008}]{DJ08}
Donoho, D. and J.~Jin (2008).
\newblock Higher criticism thresholding: Optimal feature selection when useful
  features are rare and weak.
\newblock {\em Proceedings of the National Academy of Sciences\/}~{\em
  105\/}(39), 14790--14795.

\bibitem[\protect\citeauthoryear{Donoho}{Donoho}{2000}]{donoho2000}
Donoho, D.~L. (2000).
\newblock High-dimensional data analysis: The curses and blessings of
  dimensionality.
\newblock {\em AMS Math Challenges Lecture\/}~{\em 1\/}(2000), 32.

\bibitem[\protect\citeauthoryear{Donoho and Jin}{Donoho and Jin}{2015}]{DJ14}
Donoho, D.~L. and J.~Jin (2015).
\newblock Higher criticism for large-scale inference, especially for rare and
  weak effects.
\newblock {\em Statistical Science\/}~{\em 30\/}(1), 1--25.

\bibitem[\protect\citeauthoryear{Fan, Liao, and Mincheva}{Fan
  et~al.}{2013}]{fan2013large}
Fan, J., Y.~Liao, and M.~Mincheva (2013).
\newblock Large covariance estimation by thresholding principal orthogonal
  complements.
\newblock {\em Journal of the Royal Statistical Society Series B\/}~{\em
  75\/}(4), 603--680.

\bibitem[\protect\citeauthoryear{Fan, Lou, and Yu}{Fan
  et~al.}{2024}]{fan2024latent}
Fan, J., Z.~Lou, and M.~Yu (2024).
\newblock Are latent factor regression and sparse regression adequate?
\newblock {\em Journal of the American Statistical Association\/}~{\em
  119\/}(546), 1076--1088.

\bibitem[\protect\citeauthoryear{Fan and Lv}{Fan and Lv}{2008}]{FanLv}
Fan, J. and J.~Lv (2008).
\newblock Sure independence screening for ultrahigh dimensional feature space.
\newblock {\em Journal of the Royal Statistical Society Series B\/}~{\em
  70\/}(5), 849--911.

\bibitem[\protect\citeauthoryear{Feng, Jiang, Hannig, and Marron}{Feng
  et~al.}{2018}]{feng2018angle}
Feng, Q., M.~Jiang, J.~Hannig, and J.~Marron (2018).
\newblock Angle-based joint and individual variation explained.
\newblock {\em Journal of multivariate analysis\/}~{\em 166}, 241--265.

\bibitem[\protect\citeauthoryear{Friedman and Meulman}{Friedman and
  Meulman}{2004}]{cosa}
Friedman, J.~H. and J.~J. Meulman (2004).
\newblock Clustering objects on subsets of attributes (with discussion).
\newblock {\em Journal of the Royal Statistical Society Series B\/}~{\em
  66\/}(4), 815--849.

\bibitem[\protect\citeauthoryear{Gu and Han}{Gu and Han}{2011}]{ACMfs1}
Gu, Q. and J.~Han (2011).
\newblock Towards feature selection in network.
\newblock In {\em Proceedings of the 20th ACM International Conference on
  Information and Knowledge Management}, pp.\  1175--1184.

\bibitem[\protect\citeauthoryear{Hoff, Raftery, and Handcock}{Hoff
  et~al.}{2002}]{hoff2002latent}
Hoff, P.~D., A.~E. Raftery, and M.~S. Handcock (2002).
\newblock Latent space approaches to social network analysis.
\newblock {\em Journal of the American Statistical Association\/}~{\em
  97\/}(460), 1090--1098.

\bibitem[\protect\citeauthoryear{Hofree, Shen, Carter, Gross, and
  Ideker}{Hofree et~al.}{2013}]{hofree2013network}
Hofree, M., J.~P. Shen, H.~Carter, A.~Gross, and T.~Ideker (2013).
\newblock Network-based stratification of tumor mutations.
\newblock {\em Nature methods\/}~{\em 10\/}(11), 1108--1115.

\bibitem[\protect\citeauthoryear{Hu and Wang}{Hu and
  Wang}{2024}]{hu2024network}
Hu, Y. and W.~Wang (2024).
\newblock Network-adjusted covariates for community detection.
\newblock {\em Biometrika\/}~{\em 111\/}(4), 1221–1240.

\bibitem[\protect\citeauthoryear{Janitza, Tutz, and Boulesteix}{Janitza
  et~al.}{2016}]{janitza2016random}
Janitza, S., G.~Tutz, and A.-L. Boulesteix (2016).
\newblock Random forest for ordinal responses: prediction and variable
  selection.
\newblock {\em Computational Statistics \& Data Analysis\/}~{\em 96}, 57--73.

\bibitem[\protect\citeauthoryear{Jia, Li, Carson, Wang, and Yu}{Jia
  et~al.}{2017}]{jia2017node}
Jia, C., Y.~Li, M.~B. Carson, X.~Wang, and J.~Yu (2017).
\newblock Node attribute-enhanced community detection in complex networks.
\newblock {\em Scientific Reports\/}~{\em 7\/}(1), 2626.

\bibitem[\protect\citeauthoryear{Jiang, Armour, Hu, Mei, Tian, Sharpton, and
  Jiang}{Jiang et~al.}{2019}]{jiang2019microbiome}
Jiang, D., C.~R. Armour, C.~Hu, M.~Mei, C.~Tian, T.~J. Sharpton, and Y.~Jiang
  (2019).
\newblock Microbiome multi-omics network analysis: statistical considerations,
  limitations, and opportunities.
\newblock {\em Frontiers in genetics\/}~{\em 10}, 995.

\bibitem[\protect\citeauthoryear{Jin}{Jin}{2015}]{SCORE}
Jin, J. (2015).
\newblock Fast community detection by score.
\newblock {\em The Annals of Statistics\/}~{\em 43\/}(1), 57--89.

\bibitem[\protect\citeauthoryear{Jin, Ke, and Luo}{Jin
  et~al.}{2017}]{mixedscore}
Jin, J., Z.~T. Ke, and S.~Luo (2017).
\newblock Estimating network memberships by simplex vertex hunting.
\newblock {\em arXiv preprint arXiv:1708.07852\/}~{\em 12}.

\bibitem[\protect\citeauthoryear{Jin, Ke, and Wang}{Jin
  et~al.}{2017}]{jin2017phase}
Jin, J., Z.~T. Ke, and W.~Wang (2017).
\newblock Phase transitions for high dimensional clustering and related
  problems.
\newblock {\em The Annals of Statistics\/}~{\em 45\/}(5), 2151--2189.

\bibitem[\protect\citeauthoryear{Jin and Wang}{Jin and Wang}{2016}]{IFPCA}
Jin, J. and W.~Wang (2016).
\newblock Influential features {PCA} for high dimensional clustering.
\newblock {\em The Annals of Statistics\/}~{\em 44\/}(6), 2323--2359.

\bibitem[\protect\citeauthoryear{Joseph and Yu}{Joseph and
  Yu}{2016}]{joseph2016impact}
Joseph, A. and B.~Yu (2016).
\newblock Impact of regularization on spectral clustering.
\newblock {\em The Annals of Statistics\/}~{\em 44\/}(4), 1765--1791.

\bibitem[\protect\citeauthoryear{Laurent and Massart}{Laurent and
  Massart}{2000}]{laurent2000adaptive}
Laurent, B. and P.~Massart (2000).
\newblock Adaptive estimation of a quadratic functional by model selection.
\newblock {\em The Annals of Statistics\/}, 1302--1338.

\bibitem[\protect\citeauthoryear{Lazega}{Lazega}{2001}]{lazega2001collegial}
Lazega, E. (2001).
\newblock {\em The collegial phenomenon: The social mechanisms of cooperation
  among peers in a corporate law partnership}.
\newblock OUP Oxford.

\bibitem[\protect\citeauthoryear{Li and Li}{Li and Li}{2008}]{li2008network}
Li, C. and H.~Li (2008).
\newblock Network-constrained regularization and variable selection for
  analysis of genomic data.
\newblock {\em Bioinformatics\/}~{\em 24\/}(9), 1175--1182.

\bibitem[\protect\citeauthoryear{Liu, Kuang, Gong, and Hou}{Liu
  et~al.}{2003}]{liu2003principal}
Liu, R., J.~Kuang, Q.~Gong, and X.~Hou (2003).
\newblock Principal component regression analysis with spss.
\newblock {\em Computer Methods and Programs in Biomedicine\/}~{\em 71\/}(2),
  141--147.

\bibitem[\protect\citeauthoryear{Lyzinski, Sussman, Tang, Athreya, and
  Priebe}{Lyzinski et~al.}{2014}]{RDPGfixadj}
Lyzinski, V., D.~L. Sussman, M.~Tang, A.~Athreya, and C.~E. Priebe (2014).
\newblock Perfect clustering for stochastic blockmodel graphs via adjacency
  spectral embedding.
\newblock {\em Electronic Journal of Statistics\/}~{\em 8}, 2905--2922.

\bibitem[\protect\citeauthoryear{Mao, Sarkar, and Chakrabarti}{Mao
  et~al.}{2021}]{mao2021estimating}
Mao, X., P.~Sarkar, and D.~Chakrabarti (2021).
\newblock Estimating mixed memberships with sharp eigenvector deviations.
\newblock {\em Journal of the American Statistical Association\/}~{\em
  116\/}(536), 1928--1940.

\bibitem[\protect\citeauthoryear{Niu, Ni, Pati, and Mallick}{Niu
  et~al.}{2024}]{niu2024covariate}
Niu, Y., Y.~Ni, D.~Pati, and B.~K. Mallick (2024).
\newblock Covariate-assisted bayesian graph learning for heterogeneous data.
\newblock {\em Journal of the American Statistical Association\/}~{\em
  119\/}(547), 1985--1999.

\bibitem[\protect\citeauthoryear{O'Malley and Christakis}{O'Malley and
  Christakis}{2011}]{o2011longitudinal}
O'Malley, A.~J. and N.~A. Christakis (2011).
\newblock Longitudinal analysis of large social networks: Estimating the effect
  of health traits on changes in friendship ties.
\newblock {\em Statistics in medicine\/}~{\em 30\/}(9), 950--964.

\bibitem[\protect\citeauthoryear{Pardo}{Pardo}{2018}]{pardo2018statistical}
Pardo, L. (2018).
\newblock {\em Statistical inference based on divergence measures}.
\newblock Chapman and Hall/CRC.

\bibitem[\protect\citeauthoryear{Patil and Parmigiani}{Patil and
  Parmigiani}{2018}]{patil2018training}
Patil, P. and G.~Parmigiani (2018).
\newblock Training replicable predictors in multiple studies.
\newblock {\em Proceedings of the National Academy of Sciences\/}~{\em
  115\/}(11), 2578--2583.

\bibitem[\protect\citeauthoryear{Puggini and McLoone}{Puggini and
  McLoone}{2017}]{puggini2017forward}
Puggini, L. and S.~McLoone (2017).
\newblock Forward selection component analysis: Algorithms and applications.
\newblock {\em IEEE Transactions on Pattern Analysis and Machine
  Intelligence\/}~{\em 39\/}(12), 2395--2408.

\bibitem[\protect\citeauthoryear{Raftery and Dean}{Raftery and
  Dean}{2006}]{EM1}
Raftery, A.~E. and N.~Dean (2006).
\newblock Variable selection for model-based clustering.
\newblock {\em Journal of the American Statistical Association\/}~{\em
  101\/}(473), 168--178.

\bibitem[\protect\citeauthoryear{Rubin-Delanchy, Cape, Tang, and
  Priebe}{Rubin-Delanchy et~al.}{2022}]{LatentSpec}
Rubin-Delanchy, P., J.~Cape, M.~Tang, and C.~E. Priebe (2022).
\newblock {A statistical interpretation of spectral embedding: The generalised
  random dot product graph}.
\newblock {\em Journal of the Royal Statistical Society Series B\/}~{\em
  84\/}(4), 1446--1473.

\bibitem[\protect\citeauthoryear{Rudelson and Vershynin}{Rudelson and
  Vershynin}{2013}]{hw}
Rudelson, M. and R.~Vershynin (2013).
\newblock {Hanson--Wright inequality and sub-{G}aussian concentration}.
\newblock {\em Electronic Communications in Probability\/}~{\em 18\/}(none), 1
  -- 9.

\bibitem[\protect\citeauthoryear{Shen and Wang}{Shen and Wang}{2025}]{supp}
Shen, T. and W.~Wang (2025).
\newblock Supplementary material for ``optimal network-guided covariate
  selection for high-dimensional data integration".

\bibitem[\protect\citeauthoryear{Tibshirani}{Tibshirani}{1996}]{tibshirani1996regression}
Tibshirani, R. (1996).
\newblock Regression shrinkage and selection via the lasso.
\newblock {\em Journal of the Royal Statistical Society Series B\/}~{\em
  58\/}(1), 267--288.

\bibitem[\protect\citeauthoryear{Tong, Wang, and Wang}{Tong
  et~al.}{2025}]{denoising}
Tong, X.~T., W.~Wang, and Y.~Wang (2025).
\newblock Uniform error bound for {PCA} matrix denoising.
\newblock {\em Bernoulli\/}~{\em 31\/}(3), 2251--2275.

\bibitem[\protect\citeauthoryear{Vershynin}{Vershynin}{2010}]{vershynin2010introduction}
Vershynin, R. (2010).
\newblock Introduction to the non-asymptotic analysis of random matrices.
\newblock {\em arXiv preprint arXiv:1011.3027\/}.

\bibitem[\protect\citeauthoryear{Wang and Chen}{Wang and
  Chen}{2021}]{wang2021network}
Wang, J.-H. and Y.-H. Chen (2021).
\newblock Network-adjusted kendall’s tau measure for feature screening with
  application to high-dimensional survival genomic data.
\newblock {\em Bioinformatics\/}~{\em 37\/}(15), 2150--2156.

\bibitem[\protect\citeauthoryear{Wang, Zhu, Huang, and Wang}{Wang
  et~al.}{2025}]{Wang2025refine}
Wang, L., Y.~Zhu, D.~Huang, and W.~Wang (2025).
\newblock Dual-stage network-enhanced feature screening.
\newblock {\em Manuscript\/}.

\bibitem[\protect\citeauthoryear{Wang, B{\"u}hlmann, and Guo}{Wang
  et~al.}{2023}]{wang2023distributionally}
Wang, Z., P.~B{\"u}hlmann, and Z.~Guo (2023).
\newblock Distributionally robust machine learning with multi-source data.
\newblock {\em arXiv preprint arXiv:2309.02211\/}.

\bibitem[\protect\citeauthoryear{Witten and Tibshirani}{Witten and
  Tibshirani}{2010}]{sparsekmeans}
Witten, D.~M. and R.~Tibshirani (2010).
\newblock A framework for feature selection in clustering.
\newblock {\em Journal of the American Statistical Association\/}~{\em
  105\/}(490), 713--726.

\bibitem[\protect\citeauthoryear{Wu, Zhu, and Feng}{Wu
  et~al.}{2018}]{wu2018network}
Wu, M., L.~Zhu, and X.~Feng (2018).
\newblock Network-based feature screening with applications to genome data.
\newblock {\em The Annals of Applied Statistics\/}~{\em 12\/}(2), 1250--1270.

\bibitem[\protect\citeauthoryear{Zhang}{Zhang}{2010}]{zhang2010nearly}
Zhang, C.-H. (2010).
\newblock Nearly unbiased variable selection under minimax concave penalty.
\newblock {\em The Annals of Statistics\/}~{\em 38\/}(2), 894--942.

\bibitem[\protect\citeauthoryear{Zhao, Wang, and Li}{Zhao
  et~al.}{2025}]{SpecMultiNet}
Zhao, D., W.~Wang, and J.~Li (2025).
\newblock Spectral clustering on aggregated multilayer networks with
  covariates.
\newblock {\em Statistics and Computing\/}~{\em 35}, 118.

\bibitem[\protect\citeauthoryear{Zhao, Liu, Wang, and Leng}{Zhao
  et~al.}{2022}]{zhao2022dimension}
Zhao, J., X.~Liu, H.~Wang, and C.~Leng (2022).
\newblock Dimension reduction for covariates in network data.
\newblock {\em Biometrika\/}~{\em 109\/}(1), 85--102.

\bibitem[\protect\citeauthoryear{Zhao, Levina, and Zhu}{Zhao
  et~al.}{2012}]{DCzhuji}
Zhao, Y., E.~Levina, and J.~Zhu (2012).
\newblock Consistency of community detection in networks under degree-corrected
  stochastic block models.
\newblock {\em The Annals of Statistics\/}~{\em 40\/}(4), 2266--2292.

\bibitem[\protect\citeauthoryear{Zhu, Chang, Li, and Wang}{Zhu
  et~al.}{2019}]{zhu2019portal}
Zhu, X., X.~Chang, R.~Li, and H.~Wang (2019).
\newblock Portal nodes screening for large scale social networks.
\newblock {\em Journal of Econometrics\/}~{\em 209\/}(2), 145--157.

\bibitem[\protect\citeauthoryear{Zuber and Strimmer}{Zuber and
  Strimmer}{2011}]{zuber2011high}
Zuber, V. and K.~Strimmer (2011).
\newblock High-dimensional regression and variable selection using car scores.
\newblock {\em Statistical Applications in Genetics and Molecular
  Biology\/}~{\em 10\/}(1), 1.

\end{thebibliography}

\newpage
\appendix
\begin{center}
  \LARGE\bfseries Supplementary Materials
\end{center}
\vspace{1em}
\section{Additional Simulation Studies}\label{sec:addsimulation}
In this section, we first detail some noise settings introduced in the simulation overview in Section~\ref{ex:noise}, and then present additional studies in the remaining. Section~\ref{ex:testing} evaluates the effectiveness of the NGCS testing step. Sections~\ref{ex:dcsbm}–\ref{ex:rdpg} report extended experiments on the three network models and downstream applications across more targeted parameter regimes and objectives.

\subsection{Setups of Covariate Distributions}
\label{ex:noise}
In the simulations, we consider three covariate distributions, each with a distinct noise-generation mechanism. The mixture sub-Gaussian design was only sketched in the main text; here we provide complete specifications for this case.
\begin{itemize}
    \item Rademacher$(-1, 1)$: it takes values $-1$ and $1$ with probability $1/2$ each. It has mean $0$ and variance $1$.
    \item Unif($-\sqrt{3}, \sqrt{3}$): it's uniformly distributed on $[-\sqrt{3}, \sqrt{3}]$ with mean $0$ and variance $1$.
    \item Bernoulli(0.5, centered and scaled): it takes values $0$ and $1$ with probability $1/2$ each. Then it is centered and scaled to have mean $0$ and variance $1$.
    \item $0.02\delta_{-5} + 0.02\delta_{5} + 0.96\delta_{0}$: it takes values 
    $-5,0,5$ with probabilities $0.02,0.96,0.02$, respectively. It has mean $0$ and variance $1$.
\end{itemize}

\subsection{Experiments on the Testing Step}
\label{ex:testing}
We conduct an experiment on the DCSBM model. The dimension of the latent space is $K = 3$. Let class labels $\ell(i) \in [K]$ with equal probability. The latent factors $\by_i \in \reals^3$, with $\ell(i)$-th element being 1 and others are 0. 
The network $A_{ij} \sim \text{Bernoulli}(\theta_i \theta_j \by_i^\top \bB \by_j)$, where $\theta_i \sim \text{Exp}(5)+0.06$ independently and $\bB$ has $1$ on diagonals and $a$ on off-diagonals. 
The covariates $\bX_j = \bY\bM_j + \mathcal{N}(0, \bI)$ for $j \in [p]$. Here, $\bM_j = 0$ for $j \notin \cS$ and $M_{ij} \sim \frac12 \mathcal{N}(0.07, 0.05^2) + \frac12 \mathcal{N}(-0.07, 0.05^2)$ for $j \in \cS$. Other parameter setups remain the same as the simulation section. 
We vary the network-informativeness parameter 
$a \in [0.1, 0.9]$ to examine its impact on the FDR of covariate selection.

\begin{figure}[htbp]
    \centering
    \includegraphics[width=0.8\linewidth]{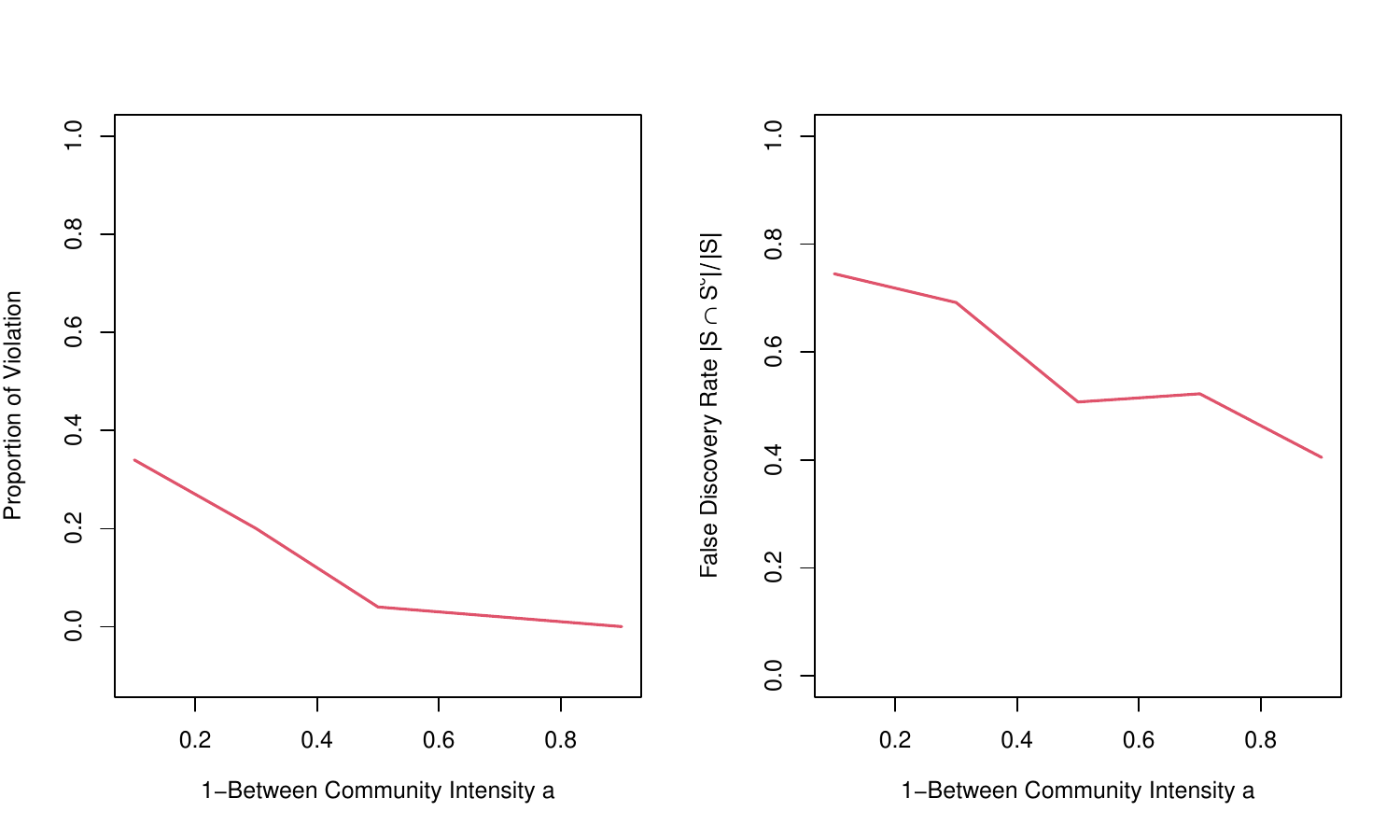}
    \caption{Summary of experimental results for verifying the testing step.}
    \label{fig:testing}
\end{figure}

As shown in Figure~\ref{fig:testing}, the red line traces the FDR of NGCS-HCT-A among repetitions without rejection in the testing step from the simulation section. Under such an extremely weak-signal regime ($\mu = 0.07$), NGCS still works when the network is informative (small $a$). As $a$ increases, the informativeness of the network decreases, having less connectivity signal between the three communities, so the FDR increases. Meanwhile, from the right panel, we could find that the number of repetitions where the maximum HC score fall below the testing threshold $\sqrt{2\log \log p}$ increases when $a$ inclines, underscoring the value of the testing step for diagnosing whether a given network can aid selection.

\subsection{Experiment on the Effect of Network Assortativity}
\label{ex:dcsbm}
In the following four experiment sections, we let $n = 1000$, $p = 1200$, and $|\cS| = 50$ informative covariates. The dimension of the latent space is $K = 3$.

We consider DCSBM. Let class labels $\ell(i) \in [K]$ with equal probability. The latent factors $\by_i \in \reals^3$, with $\ell(i)$-th element being 1 and others are 0. 
The network $A_{ij} \sim \text{Bernoulli}(\theta_i \theta_j \by_i^\top \bB \by_j)$, where $\theta_i \sim |\mathcal{N}(0.1, 0.2)|$ independently and $\bB$ has $\sin^2a$ on diagonals and $(\cos^2a)/2$ on off-diagonals. 
The covariates $\bX_j = \bY\bM_j + \mathcal{N}(0, \bI)$ for $j \in [p]$. Here, $\bM_j = 0$ for $j \notin \cS$ and $M_{ij} \sim \frac12 \mathcal{N}(0.3, 0.05^2) + \frac12 \mathcal{N}(-0.3, 0.05^2)$ for $j \in \cS$. 
We let $a \in [\pi/4,5\pi/4]$ to explore the effects of network assortativity and fixing $a = \pi/4$ to discuss the covariate selection and clustering problem.

\begin{figure}[htbp!]
    \centering
\includegraphics[width = 1\textwidth]{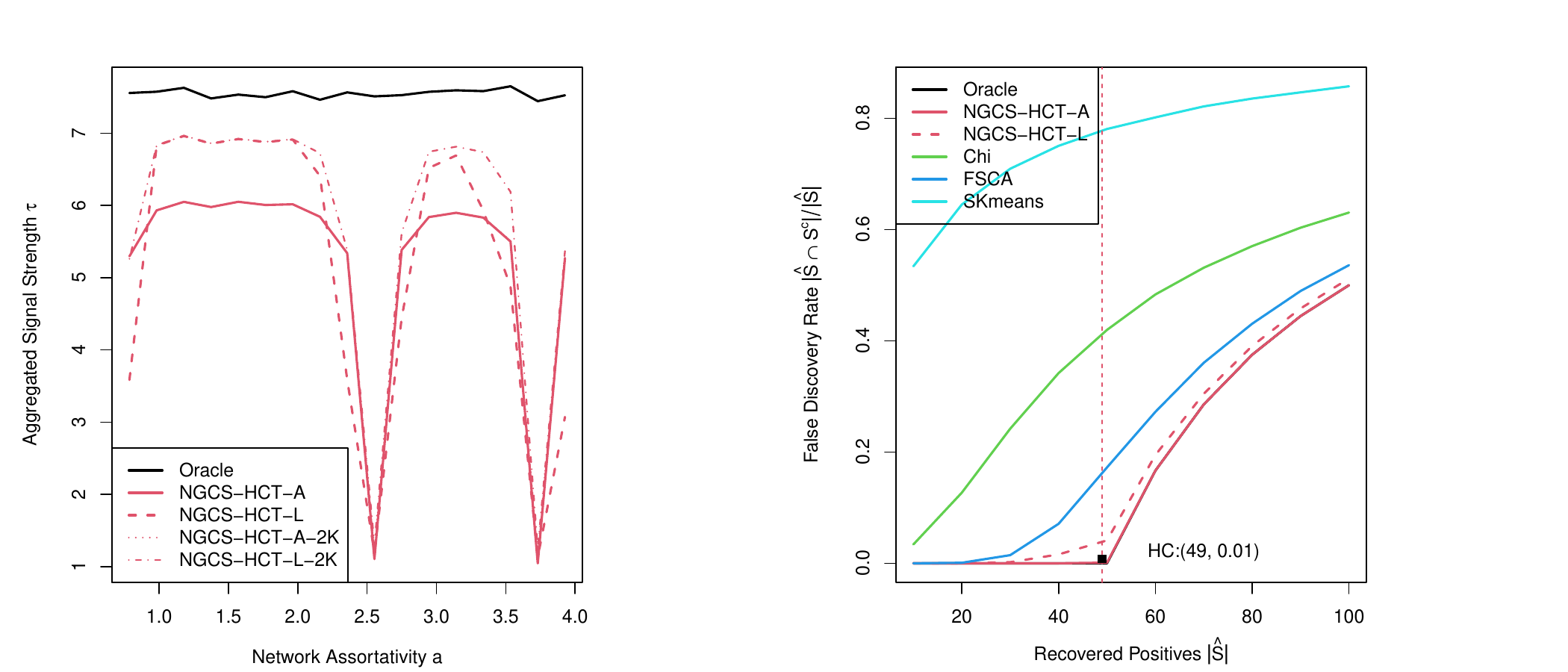}
    \caption{Summary of experimental results under DCSBM.}
    \label{fig:exp1}
\end{figure}

The left panel of Figure~\ref{fig:exp1} provides the aggregated signal strength $\signal(\bXi)$. 
In the oracle case where  $\bY$ is known, we take $\bXi = \bXi(\bY)$, denoted as Oracle. We also consider $\bXi$ from the network: $\bXi(\bA, K)$, $\bXi(\bL, K)$, $\bXi(\bA, 2K)$ and $\bXi(\bL, 2K)$. The signal strength for NGCS is close to the oracle case, except the case $\bA$ is not informative. In addition, with $\Khat \in \{K, 2K\}$, NGCS gives a consistent $\signal(\bXi)$ for all experiments, which proves the robustness with respect to $\Khat$.

We then examine FDRs for four sets of covariate selection methods: 1) NGCS with $\bY$ given (Oracle); 2) NGCS using HCT with input $(\bA, K)$ and $(\bL, K)$; 3) ranking methods based on $\bX$, including the marginal chi-square statistics (Chi), forward selection component analysis \citep{puggini2017forward} (FSCA), and sparse $k$-means \citep{sparsekmeans} (SKmeans). 
Since most methods only give a ranking of covariates, we examine the FDR when the number of selected positives $|\cShat|$ changes from 5 to 100. As can be seen from the right panel of Figure~\ref{fig:exp1}, our NGCS algorithms always outperform other approaches. Specifically, HC always yields an almost-perfect selection of threshold, with FDR around 0.05. The results demonstrate that the inclusion of network information effectively transforms the unsupervised learning problem into a supervised learning task.

\subsection{Experiment on the Effect of Mixed Membership}
Let $n = 1000$, $p = 1200$, and $|\cS| = 50$ informative covariates. The dimension of the latent space is $K = 3$.

We consider DCMM. Let $\ell(i) \in [K]$ with equal probability. Generate ${Y}_{ik} = I\{\bell(i) = k\} + \text{Unif}(0, h)$ and normalize each row as $\by_i = \by_i/\|\by_i\|_1$. The parameter $h$ denotes the mixture level. With these latent factors, the network follows $A_{ij} \sim Bernoulli(\theta_i\theta_j \by_i^\top \bB \by_j)$ where $\theta_i \sim |\mathcal{N}(0.2, 0.3)|$ and $\bB = 0.6 \bI + 0.4 {\bf 11^\top}$. 
The generation of $\bM$ and $\bX$ is the same as Experiment 1. 
We let $h \in [0,1]$ to examine the robustness of NGCS under varying levels of mixture, and fix $h = 0.3$ to explore the covariate selection. 

\begin{figure}[htbp!]
    \centering
\includegraphics[width = 1\textwidth]{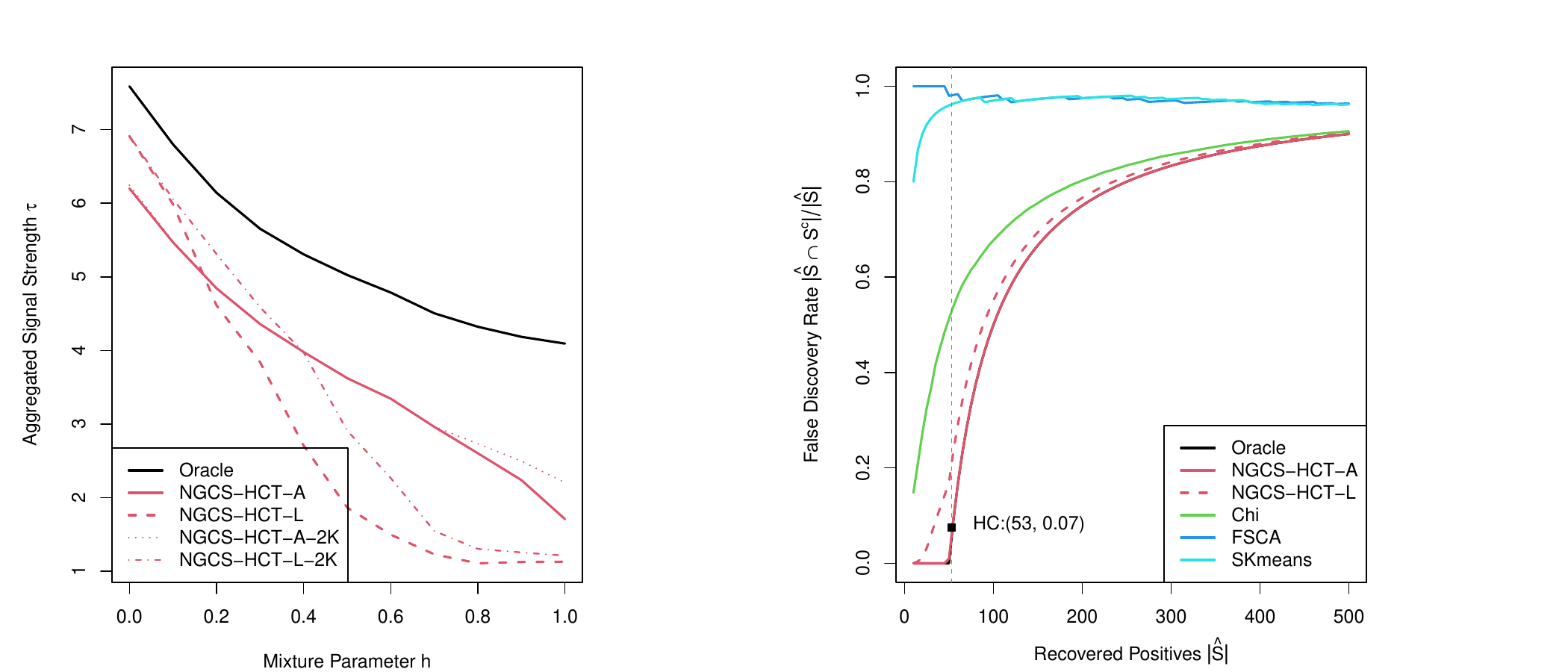}
    \caption{Summary of experimental results under DCMM.}
    \label{fig:exp2}
\end{figure}
Figure \ref{fig:exp2} gives the aggregated signal strength $\signal(\bXi)$ and the false discovery rate (FDR) over 50 repetitions. Similar with what we have found in the last experiment, here with $\Khat \in \{K, 2K\}$, NGCS also gives a consistent $\signal(\bXi)$ for all experiments. An interesting phenomenon is that using $\bA$ gives a stronger signal than using $\bL$. 
In the right figure, our NGCS algorithms always outperform other approaches while HC yields an almost-perfect selection of threshold.

\subsection{Experiment on the Effect of Signal Strength}
\label{ex:rdpg}
Let $n = 1000$, $p = 1200$, and $|\cS| = 50$ informative covariates. The dimension of the latent space is $K = 10$.

We consider RDPG. Let the latent factor $\by_i \sim \mathcal{N}(2*{\bf 1}, \bSigma)$, where $\bSigma$ is a block-diagonal matrix. Each block has diagonals as 1 and off-diagonals as $\text{Unif}(0,1)$ random variables.
The network is thus generated by $A_{ij} \sim Bernoulli(0.01\by_i^\top \by_j)$ independently. 
The covariate $\bX$ is the same with the first experiment, with $M_{ij} \sim \frac{1}{2} \text{Unif}(0.05, \mu) + \frac{1}{2} \text{Unif}(-\mu, -0.05)$ if $j \in \cS$ and 0 otherwise. 
Specifically, to investigate the power of our network-guided algorithm versus supervised learning, we generate a coefficient vector $\balpha \sim \mathcal{N}({\bf 0}, \bI)$ and a response variable for 200 samples, $z_i = \balpha^\top \by_i + \delta_i$, where $\delta_i \sim \mathcal{N}(0, 0.5)$. 
Let loading effects parameter $\mu \in [0.05, 1]$ to examine the aggregated signal strength and then fix $\mu = 0.3$ for comparison of covariate selection methods.

\begin{figure}[htbp!]
    \centering
\includegraphics[width = 1\textwidth]{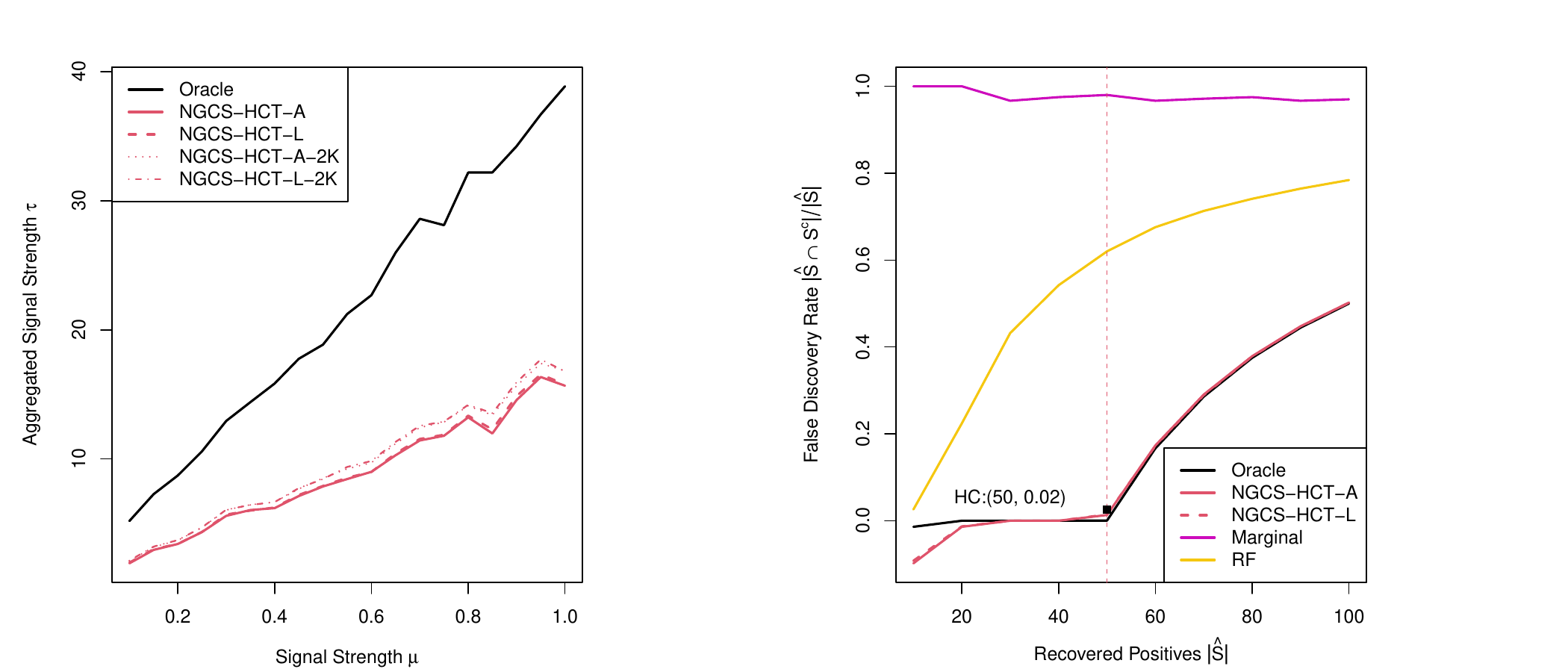}
    \caption{Summary of experimental results under RDPG.}
    \label{fig:exp3}
\end{figure}

Figure \ref{fig:exp3} gives the aggregated signal strength $\signal(\bXi)$ and the false discovery rate (FDR) over 50 repetitions. Similarly to what we have found in the last two experiments, with $\Khat \in \{K, 2K\}$, NGCS could give a consistent $\signal(\bXi)$. As for the results of FDRs, here we also consider some ranking covariate selection methods based on $\bX$ and the response $\bz$, including the marginal statistics (Marginal), and the accuracy decrease impact using random forests \citep{janitza2016random} (RF). From the records, our NGCS algorithms always outperform other approaches, even in the presence of $\bz$, and HC yields an almost-perfect selection of threshold.

\subsection{Downstream Applications on Varying Dimensions and Signal Strengths}

\begin{figure}[htbp!]
    \centering
\includegraphics[width = 1\textwidth]{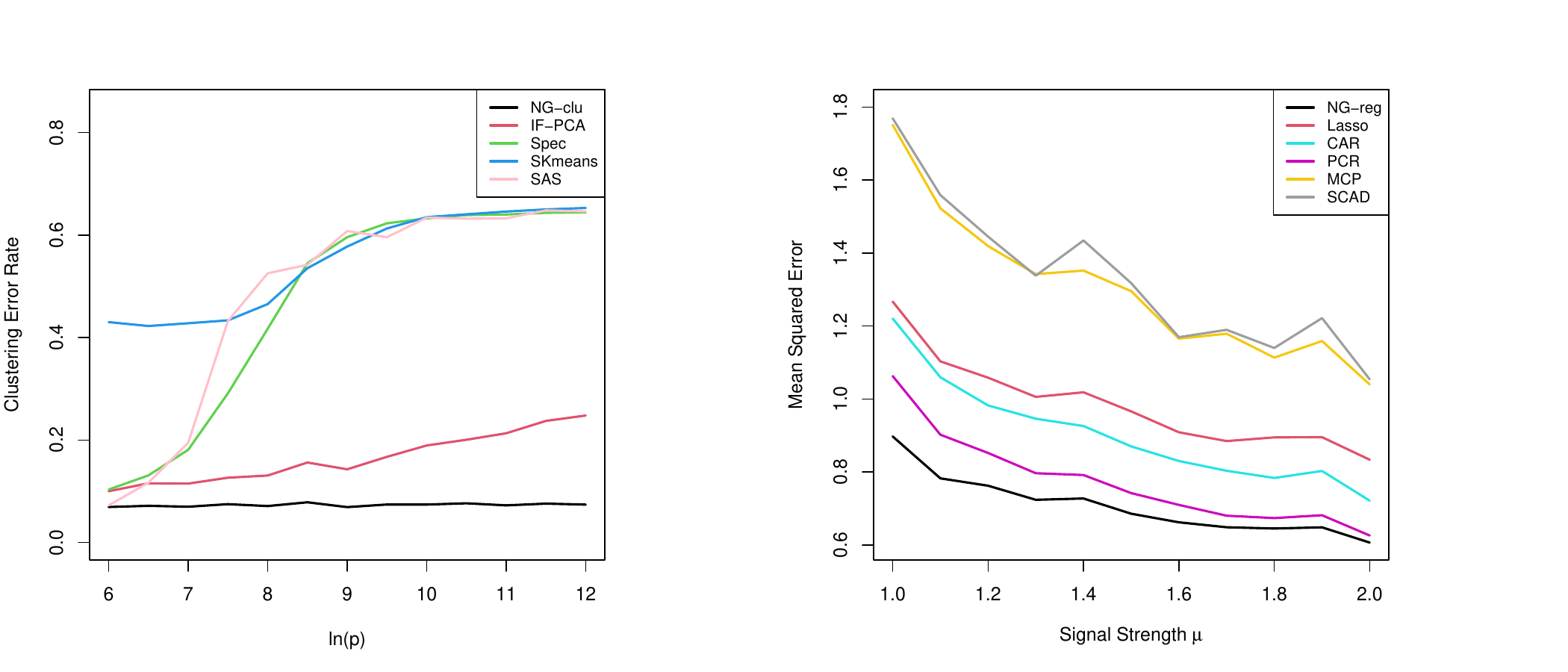}
    \caption{Summary of experimental results for downstream applications.}
    \label{fig:exp4}
\end{figure}

We first explore the clustering task, where our goal is to identify $\bell_i$ for $i \in [1000]$ under the same setup of Section~\ref{ex:dcsbm}. 
Let the number of covariates $p$ range from $e^6$ to $e^{13}$.  
We consider two sets of clustering methods: 1) NG-clu algorithm with $\bA$ and $K$; 2) $\bX$-based high-dimensional clustering methods, including spectral clustering (Spec), influential covariates PCA (IF-PCA) in \cite{IFPCA}, Sparse K-means (SKmeans) in \cite{sparsekmeans}, and Sparse Alternate Sum clustering (SAS) in \cite{arias2017simple}. 
From the left panel of Figure~\ref{fig:exp4}, it can be observed that our proposed NG-clu algorithm consistently outperforms other methods. 
The exponential increase in $p$ leads to a rapid increase in error rates for all $\bX$-based methods, whereas our approach experiences only a slight increase. It underscores the effectiveness of the NG-clu algorithm.

The regression task is explored under the setting of Section~\ref{ex:rdpg}, where our goal is to predict the responses $\bz$. 
We compare our proposed NG-reg algorithm with popular high-dimensional regression approaches, including: penalized high-dimensional regression with Lasso (Lasso), Minimax Concave Penalty (MCP), and Smoothly Clipped Absolute Deviation (SCAD) penalty, Principal Component Regression (PCR in \cite{liu2003principal}), and high-dimensional regression using correlation-adjusted marginal correlation (CAR in \cite{zuber2011high}).  
The mean squared error between the estimation $\hat{\bz}$ and $\bY\balpha$ is used to assess the performance. The right panel of Figure~\ref{fig:exp4} shows that our NG-reg algorithm outperforms all the other methods for all $\mu \in [1,2]$, with larger improvements for a smaller $\mu$. 
It highlights that enhancing data quality by additional network information is crucial for complex data.

\section{Sina$^{\text{TM}}$ Dataset Analysis}\label{sec:data} 
In the analysis of Sina Weibo data, 
we have added $2990$ noisy covariates to investigate the power of covariate selection. 
These independent noisy covariates $X_{ij}$ are generated as the logit-normal random variables with the density function 
\[
f(x) = \frac{1}{\sigma_j\sqrt{2\pi}} \frac{1}{x(1-x)}\exp[-\frac{(\log\{x/(1-x)\}-\mu_j)^2}{2\sigma_j^2}], 
\]
where $\mu_j \sim Unif(-3.5, -3.3)$, $\sigma_j \sim Unif(1.5, 2.2)$. 

The network $\bA$ is a directed network in this dataset. Hence, for NGCS, we take $\bU = \bXi(\bA; \Khat)$, where $\bXi$ contains the $\Khat$ leading left singular vectors of $\bA$. For a general directed network, $\bU$ should contain both left and right singular vectors. However, for this follower-followee network, the right singular vectors have many zeros, with little information. Therefore we only use the left singular vectors here.

In the manuscript, we mainly discuss the NG-reg algorithm. According to the two-step methodology, we can try other regression methods in the second stage. Based on the observation that there are still some noise covariates selected in NGCS, we consider the penalized regression method with Lasso. We construct the response variables with $\sigma = 0.5$, the same as the manuscript, and $\sigma = 0.2$, a new scenario with large signal-to-noise ratio.   

We consider the methods in the manuscript, and also the NGCS+Lasso method. The RMSE of each method over 50 repetitions are summarized in Table \ref{sina_classi2}. 
When the noise level is low, i.e., $\sigma = 0.2$, the response $\bz$ contains a lot of information. 
When $n_2 \leq 300$, two NG-reg algorithms outperform all other algorithms by leveraging $\bA$ and $\bX$ from the other 2000 users. 
However, when $n_2 \geq 350$, the supervised high-dimensional methods on study 2 successfully identify all the influential covariates and give a slightly better error rate, due to the supervised learning nature.  When the noise level is high, i.e., $\sigma = 0.5$, the informative component in $z$ is obscured by noise and our methods always outperform other algorithms.

\begin{table}[htbp]
\centering
\resizebox{\textwidth}{!}{
\begin{tabular}{l|lll|lll|lll|lll|lll}
\toprule
\multicolumn{16}{c}{$\sigma = 0.2$}\\
\hline
	\textbf{}   &\multicolumn{3}{c|}{\underline{\textbf{NGFS+OLS}}} &\multicolumn{3}{c|}{\underline{\textbf{NGFS+Lasso}}}  & \multicolumn{3}{c|}{\underline{\textbf{Lasso}}} & \multicolumn{3}{c|}{\underline{\textbf{MCP}}}  & \multicolumn{3}{c}{\underline{\textbf{SCAD}}}    \\
 $n_2$ & RMSE & $|\cShat|$  & $|\cS \cap \cShat|$  & RMSE & $|\cShat|$  & $|\cS \cap \cShat|$ & RMSE & $|\cShat|$  & $|\cS \cap \cShat|$ & RMSE & $|\cShat|$  & $|\cS \cap \cShat|$ & RMSE & $|\cShat|$  & $|\cS \cap \cShat|$ \\
\midrule
$100$ & 0.12 & 11.1 (1.52) & 8 (0.00) & 0.12 & 10.9 (1.37) & 8.0 (0.00) & 0.20 & 15.2 (25.79) & 0.5 (0.71) & 0.19 & 1.7 (3.77) & 0.1 (0.32) & 0.20 & 9.0 (16.28) & 0.5 (0.85)\\
$150$ & 0.10 & 10.4 (1.71) & 8 (0.00) & 0.10 & 9.6 (1.26) & 8.0 (0.00) & 0.19  & 21.1 (26.52) & 1.4 (1.07) & 0.19 & 6.8 (6.96) & 1.2 (1.14) & 0.19 & 19.5 (23.18) & 1.6 (1.84)\\
$200$ & 0.09 & 11.8 (1.81) & 8 (0.00)& 0.09 & 11.1 (1.72) & 8.0 (0.00) & 0.17 & 13.0 (9.46) & 2.2 (1.03) & 0.18 & 13.8 (12.53)& 3.0 (2.71) & 0.17 & 38.1 (31.50) & 4.1 (3.28) \\
$250$ & 0.09 & 14.3 (4.57) & 8 (0.00) & 0.09  & 13.8 (4.24) & 8.0 (0.00) & 0.17 & 32.8 (25.82) & 3.5 (0.97) & 0.16 & 18.6 (13.38) & 4.0 (2.53) & 0.14 & 67.7 (31.85) & 6.2 (2.62)\\ 
$300$ & 0.07 & 12.3 (1.94) & 8 (0.00) & 0.08 & 11.9 (1.52) & 8.0 (0.00) & 0.17 & 31.2 (15.03) & 3.8 (1.14) & 0.09 & 34.0 (9.84)& 8.6 (1.26) & 0.09 & 98.4 (19.96) & 8.8 (0.63) \\
$350$ & 0.07 & 12.0 (2.00) & 8 (0.00) & 0.07 & 11.3 (1.95) & 8.0 (0.00) & 0.17 & 47.6 (43.84) & 4.4 (1.26) & 0.07 & 25.8 (7.58)& 8.7 (0.48) & 0.07 & 72.4 (19.17) & 8.7 (0.48)\\
$400$ & 0.07 & 11.4 (0.96) & 8 (0.00) & 0.07 & 10.9 (1.45) & 8.0 (0.00) & 0.16 & 78.7 (45.33) &  6.2 (1.39) & 0.06 & 17.3 (7.20) & 8.8 (0.42) & 0.06 & 41.9 (14.69) & 8.8 (0.42)\\
$450$ & 0.07  & 12.0 (2.71) & 8 (0.00) &  0.07 & 11.4 (2.32) & 8.0 (0.00) & 0.16 & 84.7 (50.42) & 6.2 (1.03) & 0.05 & 15.6 (6.11) & 8.8 (0.42) & 0.06 & 44.9 (18.19) & 8.8 (0.42)\\
$500$ & 0.07  & 10.9 (1.37) & 8 (0.00) &  0.07 & 10.2 (1.14) & 8.0 (0.00) & 0.15 & 105.8 (72.57) & 6.6 (1.34) & 0.05 & 12.8 (4.24) & 8.8 (0.42) & 0.05 & 40.8 (35.48) & 8.8 (0.48)\\
\bottomrule
\toprule
\multicolumn{16}{c}{$\sigma = 0.5$ } \\
\hline

	\textbf{}   &\multicolumn{3}{c}{\underline{\textbf{NGFS+OLS}}} &\multicolumn{3}{c}{\underline{\textbf{NGFS+Lasso}}}  & \multicolumn{3}{c}{\underline{\textbf{Lasso}}} & \multicolumn{3}{c}{\underline{\textbf{MCP}}}  & \multicolumn{3}{c}{\underline{\textbf{SCAD}}}    \\
 $n_2$ & RMSE & $|\cShat|$  & $|\cS \cap \cShat|$  & RMSE & $|\cShat|$  & $|\cS \cap \cShat|$ & RMSE & $|\cShat|$  & $|\cS \cap \cShat|$ & RMSE & $|\cShat|$  & $|\cS \cap \cShat|$ & RMSE & $|\cShat|$  & $|\cS \cap \cShat|$ \\
\midrule
$100$ & 0.25 & 11.9 (1.79) & 8 (0.00) & 0.23 & 7.4 (4.97) & 5.2 (2.82) & 0.23 & 12.7 (23.30) & 0.2 (0.42) & 0.27 & 4.5 (7.21) & 0.2 (0.42) & 0.23 & 10.5 (16.39) & 0.2 (0.42)\\
$150$ & 0.21 & 12.5 (4.01) & 8 (0.00) & 0.19 & 8.1 (4.01) & 5.8 (2.66) & 0.20 & 14.0 (21.75) & 0.3 (0.67) &0.21 & 5.9 (6.37)& 0.3 (0.67) & 0.20 & 9.3 (16.91) & 0.5 (0.85)\\
$200$ & 0.17 & 12.2 (2.25) & 8 (0.00) & 0.18 &  8.8 (4.94) & 5.8 (2.90) & 0.20 & 2.4 (5.15) & 0.1 (0.32) & 0.20 & 1.4 (2.17)& 0.1 (0.32) & 0.19 & 3.7 (4.99) & 0.3 (0.48)\\
$250$ & 0.14 & 10.3 (1.16) & 8 (0.00) & 0.13 & 9.9 (1.10) & 7.9 (0.32) & 0.22 & 43.2 (51.47) & 0.2 (0.42) & 0.19 & 3.2 (5.79) & 0.0 (0.00) & 0.20 & 20.7 (28.23) & 0.0 (0.00)\\ 
$300$ & 0.13 & 11.0 (2.10) & 8 (0.00) & 0.14 & 9.7 (3.50) & 7.2 (2.53) & 0.19 & 11.5 (22.18) & 0.6 (0.70) & 0.19 & 5.4 (8.42)& 0.5 (0.53) & 0.20 & 13.9 (21.89) & 0.5 (0.53)\\
$350$ & 0.13 & 11.3 (1.64) & 8 (0.00) & 0.14 & 10.7 (2.75) & 7.6 (1.26) & 0.19 & 3.0 (7.15) & 0.2 (0.42) & 0.19 & 2.6 (4.86)& 0.2 (0.42) & 0.19 & 9.5 (16.66) & 0.4 (0.52)\\
$400$ & 0.11 & 12.5 (2.84) & 8 (0.00) & 0.11 & 12.2 (2.39) & 8.0 (0.00) & 0.19 & 12.9 (19.02) & 0.5 (0.97) & 0.19 & 4.6 (7.49)& 0.4 (0.70) & 0.19 & 5.1 (8.71) & 0.4 (0.70)\\
$450$ & 0.11 & 11.3 (1.77) & 8 (0.00) & 0.11 & 11.1 (1.97) & 8.0 (0.00) & 0.19 & 6.6 (7.82) &  0.7 (0.95) & 0.19 & 3.1 (4.63) & 0.5 (0.71) & 0.19 & 5.6 (6.69) & 0.6 (0.84)\\
$500$ & 0.11 & 12.2 (2.53) & 8 (0.00) &  0.11 & 11.5 (2.37) & 8.0  (0.00) & 0.19 & 10.5 (10.83) & 0.8 (0.79) & 0.19 & 6.7 (6.88) & 0.7 (0.82) & 0.19 & 13.5 (13.00) & 1.0 (1.05)\\
\bottomrule
\end{tabular}
}
\caption{Summary table for regression}
\label{sina_classi2}
\end{table}

\section{Rate-Optimality of Covariate Selection}\label{sec:csoptimal}
We say $\bX_i \sim \mathcal{SG}(\boldsymbol{\mu}, \bSigma, \sigma^2_{sg})$, if $\bX_i - \boldsymbol{\mu}$ follows a sub-Gaussian distribution with a covariance matrix $\bSigma$ and the variance proxy of each entry is bounded by $\sigma^2_{sg}$.
\subsection{Proof of Lemma: Tail Probability}
We first establish the tail probability result abut the $p$-values $\pi_j$. Based on the tail probabilities, we show the separation is clear and HCT will give a correct threshold.

\begin{lemma}\label{thm:tail}
Suppose Assumptions \ref{aspt:indep}--\ref{aspt:rw} hold.  Let $\pi_j$ be the $p$-value of $j$-th covariate based on $\bU \in \reals^{n \times \Khat}$ in the NGCS Algorithm, where $\bU^\top \bU = \bI$.  
There exists a threshold $p_0$ so that if $p > p_0$, for a constant $\delta$  that $10\Khat/\signal(\bU)^2 < \delta < 1$, there is
\[
P\bigl(\max_{j \in \cS}\pi_j\geq p^{-(1-1.2\delta) \frac{\signal(\bU)^2}{\log p}} \bigr) \leq |\cS|p^{-\tfrac18 \frac{\delta^2 \signal(\bU)^2 \sigma_{sg}^2}{\log p}}.
\]
Especially, for the Gaussian case, we have that 
\[
P\bigl(\max_{j \in \cS}\pi_j\geq p^{-(0.5-0.6\delta) \frac{\signal(\bU)^2}{\log p}} \bigr)\leq |\cS|\Phi(-\delta\signal(\bU)/2) \leq \frac{2|\cS|}{\signal(\bU)}p^{-\tfrac18 \frac{\delta^2 \signal(\bU)^2}{\log p}}.
\]
\end{lemma}
Lemma \ref{thm:tail} provides a union bound of $\pi_j$ for $j \in \cS$ in terms of the aggregated signal strength $\signal(\bU)$. In this inequality, $\delta$ is a constant to evaluate the deviation and $\tau(\bU)^2/\log p$ captures the effects of $\tau(\bU)^2$. To guarantee a separation between the $p$-values for those in $\cS$ and $\cS^c$, $\tau(\bU)^2/\log p$ must be sufficiently large, i.e. $\tau(\bU)^2 \geq C\log p$ for some constant $C > 0$. 

\begin{proof}
Let $\bU = [\bfeta_1, \bfeta_2, \cdots, \bfeta_{\Khat}]$ and $\bU_{K} = [\bfeta_1, \bfeta_2, \cdots, \bfeta_K]$. Given the latent factor matrix $\bY$, we want to control the tail probability of $t_j$ for $j \in \cS$ and $j \notin \cS$. 

Let $\bX_j$ denote the $j$-th covariate on all $n$ samples, then $\bX_j \sim \mathcal{SG}(\bY \bM_j, \bI_n, \sigma^2_{sg})$. 
The test statistic follows that 
\[
t_j = \sum_{k=1}^{\Khat} (\bfeta_k^\top \bx_j)^2.
\]

To evaluate $t_j$, we check each term in the summation. We rewrite each term as the mean and the random part, which gives that $(\bfeta_k^\top \bX_j)^2 = (\bfeta_k^\top \bY \bM_j + \bfeta_k^\top \bz_j)^2$. Here $\bz_j \sim \mathcal{SG}({\bf 0}, \bI, \sigma^2_{sg})$. Hence, we define the signal part: 
\begin{equation}
    s_j^2 = \sum_{k=1}^{\Khat} (\bfeta_k^\top \bY \bM_j)^2
    \geq \sum_{k=1}^{K} (\bfeta_k^\top \bY \bM_j)^2 = \|\bU_K^\top \bY \bM_j\|^2. 
\end{equation}
Therefore, 
\begin{eqnarray*}
    t_j & = & \sum_{k=1}^{\Khat} (\bfeta_k^\top \bx_j)^2 = \sum_{k=1}^{\Khat} (\bfeta_k^\top \bY \bM_j + \bfeta_k^\top \bz_j)^2\\
    & = & \sum_{k=1}^{\Khat} (\bfeta_k^\top \bz_j)^2 + 2\sum_{k=1}^{\Khat} (\bfeta_k^\top \bY \bM_j)(\bfeta_k^\top \bz_j) + s_j^2\\
    & = & A_j + B_j + s_j^2.
\end{eqnarray*}
Since $\|\bfeta_k\| = 1$ and orthogonal with each other, $A_j = \bz_j^\top (\sum_{k=1}^{\Khat} \bfeta_k \bfeta_k^\top) \bz_j$. By Hanson-Wright inequality for sub-Gaussian distributions \citep{hw}, 
\[
P(|A_j-\Khat|\geq t)\leq 2\exp(-c\min\{\frac{t^2}{\Khat^2\sigma^4_{sg}},\frac{t}{\sigma^2_{sg}}\}),
\]
where $c = 1/96$.
Especially, in the Gaussian case, $A_j \sim \chi^2_{\Khat}$. Now consider $B_j$. Note that $\bfeta_k^\top \bz_j \sim SG(0, 1)$ for any $k$, and $\bfeta_{k_1}^\top \bz_j$ is independent with $\bfeta_{k_2}^\top \bz_j$ for any $k_1 \neq k_2$ since $\bfeta_{k_1}^\top\bfeta_{k_2} = 0$. Hence, $B_j = 2\sum_{k=1}^{\Khat} (\bfeta_k^\top \bY \bM_j)(\bfeta_k^\top \bz_j) \sim \mathcal{SG}(0, 4s_j^2, 4s_j^2\sigma^2_{sg})$. 

It allows us to control the tail probability of $t_j$. We consider two cases that $j \in \cS$ and $j \notin \cS$. 
\begin{itemize}
    \item Case: $j \in \cS$. Note that $\signal = \signal(\bU) = \min_{j \in \cS} s_j$. 
    
    For any $\delta > 10\Khat/\signal^2 \sigma^2_{sg}> 10\Khat/s_j^2\sigma^2_{sg} = o(1)$, we have 
\begin{eqnarray*}
P(t_j < \Khat + (1-1.1\delta)s_j^2\sigma^2_{sg}) & \leq& 
P(A_j < \Khat - 0.1\delta s_j^2\sigma^2_{sg}) + P(B_j < -\delta s_j^2\sigma^2_{sg}) \\
&\leq& 0 + e^{-\delta^2 s_j^4\sigma^4_{sg}/8s_j^2 \sigma^2_{sg}} 
= e^{-\delta^2s_j^2\sigma^2_{sg}/8}.
\end{eqnarray*}
So for any $j \in \cS$, we have 
\begin{eqnarray*}
P(t_j < \Khat + (1-1.1\delta)\signal^2\sigma^2_{sg}) & \leq&  P(t_j < \Khat + (1-1.1\delta)s_j^2\sigma^2_{sg}) \\
&\leq & \exp(-\delta^2s_j^2\sigma^2_{sg}/8) \leq \exp(-\delta^2\signal^2\sigma^2_{sg}/8).
\end{eqnarray*}

Then we discuss this inequality in terms of the $p$-value $\pi_j$. 
For the Gaussian case, $\sigma^2_{sg} = 1$ and the $p$-value is calculated by $\pi_j = (\chi^2_{\Khat})^{-1}(t_j)$, so we check the $p$-value of $\chi^2_{\Khat}$ at the point $\Khat + (1-1.1\delta)\tau^2$. 
By \cite[Lemma 1 on P1325]{laurent2000adaptive},
\[
\chi^2_{\Khat}(\Khat+(1-1.1\delta)\signal^2)\leq \exp\biggl(-\frac{(1-1.1\delta)^2\signal^4}{\bigl(\sqrt{2((1-1.1\delta)\signal^2+\Khat}+\sqrt{\Khat}\bigr)^2}\biggr),
\]
which can further be bounded by $\exp(-(1-1.2\delta)\signal^2/2)$ when $\signal$ is sufficiently large. So, 
\[
P(\pi_j\geq \exp(-(0.5-0.6\delta)\signal^2))\leq \Phi(-\delta\signal/2),
\qquad 
j \in \cS.
\]

Using the HW-$\chi^2$ $p$-value, which is actually the upper bound of $p$-values, that
\[
\pi(t) = \min\{\exp(-c\min\{(t-\Khat)^2/\Khat^2\sigma^4_{sg},(t-\Khat)/\sigma^2_{sg}\}), 1\}.
\]
At the point $t = \Khat + (1 - 1.1\delta)\tau^2\sigma^2_{sg}$, there is
\begin{eqnarray*}
\pi(\Khat+(1-1.1\delta)\signal^2\sigma^2_{sg}) & = & \exp\biggl(-c\min\{\frac{(1-1.1\delta)^2\tau^4}{\Khat^2},{(1-1.1\delta)\tau^2}\}\biggr) \\
& \leq &\exp(-c(1-1.2\delta)\tau^2).    
\end{eqnarray*}

So, 
\[
P(\pi_j\geq \exp(-c_0(1-1.2\delta)\signal^2))\leq \exp(-\delta^2\signal^2\sigma^2_{sg}/8),
\qquad 
j \in \cS.
\]
\item Case: $j \notin \cS$. When $j \notin \cS$, there is $\|\bM_j\| = 0$ and hence $s_j^2 = \|\bU_{\Khat} \bY \bM_j\|^2 = 0$. As a result, $t_j = A_j$ and 
    \[
    P(t_j-\Khat\geq t)\leq \min\{\exp(-c\min\{\frac{t^2}{\Khat^2 \sigma^4_{sg}}, \frac{t}{\sigma^2_{sg}}\}), 1\}\quad j\notin \mathcal{S}.
    \]
    For the Gaussian case, it will follow $\chi^2_{\Khat}$,
    and the corresponding $p$-value follows that $\pi_j \sim {Unif}(0,1)$. For the general case, we consider the HW-$\chi^2$ $p$-value, where
    \[
    P(\pi_j\leq \exp(-c(1-1.2\delta)\signal^2))\leq P(t_j\geq \Khat+(1-1.2\delta)\signal^2\sigma_{sg}^2)
    \leq \exp(-c(1-1.2\delta)\signal^2).
    \]
For sufficiently large $\signal$, it is further bounded by $\exp(-c\tau^2)$. 
\end{itemize}

The union bound can be found. For Gaussian case, 
\[
P(\max_{j\in\cS}\pi_j\geq \exp(-(0.5-0.6\delta)\signal))\leq |\cS|\Phi(-\delta\signal/2) \leq \frac{2|\cS|}{\delta \signal}\exp(-\delta^2 \signal^2/8) \leq \frac{2|\cS|}{\signal}p^{-\frac{\delta^2 \signal^2}{8\log p}}. 
\]
For the general case,
\begin{eqnarray*}
&&P(\max_{j\in\cS}\pi_j\geq \exp(-(1-1.2\delta)\signal^2)\text{ or } \min_{j\in S^c} \pi_j\leq \exp(-(1-1.2\delta)\signal^2))\\
&\leq & |\cS|\exp(-\delta^2 \signal^2\sigma^2_{sg}/8)+
|\cS^c|\exp(-c(1-1.2\delta)\signal^2)\\
&\leq & |\cS|p^{-\frac{\delta^2 \signal^2\sigma^2_{sg}}{8\log p}}+|\cS^c|\exp(-c(1-1.2\delta)\signal^2). 
\end{eqnarray*}
\end{proof}

\subsection{Proof of Theorem \ref{thm:main}: Consistency of HCT}

\begin{proof}[Proof of Theorem \ref{thm:main}] 
The proof consists of three steps. We first show that $\pi_j$ for $j \in \cS$ and $j \notin \cS$ has a clear division with high probability. Then we show that Higher Criticism Threshold (HCT) almost achieves the optimal division, in the sense that $HC(j)$ for all $j \in \cS$ is smaller than the selected HCT and no more than $C\log p$ covariates in $\cS^c$ have lower $HC(j)$ than HCT. The proof is done for general sub-Gaussian distribution, while the results can be improved for the Gaussian case.

\textbf{Step 1: Clear division with high probability.}
Given $\bU$ and $\bY$, we define a good event that the statistics from $j \in \cS$ and $j \notin \cS$ are clearly divided, that 
\[
\cB=\{\max_{j\in \cS^c}t_j<\Khat+ (1 - 1.2\delta)\signal^2 \sigma^2_{sg} <\min_{j\in \cS}t_j\}.
\]
The boundary comes from the tail probability in Lemma \ref{thm:tail}, where $\signal = \signal(\bU)$ denotes the signal strength from the network. 

According to Lemma \ref{thm:tail}, 
the tail probability of $t_j$ when $j \in \cS$ and $j \in \cS^c$ is known. 
Let $\cF_A$ be the event that the union bound holds, then we have the probability of $\cB$: 
\begin{eqnarray*}
1-P(\cB|\cF_A) &\leq &|\cS|p^{-\frac{\delta^2 \signal^2\sigma^2_{sg}}{8\log p}}+|\cS^c|\exp(-c(1-1.2\delta)\signal^2)\\
&\leq &|\cS|p^{-\frac{\delta^2 \signal^2\sigma^2_{sg}}{8\log p}}+p\exp(-c(1-1.2\delta)\signal^2).
\end{eqnarray*}
Let $\delta=1/{\sqrt{2}}$ and then the remainder probability is of order
\[
1-P(\cB|\cF_A)\lesssim |\cS|p^{-\frac{\signal^2\sigma^2_{sg}}{16\log p}}+p^{1-c(1-1.2/\sqrt{2})\signal^2/\log p}, \quad c = 1/16.
\]
Recall that $|\cS| = p^{1-\beta}$. 
So if $\signal^2>\max\{16(1-\beta)/\sigma^2_{sg}, \frac{1}{c(1-1.2/\sqrt{2})}\}\log p\geq 16\max\{(1-\beta)/\sigma^2_{sg}, 1/(1 - 1.2/\sqrt{2})\}\log p$, then we will have $P(\mathcal{B|\cF_A})\geq 1-o(1)$.

\textbf{Step 2: Under $\cB$, all covariates from $\cS$ will be selected by HCT.}
By the NGCS algorithm, $\cShat$ is achieved by selecting all the covariates with $p$-values smaller than $T_{thre}$. According to the definition of $\cB$ in Step 1, covariates from $\cS$ always have smaller $p$-values than those in $\cS^c$. Hence, it suffices to show for any $i<|\cS|$,
\begin{equation}
\label{tmp:step2}
HC(i)\leq HC(|\cS|).  
\end{equation}
Under $\cB$, $\max_{i\in \cS} \pi_i\leq \exp(-(1-1.2\delta)\signal^2) \leq 1/p^{1+c_0}$, where $c_0 = (1-1.2/\sqrt{2})\signal^2/\log p - 1 > 0$. Introduce it into $HC(|\cS|)$, we have
\begin{equation}\label{eqn:HCSlower}
HC(|\cS|)\geq \frac{|\cS|/p-\exp(-(1-1.2\delta)\signal^2)}{\sqrt{|\cS|/p(1-|\cS|/p)}}\geq 
\frac{|\cS|/p-1/p^{1+c_0}}{\sqrt{|\cS|/p(1-|\cS|/p)}}.
\end{equation}
Meanwhile, we have $HC(i)\leq \frac{\sqrt{(|\cS|-1)/p}}{\sqrt{1-(|\cS|-1)/p}}$ for all $i\leq |\cS|-1$. 

Plug these upper bounds and \eqref{eqn:HCSlower} into \eqref{tmp:step2}. We have
\begin{align*}
[HC(i)]^2 & \leq  [HC(|\cS|)]^2\\
\frac{(|\cS|-1)/p}{1-(|\cS|-1)/p} & \leq \frac{(|\cS|/p-1/p^{1+c_0})^2}{|\cS|/p(1-|\cS|/p)}\\
(|\cS| - 1)|\cS|(p - |\cS|) & \leq  (|\cS| - p^{-c_0})^2 (p - |\cS|+1).
\end{align*}
Rearrange the terms and it suffices to show \eqref{tmp:step2} if we can show that 
\[
|\cS|(p - |\cS|)(1 - 2p^{-2c_0}) + |\cS|(|\cS| - 2p^{-c_0}) + p^{-2c_0}(p - |\cS|+1)\geq 0.
\]
When $p$ is sufficiently large, it holds. So \eqref{tmp:step2} holds for sufficiently large $p$.

\textbf{Step 3: Under $\cB$, only $C\log^2 p$ covariates from $\cS^c$ will be selected by HCT.} This is equivalent to show $HC(|\cS|+k)<HC(|\cS|)$ when $k>C\log^2 p$. We first give the proof under Gaussian assumption, and then discuss the subGaussian case.

Consider a sequence of threshold $v_k = k/|\cS^c|$ for $k > C_1\log^2 p$. 
Note that $\pi_i \sim Unif(0,1)$ for all $i \in \cS^c$. By the Bernstein inequality with $\delta_k = 4\sqrt{{v_k\log p}/{p}}$, when $C_1 > 2(p/|\cS^c|)^2$ and $|\cS^c|/p > 1/2$, we have 
\begin{eqnarray}
P\left(\biggl|\frac{1}{|\cS^c|}\sum_{i\in \cS^c} 1_{\pi_i<v_k}-v_k\biggr| > \delta_k\bigg|\cF_A\right) & \leq & \exp\left(-\frac{|\cS^c|\delta_k^2}{2(v_k(1-v_k)+\delta_k /3)}\right)\nonumber\\
& \leq & \exp\left(-\frac{|\cS^c|\delta_k^2}{4v_k(1-v_k)}\right) + e^{-{3|\cS^c|\delta_k}/{4}}\nonumber\\
& \leq & 2p^{-2}.
\end{eqnarray}
Naturally, the union bound follows, which is 
\[
P\left(\bigl|\frac{1}{|\cS^c|}\sum_{i\in \cS^c} 1_{\pi_i<v_k}- v_k\bigr|>\delta_k, \text{ for any $k \geq C_1\log^2 p$}\right)\leq \frac{1}{p}. 
\]
Hence, the complementary event happens with probability $1 - O(p^{-1})$. Define it as 
\[
\cC:=\left\{\bigl|\frac{1}{|\cS^c|}\sum_{i\in \cS^c} 1_{\pi_i<v_k}- v_k\bigr|<\delta_k, \text{ for any $k \geq C_1\log^2 p$}\right\}. 
\]

Denote the ordered $\pi_i$ as $\pi_{(1)}\leq \pi_{(2)}\leq\ldots\leq \pi_{(p)}$. Under the event $\cB$,  $\pi_{(|\cS|+k)} = \pi^0_{(k)}$ where $\pi^0_{(\cdot)}$ are the ordered $p$-values of $i \in \cS^c$. Now we want to derive a lower bound of $\pi_{(k)}$. 

Under $\cC$, recall that $v_j = j/|\cS^c|$ and $\delta_j = 4\sqrt{v_j \log p/p}$, when $j > C_1\log^2p$, 
\[
\sum_{i \in \cS^c} 1_{\pi_i < v_j} < |\cS^c|(v_j + \delta_j) < j + 4\sqrt{j \log p}.
\]
It means $\pi^0_{(j + 4\sqrt{j \log p})} \geq v_j$. 
Now let $k$ be the smallest integer so that $k \geq j + 4\sqrt{j\log p}$, then it follows $j \geq k-1 - 4\sqrt{k \log p}$. 
Hence, under $\cC \cap \cB$, for $k \geq 3C_1\log^2p$,
\[
\pi_{(|\cS|+j + 4\sqrt{j \log p})} \geq v_j\Longrightarrow  \pi_{(|\cS|+k)} \geq (k-1)/p - 4\sqrt{k \log p}/p,
\]
For the subGaussian case, the same derivation works for $p_j$ being the exact $p$-value under null hypothesis. Further, since the HW-$\chi^2$ statistic $\pi_j \geq p_j$ according to the definition, so $\pi_{|\cS|+k} \geq (k-1)/p - 4\sqrt{k \log p}/p$ holds for HW-$\chi^2$ statistic, too.
This further leads to 
\begin{align*}
HC(|\cS|+k)=\frac{\frac{|\cS|+k}{p}-\pi_{(|\cS|+k)}}{\sqrt{\frac{(|\cS|+k)}{p}(1-\frac{|\cS|+k}{p})}}\leq \frac{\frac{|\cS|+k}{p}-\frac{k-1}{p}+\frac{4\sqrt{k\log p}}{p}}{\sqrt{\frac{(|\cS|+k)}{p}(1-\frac{|\cS|+k}{p})}}.
\end{align*}

Combine it with \eqref{eqn:HCSlower}. To conclude our claim, we need to show 
that if $0.25p>k\geq 3C_1\log^2 p$, there is  
\begin{align*}
 \frac{(|\cS|-p^{-c})^2}{{|\cS|(p-|\cS|)}}>    \frac{(|\cS|+1+4\sqrt{k\log p})^2}{(|\cS|+k)(p-|\cS|-k)}.
\end{align*}
This can be shown if 
\[
(|\cS|-p^{-c})^2(|\cS|+k)(p-|\cS|-k)>(|\cS|+1+4\sqrt{k\log p})^2(|\cS|+k)(p-|\cS|-k).
\]
When $|\cS| \gg \log p$, it can further be simplified as 
\[
p|S|^3+kp|S|^2>p|S|^3+8p\sqrt{k\log p}|S|^2+\text{lower order terms}. 
\]
It holds when $k\geq 3C_1\log^2 p$ and $p$ large enough. In other words, $HC(|\cS|+k)<HC(|\cS|)$ for $k > C\log^2 p$ where $C = 3C_1$. This gives our second bound.

\end{proof}

\subsection{Proof of Theorem \ref{thm:lowerbound}: Lower Bound}

\begin{lemma}
\label{lem:Hellinger}
Let $\delta_H(f_1,f_2)$ be the Hellinger distance between $f_1$ and $f_2$. 
Let $\delta_{tv}(f_1,f_2)$ be the total variation distance between $f_1$ and $f_2$.
The following hold 
\[
\delta_H(f_1,f_2)\leq 1-\sqrt{1-\frac14\delta_{tv}(f_1,f_2)^2}. 
\]
\end{lemma}
\begin{proof}
Denote the density of $f_i$ to be $p_i$, so we have 
\[
\delta_H(f_1,f_2)=\frac12\int (\sqrt{p_1}(x)-\sqrt{p_2}(x))^2dx=2-\frac12\int (\sqrt{p_1}(x)+\sqrt{p_2}(x))^2dx
\]
Therefore by the Cauchy-Schwartz inequality, 
\begin{align*}
    \delta_{tv}(f_1,f_2)^2&\leq  \int (\sqrt{p_1}(x)-\sqrt{p_2}(x))^2dx \cdot \int (\sqrt{p_1}(x)+\sqrt{p_2}(x))^2dx\\
    &=2\delta_H(f_1,f_2)(4-2\delta_H(f_1,f_2)).
\end{align*}
So 
\[
\sqrt{1-\frac14\delta_{tv}(f_1,f_2)^2}\geq 1-\delta_H(f_1,f_2). 
\]
and we have the result. 
\end{proof}

\begin{proof}[Proof of Theorem \ref{thm:lowerbound}]
Let $\theta_i \in \{1, -1\}$ denote whether $i \in \cS$ or not. Hence, any estimator $\cShat$ can be seen as the collection that $\cShat = \{j: \hat\theta(\bX,\bA) = 1\}$, where $\hat\theta(\bX,\bA) \in \{-1, 1\}$. Our goal is to find the lower bound of $\hat\theta(\bX, \bA)$. 

If there exists $\hat\theta(\bX, \bA)$ so that \eqref{eqn:FDR} holds, then 
\[
\E[ (\hat\theta_i(\bX, \bA)-\theta_i)^2]\leq \epsilon P(\hat\theta_i(\bX, \bA)=-1|\theta_i=1)
+(1-\epsilon)P(\hat\theta_i(\bX, \bA)=1|\theta_i=-1)=4(q_1+\epsilon q_2). 
\]
Note that the likelihood of $\bX, \bA$ are given by 
\[
l(\bA|\bY)*l(\bX|\bY) = C*l(\bA|\bY)\prod_{i=1}^p\exp(-\frac12 \|\bX_i-\bY \bM_i\|^2 ).
\]
So $(\bY, \bX_i)$ are sufficient statistics for the estimation of $\theta_i$. Define $G(\bX_i|\bY)=\E[\hat\theta_i(\bX, \bA)|\bX_i,\bY]$, then $G$ is a bounded function between $-1$ and $1$. By Blackwell theorem, 
\begin{equation}
\label{tmp:lower1}
\E[ (G(\bX_i|\bY)-\theta_i)^2]\leq \E[ (\hat\theta_i(\bX, \bA)-\theta_i)^2]=4(q_1 + \epsilon q_2).
\end{equation}

Denote 
$Z_1(\bY)=\E[G(\bX_i|\bY) |\bY, \theta_i=1]$ and 
$Z_2(\bY)=\E[G(\bX_i|\bY) |\bY, \theta_i=-1]$.
Denote $\Delta=Z_1-Z_2\leq 2$, there is 
\begin{align*}
\E[ (G(\bX_i|\bY)-\theta_i)^2]&=\E[\epsilon\E[ (G(\bX_i|\bY)-1)^2|\bY,\theta_i=1]+(1-\epsilon)\E[(G(\bX_i|\bY)+1)^2|\bY,\theta_i=-1]]\\
&\geq  \E[ \epsilon(1-Z_1)^2+(1-\epsilon)(Z_2+1)^2]\\
&\geq  \E[ Z_2^2+(2-4\epsilon+2\Delta\epsilon )Z_2+1 -2\Delta\epsilon+\Delta^2 \epsilon ],\quad \text{by minimizing }Z_2\\
&\geq \epsilon(1-\epsilon)\E[(2-\Delta)^2]\geq  \epsilon(1-\epsilon)\E[2-\Delta]^2.
\end{align*}
Denote the law of $\bX_i$ conditional on $\bY$ and $\theta_i=1$ as $f_1$, which is $\mathcal{N}(\bY \bM_i, \bI)$; and the law of $\bX_i$ conditional on $\bY,\theta_i=-1$ as $f_2$, that $\mathcal{N}({\bf 0}, \bI)$. Let $\delta_H(f_1,f_2)$ be the Hellinger distance between $f_1$ and $f_2$.
Meanwhile, $\|\bY \bM_i\|\leq \sqrt{n} \|\bM_i\|\leq \sqrt{2r\log p}$ and by (\cite[Page 51 and 46]{pardo2018statistical}), we have
\begin{align*}
\delta_H(f_1,f_2)^2=1-\exp(-\|\bY \bM_i\|^2/8)\leq 1-p^{-\frac{1}{4}r}.
\end{align*}
By Lemma \ref{lem:Hellinger}, we have 
\begin{align*}
[\E |\Delta|]^2 \leq \E \Delta^2\leq 2\E [\delta_H(f_1,f_2)(4-2\delta_H(f_1,f_2))]\leq (2-p^{-\frac14r})^2. 
\end{align*}
Combining them together, we find 
\[
4(q_1+\epsilon q_2)\geq \epsilon(1-\epsilon) p^{-\frac12 r}. 
\]
It follows that $r>-{\log (q_1/\epsilon+q_2)}/{\log p}$.  
When $r<4\beta$, it is impossible to recover $\cS$. 
\end{proof}

\section{Consistency of Downstream Applications}\label{sec:downstream}
\subsection{Information in the eigenvectors}\label{subsec:eigen}
In this section, we prove the lower bound of signal $\signal(\bXi_{\Khat})$ under both the DCSBM and the RDPG when regular conditions hold. When the degree and the spectrum of $\E[\bA]$ can be well-bounded, the analysis for both models is the same. 

To prove this, we need a lemma on the condition number of the matrix product.
\begin{lemma}
\label{lem:condprod}[Condition number of matrix product]
Suppose $\bA\in \reals^{n\times K}, \bB\in \reals^{K\times p}$ and $rank(\bA) = rank(\bB) = K$, then 
\[
\frac{\lambda_{K}(\bA\bB)}{\lambda_1(\bA\bB)}\geq \frac{\lambda_K(\bA)\lambda_K(\bB)}{\lambda_1(\bA)\lambda_1(\bB)}. 
\]
\end{lemma}

\begin{proof}
We write $\bB = \bU\bLambda \bV^{\top}$, then 
\begin{align*}
\lambda_1(\bA\bB)  =  \lambda_1(\bA \bU\bLambda \bV^{\top})& =\sqrt{\lambda_1(\bA\bU\bLambda^2\bU^{\top} \bA^{\top})}\\
&\leq \lambda_1(\bB) \sqrt{\lambda_1(\bA\bU\bU^{\top} \bA^{\top})}=
\lambda_1(\bB) \sqrt{\lambda_1(\bU^{\top} \bA^{\top}\bA\bU)}=\lambda_1(\bB)\lambda_1(\bA).    
\end{align*}

Following the same line, we have
\begin{align*}
\lambda_K(\bA\bB)=\lambda_K(\bA \bU\bLambda \bV^{\top})&=\sqrt{\lambda_K(\bA\bU\bLambda^2\bU^{\top} \bA^{\top})}\\
&\geq \lambda_K(\bB) \sqrt{\lambda_K(\bA\bU\bU^{\top}\bA^{\top})}\geq \lambda_K(\bB)\lambda_K(\bA).
\end{align*}
The lemma then follows.
\end{proof}

\begin{proof}[Proof of Theorems \ref{thm:dcsbm} and \ref{thm:lp}]
The proof of the adjacency matrix and the Laplacian is very similar. We focus on the Laplacian $\bL =\bD^{-1/2} \bA \bD^{-1/2}=\bXi \bLambda \bXi^{\top}$, where the diagonals of $\bLambda$ are ordered so that the magnitude is decreasing. 
The results can be naturally generalized to $\bA$, which is easier without the perturbation of $\bD$. 

Suppose we have prior knowledge that the number of true classes is no larger than $\Khat$, then we take $\bXi_{\Khat} = [\bxi_1,\bxi_2,\ldots,\bxi_\Khat] \in \reals^{n\times \Khat}$. The signal strength is then 
\begin{equation}\label{eqn:cam}
\signal(\bXi_{\Khat}) = \min_{j \in \cS} s_j, 
\end{equation}
where $s_j^2=\|\bXi^{\top}_{\Khat} \bY \bM_j\|^2\geq \|\bXi^{\top}_K \bY \bM_j\|^2 \geq 
\lambda_K(\bY^{\top}\bXi_K \bXi_K^{\top}\bY)\|\bM_j\|^2$ and $\bY$ is the latent factor matrix. 
Recall that $\kappa = \min_{j \in \cS} \|\bM_j\|$. 
To assure the signals are exactly recovered, by Theorem \ref{thm:main} and the above inequality, it suffices to prove that with high probability, 
\begin{equation}\label{eqn:eigenresult}
\kappa^2 \lambda_K(\bY^{\top}\bXi_K \bXi_K^{\top}\bY) \geq \max\{16-16\beta, 14\}\log p.
\end{equation}
In other words, we have to show $\bXi_K$ is closely related to the latent information $\bY$.

To proceed, we define the partial-oracle Laplacian, which is  
\begin{equation}\label{eqn:oracleL}
\bar{\bL} =\bD^{-1/2}\E[\bA] \bD^{-1/2}, 
\end{equation}
where $\E[\bA] = \bTheta \bY \bB \bY^{\top} \bTheta$ under DCSBM and $\E[\bA] = \rho_n \bY \bY^{\top}$ under RDPG. 
We say it is a partial oracle, because it assumes knowledge of $\E[\bA]$ but still uses the observed $\bD$. 
Let the eigen-decomposition of $\bar{\bL}$ be $\bar{\bL} = \bar{\bXi} \bar{\bLambda} \bar{\bXi}^{\top}$, and $\bar{\bXi}_K = [\bar{\bxi}_1, \cdots, \bar{\bxi}_K]$. 
We want to show that $\bXi$ and $\bar{\bXi}$ are close, so we can use the lower bound of $\lambda_K(\bY^{\top} \bar{\bXi}_K \bar{\bXi}_K^{\top} \bY)$ and the difference between them to prove \eqref{eqn:eigenresult}.

Now we show that $\lambda_K(\bY^{\top} \bar{\bXi}_K \bar{\bXi}_K^{\top} \bY)$ and $\lambda_K(\bY^{\top} {\bXi}_K {\bXi}_K^{\top} \bY)$ are close. Let $d_m = n\rho_n$ under RDPG and $d_m = n\theta_0^2$ under DCSBM. By Lemmas \ref{lem:diRDPG} and \ref{lem:di} with $\delta = 0.1$, with probability $1-O(n^{-1})$, there exists a constant $c$ so that the degrees follow $0.9c d_m \leq d_i \leq 1.1 d_m$ for all $i \in [n]$ for both models. By Lemma \ref{lem:Anorm}, there is $\|\bA - \E[\bA]\| \leq 6\sqrt{d_m}$ with probability $1 - O(n^{-1})$. 
Hence, with probability $1 - O(n^{-1})$,  
\[
\|\bL-\bar{\bL}\|=\|\bD^{-1/2}(\bA-\E[\bA]) \bD^{-1/2}\|\leq \frac{\|\bA - \E[\bA]\|}{\min_i d_i} \leq C\frac{6\sqrt{d_m}}{0.9 c d_m} \leq C/\sqrt{d_m}.
\]
When $d_m \to \infty$, $\|\bL - \bar{\bL}\| \to 0$, which indicates the similarity in their eigenvalues. By Lemmas \ref{lem:LArdpg} and \ref{lem:LA}, there exists a constant $c > 0$ that depends on $c_{Y}$, $c_{B}$, and $\rho$, so that $\lambda_K(\bar{\bL}) \geq c$ for both models. 
Combining it with the Davis--Kahan Theorem in \cite{daviskahan},  with probability $1 - O(n^{-1})$,  the distance between the eigen-spaces follows:
\[
\|\bXi_K \bO-\bar{\bXi}_K\|\leq \frac{C}{\sqrt{d_m}\lambda_K(\bar{\bL})}\leq \frac{C/c}{\sqrt{d_m}} \leq C/\sqrt{d_m}.
\]

Since $\|\bXi_K\|, \|\bar{\bXi}_K\| \leq 1$, there is $\|\bXi_K \bXi_K^{\top} - \bar{\bXi}_K \bar{\bXi}_K^{\top}\| \leq \|\bXi_K \bO - \bar{\bXi}_K\|$. Therefore, the goal follows that 
\begin{equation}\label{eqn:diffgoal}
\|\bY^{\top}\bXi_K \bXi_K^{\top} \bY-\bY^{\top}\bar{\bXi}_K\,\bar{\bXi}_K^{\top} \bY\|\leq {C\|\bY^{\top}\bY\|}/{\sqrt{d_m}}. 
\end{equation}
Hence, the difference in the $K$-th eigenvalue $|\lambda_{K}(\bY^{\top}\bar{\bXi}_K\,\bar{\bXi}_K^{\top} \bY)-\lambda_K(\bY^{\top}\bXi_K \bXi_K^{\top} \bY)|$ has the same bound. 
We only need to study the partial-oracle case $\lambda_K(\bY^{\top}\bar{\bXi}_K\,\bar{\bXi}_K^{\top} \bY)$. 

Now we study $\lambda_K(\bY^{\top}\bar{\bXi}_K\,\bar{\bXi}_K^{\top} \bY)$. 
By Lemma \ref{lem:LA} on the largest eigenvalue of $\bar{\bL}$, 
\[
\bY^{\top}\bar{\bXi}_K\bar{\bLambda}_K\bar{\bXi}_K^{\top}\bY
\preceq {{\lambda}_1(\bar{\bL})}\bY^{\top}\bar{\bXi}_K\,\bar{\bXi}_K^{\top} \bY \preceq 1.1\bY^{\top}\bar{\bXi}_K\,\bar{\bXi}_K^{\top} \bY/\rho c_B.
\]
So $\lambda_K(\bY^{\top}\bar{\bXi}_K\bar{\bLambda}_K\bar{\bXi}_K^{\top}\bY)$ gives a lower bound up to the constants. It can be derived by the relationship that 
$\bY^{\top}\bar{\bXi}_K\bar{\bLambda}_K\bar{\bXi}_K^{\top}\bY=\bY^{\top} \bar{\bL} \bY. $
We discuss it separately for the two models. 
\begin{itemize}
    \item DCSBM. Under DCSBM, $\bar{\bL} = \bD^{-1/2} \bTheta \bY \bB \bY^{\top} \bTheta \bD^{-1/2}$. Substituting into $\bY^{\top} \bar{\bL} \bY$, 
    \[
    \bY^{\top}\bD^{-1/2}\bTheta \bY \bB \bY^{\top} \bTheta \bD^{-1/2}\bY \succeq 
    \lambda_K(\bB)(\bY^{\top}\bD^{-1/2}\bTheta \bY)^2
    \succeq \lambda_K(\bB)\lambda_K^2(\bY^{\top}\bD^{-1/2}\bTheta \bY) \bI.
    \]
    By assumptions, $\lambda_K(\bB) \geq c_B$ is a non-zero constant. 
    The matrix $\bY^{\top}\bD^{-1/2}\bTheta \bY$ is diagonal, with $(k,k)$-th entry being $\sum_{i: \ell(i)=k} {\theta_i}/{\sqrt{d_i}}$. By Lemma \ref{lem:di}, with probability $1 - O(n^{-1})$,
    \[
    \sum_{i: \ell(i)=k} \frac{\theta_i}{\sqrt{d_i}}
    \geq \sum_{i: \ell(i)=k}\frac{\theta_i}{0.9\sqrt{\theta_i} \sqrt{\sum_j \theta_j }}\geq C\rho \sqrt{n} \geq C\sqrt{n}.
    \]
    Therefore, $\lambda_K(\bY^{\top}\bD^{-1/2}\bTheta \bY)\geq C \sqrt{n}$, and 
    \[
    \lambda_K(\bY^{\top} \bar{\bL} \bY) = \lambda_K(\bY^{\top}\bD^{-1/2}\bTheta \bY \bB \bY^{\top} \bTheta \bD^{-1/2}\bY)
    \geq C^2\lambda_K(\bB) n. 
    \]
    Combining this with \eqref{eqn:diffgoal}, where $d_m = n\tmax^2 \geq c_{\theta}\log n$, we conclude that with probability $1 - O(1/n)$, when $n \geq n_0$ for some given $n_0$, 
    \[
    \lambda_K(\bY^{\top} \bXi_K \bXi_K^{\top} \bY) \geq (C^2 \lambda_K(\bB)  - C/\sqrt{c_{\theta}\log n})n 
    \geq c_D n.
    \]
    \item RDPG. Under RDPG, $\bar{\bL} = \rho_n \bD^{-1/2} \bY \bY^{\top} \bD^{-1/2}$. For this case, by Lemma \ref{lem:diRDPG}, the lower bound can be determined with probability $1 - O(n^{-1})$:
    \[
    \bY^{\top}\bD^{-1/2}\bY\bY^{\top} \bD^{-1/2}\bY\succeq 
    \lambda_K^2(\bY^{\top}\bD^{-1/2}\bY) \bI\succeq  {\lambda_K^4(\bY)}\bI/1.1d_m. 
    \]
    By Lemma \ref{lem:ymspectrum}, we have 
    $\lambda_K(\bY^{\top} \bY) \geq c^2 n$ under Assumption \ref{aspt:latentrandomnew}. 
    Therefore, 
    \[
    \lambda_K(\bY^{\top} \bar{\bL} \bY) = 
    \lambda_K(\bY^{\top}\bD^{-1/2}(\rho_n \bY\bY^{\top}) \bD^{-1/2}\bY)
    \geq {c^2n^2\rho_n }/{d_m} = c^2n. 
    \]
    Combining this with \eqref{eqn:diffgoal},with probability $1 - O(1/n)$, when $n \geq n_0$, 
    \[
    \lambda_K(\bY^{\top} \bXi_K \bXi_K^{\top} \bY) \geq c^2n - Cn/\sqrt{c_{\rho}\log n} 
    \geq c_{lp} n.
    \]
\end{itemize}
With \eqref{eqn:cam}, it further leads to our claim when $\bXi_{\Khat}$ is used:
\[
\signal(\bXi_{\Khat})^2 \geq \left\{\begin{array}{ll}
c_D n \min_{j\in \cS}\|\bM_j\|^2, & \mbox{for DCSBM},\\
c_{lp} n \min_{j\in \cS}\|\bM_j\|^2, & \mbox{for RDPG},
\end{array}
\right.
\]

For the case when eigenvectors of $\bA$ are used, the same proof can be recycled. All we need to do is replace $\bD$ with a diagonal matrix $\tilde{\bD} = (\log n) \bI$. It is easy to check all arguments above hold for $\tilde{\bD}^{-1/2}\bA \tilde{\bD}^{-1/2}$, which has the same eigenvectors that $\bA$ has.  
\end{proof}

\subsection{Proof of Theorem \ref{thm:cluster}: Clustering Consistency}
When $\cS$ is recovered with small errors, it is not surprising that the clustering problem can be solved. The accuracy will depend on the accuracy of $\cShat$, which is demonstrated in Theorem \ref{thm:cluster}. In this section, we will show the proof of this result. 

Recall that $\bX$ is the full data matrix consisting of samples in the network and samples out of the network. Let $\cShat$ be the selected covariates by Algorithm \ref{tab:alg1}. 
Let $\bY$ be the full label matrix and $\bM^{\cShat}$ be the sub-matrix of $\bM$ restricted on $\cShat$ columns. 
For the post-selection matrix, $\bY \bM^{\cShat}$ is the signal part.
In the proof, we only consider the post-selection matrix $\bX^{\cShat}$, so we ignore the superscript $\cShat$ for $\bX^{\cShat}$ and $\bM^{\cShat}$ throughout this section. 

To start with, we need the following results on post-selection random matrix theory. 
\begin{lemma}[Operator norm of the post-selected matrix]
\label{prop:post-selection-operator-norm}
Suppose Assumptions \ref{aspt:indep}, \ref{aspt:rw}--\ref{aspt:dcsbm}, and \ref{aspt:covcluster}--\ref{aspt:cluster} hold, and $\epsilon = p^{-\beta}$ for $0 < \beta < 1$. If the signal strength $\kappa > \sqrt{\max\{16-16\beta, 14\}\log p/(c_Dn)}$. 
Define the selection set
$\cShat
= 
\{\, j \in \{1,\dots,p\} : \|\bU^\top \bZ_j\|^2 \;\ge\; K + c \log p \}$, and let $\bZ^{\cShat} \in \mathbb{R}^{n\times |S|}$ be the post-selected matrix.
Then, there exists absolute constants $c, C > 0$, so that when $N \gg \sqrt{\Khat\log p}$, with probability $1 - p^{-c}$, there is 
\begin{equation}
\label{eq:Z_S-operator-norm}
\|\bZ^{\cShat}\|
\;\le\; 2(\sqrt{|\cShat|} + \sqrt{CN}\log p).
\end{equation}
\end{lemma}

\begin{proof}
Note that by Theorem 5.39 of \cite{vershynin2010introduction}, for each $i\in[p]$,  $Z e_i$ is a Gaussian vector and $\max_{i\in [p]}\|Z e_i\|\leq 2\sqrt{N}$ with probability $1-p\exp(-cN)$. 
\begin{equation}
\|\bZ^{\cShat\backslash \cS}\|
\;\le\; 
\sqrt{|\cShat\backslash \cS|}
\{\sqrt{N}
+ C\sqrt{K\log p}
+ C\sqrt{\log p}\} \leq \sqrt{CN|\cShat\backslash \cS|\log p}.
\end{equation}
Consider the post-selection noise matrix $\bZ$ and the good set $\cG = \{\cS \subset \cShat, |\cShat\backslash\cS| \leq C\log^2 p\}$, which happens with probability at least $1 - O(1/p)$ by Corollary \ref{cor:main}. 
Therefore, $|\cShat\backslash \cS| \leq C\log^2 p$. contribute $\sqrt{CN\log^2 p}$. Meanwhile, $\|Z^{\cS}\|\leq 2C(\sqrt{N}+\sqrt{|\cS|})$ takes place with probability $1-\exp(-cN)$. 
Then note that for any $\cShat$, with probability $1- O(1/p)$, 
\[
\|Z^{\cShat}\|\leq \|Z^{\cS}\|+\sum_{i\in \cShat/\cS}\|Z e_i\|
\leq 2(\sqrt{|\cShat|} + \sqrt{CN}\log p).
\] 
In combination we have our claim.
\end{proof}

\begin{proof}[Proof of Theorem \ref{thm:cluster}]
In the proof, we always denote the SVD of $\bX = \hat{\bU} \hat{\bLambda} \hat{\bV}^{\top}$ and $\bY \bM = \bU \bLambda \bV^{\top}$. Since $\text{rank}(\bM) = r \leq K$, we have $\bU \in \reals^{N \times r}$, $\bLambda \in \reals^{r \times r}$, and $\bV \in \reals^{|\cShat| \times r}$. 
Let $\hat{\bU}_K$ and $\hat{\bV}_K$ be the matrices consisting of the first $K$ columns of $\hat{\bU}$ and $\hat{\bV}$, respectively. Let $\hat{\bLambda}_K$ be the diagonal matrix consisting of the largest $K$ singular values.

\textbf{Step 1: The ideal spectrum and the empirical spectrum.} 
After selection, we have $\bX = \bY \bM + \bZ$, where the noise $\bZ$ is no longer Gaussian distributed. 
By Lemma \ref{prop:post-selection-operator-norm}, we have the control on post-selection noise matrix that 
\[
\|\bX - \bY \bM\| \leq 2(\sqrt{|\cShat|} + \sqrt{CN}\log p).
\]
By Proposition \ref{prop:davis}, a variant of the Davis--Kahan theorem when $K \geq r = \text{rank}(\E[\bX])$ singular vectors are considered. There exists an orthogonal matrix $\bO \in \reals^{r \times r}$ such that 
\begin{align}\label{tmp:D1}
\|\hat{\bU}_K \hat{\bLambda}_K - [\bU \bLambda \bO, {\bf 0}] \|_F &\leq \left(\frac{\lambda_1(\bY \bM)}{\lambda_r(\bY \bM)} + 2\right)\sqrt{K}\|\bX - \bY \bM\| \nonumber\\
&\leq \frac{6\sqrt{K}}{\rho c_M}(\sqrt{|\cShat|} + \sqrt{CN}\log p) =: D_1.
\end{align}
Here, the last inequality follows by Lemma \ref{lem:condprod} and Assumption \ref{aspt:cluster}. In detail, 
\[
\frac{\lambda_1(\bY \bM)}{\lambda_r(\bY \bM)} \leq \frac{\lambda_1(\bY) \lambda_1(\bM)}{\lambda_r(\bY) \lambda_r(\bM)} 
\leq \frac{1}{\rho c_M}.
\]

Denote the rows of $\hat{\bU}_K \hat{\bLambda}_K$ by $\hat{\bfeta}_i$ and the rows of $[\bU \bLambda \bO, {\bf 0}]$ by $\bfeta_i$, for $i \in [N]$. Then, 
\[
\bfeta_i = \bfe_i^{\top}[\bU \bLambda \bO, {\bf 0}] = \bfe_i^{\top}[\bY \bM \bV \bO, {\bf 0}] = [\bmu_{\ell(i)}^{\top} \bV \bO, {\bf 0}],
\]
where $\ell(i)$ is the class label of the $i$-th data point and $\bmu_{\ell(i)}^{\top}$ is the $\ell(i)$-th row of $\bM$. Naturally, all the observations with the same label share identical $\bfeta_i$. If we apply $k$-means on the ideal spectrum, we will achieve perfect clustering results. 

Finally, according to Assumption \ref{aspt:cluster} and the relationship between $\bfeta$ and $\bmu$, we have 
\[
\|\bfeta_i\| \geq \kappa c_M \sqrt{|\cS|}, \quad \|\bfeta_i - \bfeta_j\| \geq c_M \kappa \sqrt{|\cS|} =: c_\eta, \quad \text{if } \ell(i) \neq \ell(j).
\]


\textbf{Step 2: The estimated centers $\hat{\bm}_k$ by $k$-means match the true centers.} 
For any estimated label $\hat{\bell}$ and centers $\hat{\bm}_k$, define the within-cluster distance as 
$L(\hat{\bell}, \hat{\bm}) = \sum_{i=1}^n \| \hat{\bfeta}_i - \hat{\bm}_{\hat{\ell}(i)}\|^2$.
The $k$-means algorithm aims to find $\hat{\bell}$ and $\hat{\bm}$ that minimize $L(\hat{\bell}, \hat{\bm})$. 

Let $\bm_k = [\bmu_k^{\top} \bV \bO, {\bf 0}]$, then $\bfeta_i = \bm_{\ell(i)}$. For the true labels $\bell$ and centers $\bm_k$, we have 
$L(\bell, \bm) = \sum_{i=1}^n \| \hat{\bfeta}_i - \bm_{\ell(i)} \|^2$,
which results in a loss:
\begin{align}
\label{tmp:Lcontrol}
L(\bell, \bm) &= \sum_{i=1}^n \| \hat{\bfeta}_i - \bm_{\ell(i)} \|^2 = \sum_{i=1}^n \| \hat{\bfeta}_i - \bfeta_i \|^2 \leq D_1^2.
\end{align}
By the optimality of $\hat{\ell}$ and $\hat{\bm}$, we have the upper bound:
\begin{align}
\label{tmp:Lupp}
L(\hat{\bell}, \hat{\bm}) \leq L(\bell, \bm).
\end{align}

Now, we match $\hat{\bm}$ with $\bm$. For any class $k$, define the permutation as
$\pi(k) = \arg\min_{1 \leq j \leq K} \| \bm_k - \hat{\bm}_j \|$.
Hence, $\pi(k)$ is the class where the estimated center is closest to $\bm_k$. We consider the within-cluster distance for class $k$, which gives:
\begin{align}
\label{tmp:Llower}
L(\hat{\bell}, \hat{\bm}) &\geq \sum_{\ell(i) = k} \| \hat{\bfeta}_i - \hat{\bm}_{\hat{\ell}(i)} \|^2 \nonumber \\
&\geq \sum_{\ell(i) = k} \left(- \| \hat{\bfeta}_i - \bm_k \|^2 + \frac{1}{2} \| \bm_k - \hat{\bm}_{\hat{\ell}(i)} \|^2 \right) \nonumber \\
&\geq - \sum_{i=1}^n \| \hat{\bfeta}_i - \bm_{\ell(i)} \|^2 + \frac{1}{2} \sum_{\ell(i) = k} \| \bm_k - \hat{\bm}_{\hat{\ell}(i)} \|^2 \nonumber \\
&\geq - L(\bell, \bm) + \frac{1}{2} N_k \| \bm_k - \hat{\bm}_{\pi(k)} \|^2.
\end{align}

Combining \eqref{tmp:Lupp} and \eqref{tmp:Llower}, we obtain:
\begin{equation}\label{eqn:mupp}
\| \bm_k - \hat{\bm}_{\pi(k)} \|^2 \leq \frac{4L(\bell, \bm)}{N_k} \leq \frac{4D_1^2}{\rho N}.
\end{equation}

Recall that the centers $\{ \bm_k, k = 1, \ldots, K \}$ are apart by Step 1. Under our assumption 
$\kappa > \frac{6(r+K)}{\rho c_M^2 \sqrt{\rho}} \frac{\sqrt{|\cShat|} + \sqrt{N}}{\sqrt{N |\cS|}}$ and the definition of $D_1$ in \eqref{tmp:D1},
the distance follows that 
\[
c_\eta = c_M \kappa \sqrt{|\cS|} > \frac{6\sqrt{K}}{\rho c_M \sqrt{\rho}} \frac{\sqrt{|\cShat|} + \sqrt{N}}{\sqrt{N}} = \frac{4D_1}{\sqrt{\rho N}}.
\]
Thus, each $\bm_k$ is paired with a unique $\hat{\bm}_{\pi(k)}$. 

\textbf{Step 3: The mis-classification rate is small.} 
To simplify the notations, we assume $\pi(k) = k$ without loss of generality. Then the misclassified nodes are $B = \{i: \ell(i) \neq \hat{\ell}(i)\}$. The misclassification rate is $Err_n = |B|/N$. 
\begin{align*}
 L(\hat{\bell}, \hat{\bm}) = \sum_{i=1}^N \| \hat{\bfeta}_i - \hat{\bm}_{\hat{\ell}(i)}\|^2 &\geq -\sum_{i=1}^N \| \hat{\bfeta}_i - \bm_{\ell(i)}\|^2 + \frac{1}{2} \sum_{i=1}^N \| \bm_{\ell(i)} - \hat{\bm}_{\hat{\ell}(i)}\|^2 \\
&\geq - L(\bell, \bm) + \frac{1}{2} \sum_{i \in B} \| \bm_{\ell(i)} - \hat{\bm}_{\hat{\ell}(i)}\|^2 \\
\mbox{by \eqref{eqn:mupp}}\qquad &\geq - L(\bell, \bm) + \frac{1}{2} \sum_{i \in B} (\| \bm_{\ell(i)} - \bm_{\hat{\ell}(i)} \| - 2D_1 / \sqrt{\rho N})^2 \\
&\geq - L(\bell, \bm) + c_\eta^2 |B|/8.
\end{align*}

Combining this with $L(\hat{\bell}, \hat{\bm}) \leq L(\bell, \bm)$ in (\ref{tmp:Lupp}) and $L(\bell, \bm) \leq D_1^2$ in (\ref{tmp:Lcontrol}), we obtain:
\[
Err_n = \frac{|B|}{N} \leq \frac{16}{N c_\eta^2} D_1^2 \leq C \frac{|\cShat| + N \log^2 p}{N |\cS| \kappa^2}.
\]
So, Theorem \ref{thm:cluster}  is proved. 
\end{proof}

\begin{prop}
\label{prop:davis}
Suppose $\bX \in \reals^{n_1 \times n_2}$ has rank $r$ and its SVD is given by $\bX = \bU \bLambda \bV^{\top}$. Denote $\hat{\bX} = \bX + \bE$. For any $K \geq r$, let $\hat{\bU}_K$ and $\hat{\bLambda}_K$ consist of the first $K$ left singular vectors and singular values of $\hat{\bX}$. Then there is an orthogonal matrix $\bO$, such that 
\begin{align*}
&\|\hat{\bU}_K \hat{\bLambda}_K - [\bU \bLambda \bO^{\top}, {\bf 0}]\| \leq \left(\frac{\lambda_1(\bX)}{\lambda_r(\bX)} + 2\right)\|\bE\|,\\
&\|\hat{\bU}_K \hat{\bLambda}_K - [\bU \bLambda \bO^{\top}, {\bf 0}]\|_F \leq \sqrt{K}
\left(\frac{\lambda_1(\bX)}{\lambda_r(\bX)} + 2\right)\|\bE\|.
\end{align*}
\end{prop}

\begin{proof}
We write the SVD of 
\[
\hat{\bX} = \hat{\bU} \hat{\bLambda} \hat{\bV}^{\top} = [\hat{\bU}_r, \hat{\bU}_-]
\begin{bmatrix}
\hat{\bLambda}_r & {\bf 0} \\
{\bf 0} & \hat{\bLambda}_-
\end{bmatrix}
[\hat{\bV}_r, \hat{\bV}_-]^{\top},
\]
where $\hat{\bU}_r \in \reals^{n_1 \times r}$, $\hat{\bU}_- \in \reals^{n_1 \times (\min\{n_1, n_2\} - r)}$, and $\hat{\bLambda}_r \in \reals^{r \times r}$ is the upper-left $r \times r$ submatrix of $\hat{\bLambda}$. $\hat{\bV}_r$ and $\hat{\bV}_{-}$ are defined similarly. 

We first consider the largest $r = \text{rank}(\bX)$ singular vectors. Note that 
\[
\bU \bLambda = \bX \bV, \quad [\hat{\bU}_r \hat{\bLambda}_r, \hat{\bU}_- \hat{\bLambda}_-] = \hat{\bU} \hat{\bLambda} = \hat{\bX} \hat{\bV} = (\bX + \bE)[\hat{\bV}_r, \hat{\bV}_-]. 
\]

By the Davis–Kahan Theorem, there is an orthogonal matrix $\bO$ such that $\|\hat{\bV}_r - \bV \bO^{\top}\| \leq {\|\bE\|}/{\lambda_r(\bX)}$. Therefore, 
\[
\hat{\bU}_r \hat{\bLambda}_r - \bU \bLambda \bO^{\top} = (\bX + \bE)\hat{\bV}_r - \bX \bV \bO^{\top} = 
(\bX + \bE)(\hat{\bV}_r - \bV \bO^{\top}) + \bE \bV \bO^{\top}.
\]
Hence, the norm can be controlled by
\[
\|\hat{\bU}_r \hat{\bLambda}_r - \bU \bLambda \bO^{\top}\| \leq \|\bX\| \|\hat{\bV}_r - \bV \bO^{\top}\| + \|\bE\| \|\hat{\bV}_r\| \leq \left(\frac{\lambda_1(\bX)}{\lambda_r(\bX)} + 1\right)\|\bE\|.
\]

Now we consider the first $K$ singular vectors, where $K \geq r$. The additional singular vectors and values are denoted as $\hat{\bU}_{r:K} = [\hat{\bu}_{r+1}, \hat{\bu}_{r+2}, \ldots, \hat{\bu}_K]$ and $\hat{\bLambda}_{R:K}$ with diagonals $\lambda_{r+1}, \ldots, \lambda_K$. Since $\bX$ has rank $r$, it follows that $\|\hat{\bLambda}_{r:K}\| \leq \lambda_{r+1}(\hat{\bX}) \leq \|\bE\|$, and 
\[
\|\hat{\bU}_K \hat{\bLambda}_K - [\bU \bLambda \bO^{\top}, {\bf 0}]\| \leq \|\hat{\bU}_r \hat{\bLambda}_r - \bU \bLambda \bO^{\top}\| + \|\hat{\bU}_{r:K} \hat{\bLambda}_{r:K}\| \leq \left(\frac{\lambda_1(\bX)}{\lambda_r(\bX)} + 2\right)\|\bE\|.
\]

Since $\hat{\bU}_K \hat{\bLambda}_K - [\bU \bLambda \bO^{\top}, {\bf 0}]$ has rank at most $K$, we have the Frobenius norm:
\[
\|\hat{\bU}_K \hat{\bLambda}_K - [\bU \bLambda \bO^{\top}, {\bf 0}]\|_F \leq \sqrt{K} \|\hat{\bU}_K \hat{\bLambda}_K - [\bU \bLambda \bO^{\top}, {\bf 0}]\| \leq \left(\frac{\lambda_1(\bX)}{\lambda_r(\bX)} + 2\right)\sqrt{K}\|\bE\|.
\]
\end{proof}

\subsection{Proof of Theorem \ref{thm:regression}: Consistency of Regression}

\begin{proof} 
For simplification, we ignore the superscript $\cShat$ for $\bX^{\cShat}_{(2)}$, $\bx^{\cShat}_{new}$, and $\bM^{\cShat}$ throughout this proof.
Recall that $s = |\cS|$ and $\hat{s} = |\cShat|$. We take $\bX_{(2)} = \hat{\bU}\hat{\bLambda} \hat{\bV}^{\top}$ and $\E[\bX_{(2)}] = \bY_{(2)} \bM = \bU \bLambda \bV^{\top}$ to differentiate the ideal spectrum and the empirical spectrum. Hence, by the NG-reg Algorithm, we have $\hat{\bgamma} = \hat{\bV}_{\Khat} \hat{\bLambda}_{\Khat}^{-1} \hat{\bU}_{\Khat}^{\top} \bz$. 

The prediction error is given by 
\[
\hat{\bgamma}^{\top}(\bM^{\top} \by_{new} + \beps_{new}) - \bbeta^{\top} \by_{new} = (\hat{\bgamma}^{\top}\bM^{\top} - \bbeta^{\top})\by_{new} + \hat{\bgamma}^{\top}\beps_{new}.
\]
We have to bound $\|\bM\hat{\bgamma} - \bbeta\|$ and $\|\hat{\bgamma}\|$. 
Introduce $\hat{\bgamma} = \hat{\bV}_{\Khat} \hat{\bLambda}_{\Khat}^{-1} \hat{\bU}_{\Khat}^{\top}(\bY_{(2)} \bbeta + \bdelta)$ into the error term. Since $\beps \sim \mathcal{N}({\bf 0}, \bI)$, we have the following with high probability:
\[
\|\bM\hat{\bV}_{\Khat} \hat{\bLambda}_{\Khat}^{-1} \hat{\bU}_{\Khat}^{\top}(\bY_{(2)} \bbeta + \bdelta) - \bbeta\|
\lesssim \|\bM \hat{\bV}_{\Khat} \hat{\bLambda}_{\Khat}^{-1} \hat{\bU}_{\Khat}^{\top} \bY_{(2)} - \bI\|\|\bbeta\| + \|\bM \hat{\bV}_{\Khat} \hat{\bLambda}_{\Khat}^{-1} \hat{\bU}_{\Khat}^{\top}\|\sigma_{\delta}.
\]

\textbf{Bounding the first term of $\|\bM\hat{\bgamma} - \bbeta\|$.} The coefficient vector $\bbeta$ is fixed and $\|\bbeta\| = O(1)$ in the assumption. To bound the first term, we only need to consider $\|\bM \hat{\bV}_{\Khat} \hat{\bLambda}_{\Khat}^{-1} \hat{\bU}_{\Khat}^{\top} \bY_{(2)} - \bI\|$. Note that although we consider the first $\Khat$ singular vectors, this matrix is a $K \times K$ matrix. By Lemma \ref{lem:prod}, we have 
\[
\|\bM \hat{\bV}_{\Khat} \hat{\bLambda}_{\Khat}^{-1} \hat{\bU}_{\Khat}^{\top} \bY_{(2)} - \bI\| \leq 
\frac{\|\bY_{(2)}(\bM \hat{\bV}_{\Khat} \hat{\bLambda}_{\Khat}^{-1} \hat{\bU}_{\Khat}^{\top} \bY_{(2)} - \bI)\|}{\lambda_K(\bY)}. 
\]
We decompose $\hat{\bU}_{\Khat}$ into two parts: $\hat{\bU}_K$ that contains the first $K$ columns of $\hat{\bU}_{\Khat}$ and $\hat{\bU}_{\bot}$ that consists of the last $\Khat - K$ columns of $\hat{\bU}_{\Khat}$. Similarly, we define $\hat{\bV}_K$, $\hat{\bV}_{\bot}$, $\hat{\bLambda}_K$, and $\hat{\bLambda}_{\bot}$. According to the properties of singular values and vectors, there is:
\begin{align}\label{eqn:term1}
& \|\bY_{(2)}(\bM \hat{\bV}_{\Khat} \hat{\bLambda}_{\Khat}^{-1} \hat{\bU}_{\Khat}^{\top} \bY_{(2)} - \bI)\| 
 \leq 
\|\bY_{(2)}(\bM \hat{\bV}_K \hat{\bLambda}_K^{-1} \hat{\bU}_K^{\top} \bY_{(2)} - \bI)\|
+ 
\|\bY_{(2)} \bM \hat{\bV}_{\bot} \hat{\bLambda}_{\bot}^{-1} \hat{\bU}_{\bot}^{\top} \bY_{(2)}\|\nonumber\\
&\qquad \leq  {\|(\bY_{(2)}\bM - \hat{\bU}_K \hat{\bLambda}_K \hat{\bV}_K^{\top})\hat{\bV}_K \hat{\bLambda}_K^{-1}\hat{\bU}_K^{\top} \bY_{(2)}\|}
+{\|(\hat{\bU}_K\hat{\bU}_K^{\top} - \bI)\bY_{(2)}\| + 
\|\bY_{(2)} \bM \hat{\bV}_{\bot} \hat{\bLambda}_{\bot}^{-1} \hat{\bU}_{\bot}^{\top} \bY_{(2)}\|}\nonumber\\
& \qquad = (a) + (b) + (c). 
\end{align}
We then discuss each part.
\begin{itemize}
\item Part (a) can be further decomposed into $\|\bY_{(2)}\bM - \hat{\bU}_K \hat{\bLambda}_K \hat{\bV}_K^{\top}\|$ and $\|\hat{\bV}_K \hat{\bLambda}_K^{-1}\hat{\bU}_K^{\top} \bY_{(2)}\|$. By the random matrix theory in \cite{vershynin2010introduction}, with probability $1 - o((N-n)^{-4})$, there is $\|\bX_{(2)} - \bY_{(2)}\bM\| \leq 2(\sqrt{N-n} + \sqrt{\cshat})$. Hence,  with probability $1 - o((N-n)^{-4})$, 
\begin{align*}
\|\bY_{(2)}\bM - \hat{\bU}_K \hat{\bLambda}_K \hat{\bV}_K^{\top}\|  \leq \|\bY_{(2)}\bM - \bX_{(2)}\| + \|\bX_{(2)} - \hat{\bU}_K \hat{\bLambda}_K \hat{\bV}_K^{\top}\|\leq 4(\sqrt{N-n} + \sqrt{\cshat}). 
\end{align*}
By Lemma \ref{lem:ymspectrum}, we have $\|\bY_{(2)}\|/\lambda_K(\bX_{(2)}) \leq \sqrt{n}/(c\sqrt{N-n}\|\bM\| - 2(\sqrt{\cshat} + \sqrt{N-n}))$. 

\item Consider Part (b). According to Lemma \ref{lem:ymspectrum}, $\lambda_K(\bY_{(2)}\bM) \geq C\kappa\sqrt{(N-n)s}$. 
By the Davis--Kahan Theorem and noting that $\bY_{(2)}\bM = \bU \bLambda \bV^{\top}$, there is an orthogonal matrix $\bO_1$
so that 
\[
\|\hat{\bU}_K - \bU\bO_1\|\leq C\frac{\sqrt{N-n}+\sqrt{\cshat}}{\lambda_K(\bY_{(2)}\bM)} \leq C\frac{\sqrt{N-n}+\sqrt{\cshat}}{\kappa\sqrt{(N-n)s}}
. 
\]
Note that $\bU \bU^{\top} \bY_{(2)} = \bY_{(2)}$. 
Since $\|\hat{\bU}_K\|, \|\bU\|\leq 1$, we have 
\[
\|(\hat{\bU}_K\hat{\bU}_K^{\top} - \bI)\bY_{(2)}\|=
\|(\hat{\bU}_K\hat{\bU}_K^{\top} - \bU\bU^{\top})\bY_{(2)}\|
\leq 2\|\hat{\bU}_K - \bU \bO_1\|\|\bY_{(2)}\| \leq 
C\frac{\sqrt{N-n}+\sqrt{\cshat}}{\kappa\sqrt{s}}.
\]

\item Now we check Part (c). 
By Lemma \ref{lem:semi}, with probability $1 - O((N-n)^{-4})$, the $\Khat$-th singular value follows $\lambda_{\Khat}(\bX) \geq \sqrt{N-n+\cshat-K}$. By Lemma \ref{lem:ymspectrum}, $\lambda_K(\bY_{(2)}\bM)\geq C\|\bY_{(2)}\bM\|$. By the Davis--Kahan Theorem, 
\begin{align*}
&\|\bY_{(2)}\bM \hat{\bV}_{\bot} \hat{\bLambda}_{\bot}^{-1} \hat{\bU}_{\bot}^{\top} \bY_{(2)}\|\leq \frac{\|\bY_{(2)}\bM \hat{\bV}_{\bot}\|\|\bY_{(2)}^{\top}\hat{\bU}_{\bot}\|}{\sqrt{N-n + \cshat - K}} \\
&\qquad \leq 
\frac{C\|\bY_{(2)}\bM\|\|\bY_{(2)}\|(\sqrt{N-n}+\sqrt{\cshat})^2}{\sqrt{N-n+\cshat - K}\lambda_K^2(\bY_{(2)}\bM)}\leq \frac{C(\sqrt{N-n}+\sqrt{\cshat})}{\kappa\sqrt{s}}.
\end{align*}
\end{itemize}

Finally, we combine parts (a), (b), (c) with Assumption \ref{aspt:latentrandomnew} that $\lambda_K(\bY_{(2)}) \geq C\sqrt{N-n}$. Note that $\|\bM\| \geq \kappa \sqrt{s/K}$. 
It follows that with probability at least $1 - O((N-n)^{-2})$, 
\begin{align}\label{eqn:firstone}
 &   \|(\bM \hat{\bV}_{\Khat} \hat{\bLambda}_{\Khat}^{-1} \hat{\bU}_{\Khat}^{\top} \bY_{(2)} - \bI)\|  \leq 
\frac{\|\bY_{(2)}(\bM \hat{\bV}_{\Khat} \hat{\bLambda}_{\Khat}^{-1} \hat{\bU}_{\Khat}^{\top} \bY_{(2)} - \bI)\|}{\lambda_K(\bY_{(2)})}\nonumber\\
&\qquad \leq \frac{C}{C\sqrt{N-n}}\left(\frac{4(\sqrt{N-n}+\sqrt{\cshat})\sqrt{N-n}}{c\sqrt{N-n}\|\bM\| - 2(\sqrt{N-n}+\sqrt{\cshat})} + \frac{\sqrt{N-n}+\sqrt{\cshat}}{\kappa\sqrt{s}} + \frac{\sqrt{N-n}+\sqrt{\cshat}}{\kappa\sqrt{s}}\right)\nonumber\\
&\qquad \leq \frac{C(\sqrt{N-n}+\sqrt{\cshat})}{\sqrt{(N-n)s}\kappa}.
\end{align}

\textbf{Bounding the second term of $\|\bM\hat{\bgamma} - \bbeta\|$.} 
The second term can be written as 
\begin{align*}
\|\bM\hat{\bV}_{\Khat}\hat{\bLambda}_{\Khat}^{-1}\hat{\bU}_{\Khat}^{\top}\|\sigma_{\delta}
& \leq \frac{1}{\lambda_K(\bY_{(2)})}\|\bM\hat{\bV}_{\Khat}\hat{\bLambda}_{\Khat}^{-1}\hat{\bU}_{\Khat}^{\top}\bY_{(2)}\|\sigma_{\delta}\\
& \leq \frac{\sigma_{\delta}}{\lambda_K(\bY_{(2)})}(\|\bM\hat{\bV}_{\Khat}\hat{\bLambda}_{\Khat}^{-1}\hat{\bU}_{\Khat}^{\top}\bY_{(2)} - \bI\| + 1) 
\end{align*}
In \eqref{eqn:firstone}, the upper bound of $\|\bM\hat{\bV}_{\Khat}\hat{\bLambda}_{\Khat}^{-1}\hat{\bU}_{\Khat}^{\top}\bY_{(2)} - \bI\|$ is given. Introduce it in, then  
\begin{equation}\label{eqn:secondone}
    \|\bM\hat{\bV}_{\Khat}\hat{\bLambda}_{\Khat}^{-1}\hat{\bU}_{\Khat}^{\top}\|\sigma_{\delta}
    \leq \frac{C\sigma_{\delta}}{\sqrt{N-n}} \left(1 + \frac{C(\sqrt{N-n}+\sqrt{\cshat})}{\sqrt{(N-n)s}\kappa}\right)
\end{equation}

\textbf{Bounding $\|\hat{\bgamma}\|$.}
We note that 
\[
\|\hat{\bgamma}\| = \|\hat{\bV}_{\Khat} \hat{\bLambda}_{\Khat}^{-1} \hat{\bU}_{\Khat}^{\top} (\bY_{(2)}\bbeta + \bdelta)\| \leq 
\|\bbeta\|\|\hat{\bV}_{\Khat} \hat{\bLambda}_{\Khat}^{-1} \hat{\bU}_{\Khat}^{\top} \bY_{(2)}\| + \|\hat{\bV}_{\Khat} \hat{\bLambda}_{\Khat}^{-1} \hat{\bU}_{\Khat}^{\top} \|\sigma_{\delta}.
\]
For the first term, again we apply Lemma \ref{lem:prod} and \eqref{eqn:firstone}, and there is 
\begin{align*}
\|\hat{\bV}_{\Khat}\hat{\bLambda}_{\Khat}^{-1}\hat{\bU}_{\Khat}^{\top}\bY_{(2)}\|
& \leq \frac{1}{\lambda_K(\bM)}\|\bM\hat{\bV}_{\Khat}\hat{\bLambda}_{\Khat}^{-1}\hat{\bU}_{\Khat}^{\top}\bY_{(2)}\|\\
& \leq \frac{1}{\lambda_K(\bM)}\left(\|\bM\hat{\bV}_{\Khat}\hat{\bLambda}_{\Khat}^{-1}\hat{\bU}_{\Khat}^{\top}\bY_{(2)} - \bI\| + 1\right)  \leq \frac{C}{\kappa\sqrt{s}}\left(\frac{\sqrt{N-n} + \sqrt{\cshat}}{\kappa\sqrt{(N-n)s}} + 1\right).
\end{align*}
The last term is bounded by $1/\lambda_{\Khat}(\bX_{(2)}) \leq \frac{1}{\sqrt{N-n+\cshat-K}}$. 

Putting them together, we have 
\begin{equation}\label{eqn:gammabound}
    \|\hat{\bgamma}\| \leq \frac{C\|\bbeta\|}{\kappa\sqrt{s}}\left(\frac{\sqrt{N-n} + \sqrt{\cshat}}{\kappa\sqrt{(N-n)s}} + 1\right) + \frac{\sigma_{\delta}}{\sqrt{N-n+\cshat - K}}. 
\end{equation}

{\bf Conclusion.} We put \eqref{eqn:firstone}, \eqref{eqn:secondone} and \eqref{eqn:gammabound} together with the prediction error. Under the assumption $\kappa\sqrt{(N-n)s} \geq \sqrt{N-n}+\sqrt{\cshat}$, there is 
\begin{align*}
|\hat{\bgamma}^{\top} \bx_{new} - \bbeta^{\top} \by_{new}|
&\leq C\left[\frac{\|\bbeta\|(\sqrt{N-n}+\sqrt{\cshat})}{\kappa\sqrt{(N-n)s}} + \frac{\sigma_{\delta}}{\sqrt{N-n}} \right]\|\by_{new}\| 
+ C\left[\frac{\|\bbeta\|}{\kappa\sqrt{s}} + \frac{\sigma_{\delta}}{\sqrt{N-n+\cshat - K}}\right] \sigma_{\beps}\\
&\leq \frac{C\|\bbeta\|}{\kappa\sqrt{s}} \left(\frac{\sqrt{N-n}+\sqrt{\cshat}}{\sqrt{N-n}}\|\by_{new}\| + \sigma_{\beps}\right) 
+ \frac{\sigma_{\delta}(\|\by_{new}\|+\sigma_{\beps})}{\sqrt{N-n}}.
\end{align*}
When $\sigma_{\beps} = 1$ and the new data $\|\by_{new}\| = O(1)$, the conclusion follows that 
\[
|\hat{\bgamma}^{\top} \bx^{\cShat}_{new} - \bbeta^{\top} \by_{new}|
\lesssim \frac{\sqrt{N-n}+\sqrt{\cshat}}{\kappa\sqrt{(N-n)s}}\|\bbeta\|
+ \frac{C\sigma_{\delta}}{\sqrt{N-n}}.
\]
Theorem \ref{thm:regression} is proved. 
\end{proof}

\begin{lemma}\label{lem:ymspectrum}[Spectrum of $\bY\bM$ and $\bX$] 
Consider $\bY_{(2)}$, $\bM^{\cShat}$, and $\bX_{(2)}^{\cShat}$, where $\bx^{\cShat}_i \sim \mathcal{N}(\by_i^T \bM^{\cShat}, \bI)$ and $\cshat \geq p^{c}$ for a constant $c > 0$. Under Assumption \ref{aspt:latentrandomnew} and \ref{aspt:regression}, 
\begin{itemize}
    \item $\|\bY\| \leq \sqrt{n}$, $\lambda_K(\bY) \geq c_Y\|\bY\|$ for a constant $c_Y > 0$. Similarly, $\|\bY_{(2)}\| \leq \sqrt{N-n}$, $\lambda_K(\bY_{(2)}) \geq c_Y\|\bY_{(2)}\|$ for a constant $c_Y > 0$; 
    \item $\|\bM^{\cShat}\| \geq \kappa\sqrt{s/K}$; 
    \item there is a constant $c > 0$, such that $\lambda_K(\bY_{(2)}\bM^{\cShat}) \geq c\|\bY_{(2)}\bM^{\cShat}\|$;
    \item with probability at least $1 - O(n^{-4})$, 
    \[
    \|\bX^{\cShat}_{(2)}\| \leq \|\bY_{(2)}\bM^{\cshat}\| + 2(\sqrt{N-n}+\sqrt{\cshat}), \quad 
    \lambda_K(\bX^{\cShat}_{(2)}) \geq c\|\bY_{(2)}\bM^{\cShat}\| - 2(\sqrt{N-n}+\sqrt{\cshat}). 
    \]
\end{itemize}
\end{lemma}

\begin{lemma}\label{lem:prod}[Inequality of matrix norm for matrix product] 
Consider a matrix $\bA \in \reals^{(N-n) \times K}$ and a matrix $\bB \in \reals^{K \times n}$, then there is 
\[
\|\bA\bB\| \geq \|\bA\| \lambda_K(\bB). 
\]
\end{lemma}
\begin{proof}
    We write $\bB = \bU\bLambda \bV^{\top}$, where $\bLambda$ contains the singular values of $\bB$, and the minimum is $\lambda_K(\bB)$. Hence, we have 
\begin{align*}
\|\bA\bB\| = \lambda_1(\bA\bB)  & =  \lambda_1(\bA \bU\bLambda \bV^{\top}) = \sqrt{\lambda_1(\bA\bU\bLambda^2\bU^{\top} \bA^{\top})}\\
&\geq \lambda_K(\bB) \sqrt{\lambda_1(\bA\bU\bU^{\top} \bA^{\top})}=
\lambda_K(\bB) \sqrt{\lambda_1(\bU^{\top} \bA^{\top}\bA\bU)} = \|\bA\|\lambda_K(\bB).    
\end{align*}
The result is proved. 
\end{proof}

\begin{lemma}
\label{lem:semi}
Consider $\bZ = \bX + \bE$, where $\bE \in \reals^{n \times m}$ has entries being i.i.d. $\mathcal{N}(0,1)$ and $\bX \in \reals^{n \times m}$ is a matrix with rank $r$. Then for any $K \geq r$, $K = O(1)$, with probability $1 - O(n^{-4})$, we have the following 
\[
\lambda_K^2(\bX + \bE) \geq n - r + m.
\]
\end{lemma}
\begin{proof}
Without loss of generality, we assume $n > m$. Otherwise, we consider the transpose. 

Since $rank(\bX) = r < K$, the intuition is that $\lambda_K(\bX + \bE) \geq \lambda_K(\bE)$, which can be bounded by the semi-circle law. We show the mathematical analysis below. 

Denote $\bP \in \reals^{(n-r) \times n}$ as the projection onto the subspace orthogonal to the space spanned by the columns of $\bX$. Therefore, $\bP\bE \in \reals^{(n-r) \times m}$ has entries being i.i.d. $\mathcal{N}(0,1)$.
By \cite{alex2014isotropic}, with probability $1 - O(n^{-4})$, for a small constant $\epsilon > 0$, 
\[
\lambda_K^2(\bP\bE) \geq (\sqrt{n-r}+\sqrt{m})^2 - n^{\epsilon + 1/2}m^{-1/6}.
\]

Now we check the singular value of $\bX + \bE$ and $\bE$. For any vector $\bu$, there is 
\[
\|(\bX + \bE)\bu\|^2 \geq \|\bP(\bX + \bE)\bu\|^2 = \|\bP\bE \bu\|^2. 
\]
By the Courant-Fischer Theorem, 
$\lambda_K(\bP\bE) = \max_{S: \dim(S)=K}\min_{\bu \in S, \|\bu\|=1} \|\bP\bE \bu\|$.
The same holds for $\lambda_K(\bX + \bE)$. So, let $S$ be the subspace of dimension $K$ such that the maximum of $\min \|\bP\bE \bu\|$ is achieved, then there is 
\[
\lambda_K^2(\bX + \bE) \geq \min_{\bu \in S, \|\bu\| = 1} \|(\bX + \bE)\bu\|^2 \geq \min_{\bu \in S, \|\bu\| = 1} \|\bP\bE \bu\|^2 = \lambda_K^2(\bP\bE). 
\]

Introduce in the lower bound of $\lambda_K(\bP\bE)$ with a small $\epsilon$ such that $n^{\epsilon + 1/2}m^{-1/6} \leq \sqrt{(n-r)m}$, then 
$\lambda_K^2(\bX + \bE) \geq \lambda_K^2(\bP\bE) \geq (\sqrt{n-r}+\sqrt{m})^2 - n^{\epsilon + 1/2}m^{-1/6} \geq n - r + m. $
\end{proof}

\section{Proof of Secondary Results}\label{sec:supp}
We provide a series of secondary results in this section. The first subsection is about the adjacency matrix $\bA$. In detail, Lemmas \ref{lem:Anorm} present general results for any Bernoulli distributed matrix. Lemmas \ref{lem:diRDPG} and \ref{lem:di}  demonstrate the distribution of degrees under the RDPG and the DCSBM, respectively. Lemmas \ref{lem:LArdpg} and \ref{lem:LA} gives the spectrum of the adjacency matrix and the Laplacian under the RDPG and the DCSBM, respectively. The second subsection gives the proof of Lemma \ref{lem:ymspectrum}.

\begin{lemma}
\label{lem:Anorm}
Consider a symmetric Bernoulli matrix $\bA \in \reals^{n \times n}$ with $\bar{\bA}$ as the conditional expectation when the latent factor, i.e., $\bY$ in DCSBM and RDPG, is given. 
Then 
\[
\frac{1}{n} \sum_{1 \leq i,j \leq n} \bA_{ij} \leq \|\bA\| \leq \max_{i}\sum_j \bA_{ij}.
\]
Further, let $d_m = \max_{i \in [n]} \E[d_i]$ be the maximum expected degree and $\bW = \bA - \bar{\bA}$. 
Then there are constants $C, c > 0$, such that 
\[
\Prob(\|\bW\| \geq 6\sqrt{d_m}) \leq \exp(-c d_m).
\]
\end{lemma}
\begin{proof}
Since $\bA = \bA^{\top}$, $\|\bA\|_{1} = \|\bA\|_{\infty}$, and all the entries are non-negative. Hence, we have 
\[
\|\bA\| \leq \sqrt{\|\bA\|_{\infty}\|\bA\|_{1}} = \|\bA\|_{\infty} = \max_{i} \sum_j \bA_{ij}.
\]
Meanwhile, denote $\mathbf{1}/\sqrt{n}$ the vector with all entries being $1/\sqrt{n}$, then 
\[
\|\bA\| \geq  \frac{\mathbf{1}^{\top}}{\sqrt{n}} \bA \frac{\mathbf{1}}{\sqrt{n}} = 
\frac{1}{n} \sum_{1 \leq i,j \leq n} \bA_{ij} = \frac{1}{n} \sum_{i=1}^n d_i. 
\]

Now we check $\bW$. Since it follows a centered Bernoulli distribution, there is $\E[\bW] = 0$ and $\text{Var}(\bW_{ij}) = \bar{\bA}_{ij} (1 - \bar{\bA}_{ij}) \leq 1$. 
According to \cite[Corollary 3.12, Remark 3.13]{bandeira2016sharp}, there is a universal constant $c$ such that for any $t \geq 0$,  
\[
\Prob(\|\bW\| \geq 3\sqrt{2}\sigma + t) \leq \exp(-ct^2), \mbox{ where }
\]
\[
\sigma^2 = \max_{i}\sum_{j = 1}^n \text{Var}(\bW_{ij}) = \max_i \sum_{j} \bar{\bA}_{ij} (1 - \bar{\bA}_{ij}) \leq \max_i \sum_{j=1}^{n} \bar{\bA}_{ij} = d_m.
\]
Let $t = \sigma \leq \sqrt{d_m}$. Hence, we have 
$\Prob(\|\bW\| \geq 6\sigma) \leq \exp(-c d_m)$. 
Result proved. 
\end{proof}

\begin{lemma}
\label{lem:di}[Concentration of degrees for DCSBM]
Under Assumptions \ref{aspt:dcsbm}, there exists $n_0$ and $c_0$, so that when $n>n_0$ and $c_{\theta} > c_0$, with probability $1 - O(n^{-1})$,  
\[
\rho c_B n\tmax^2\leq \E[d_i]\leq n\tmax^2, 
\quad |d_i/\E[d_i]-1|\leq \delta, \text{ for all } i\in[n].
\]
\end{lemma}
\begin{proof}
Under the assumptions of DCSBM, $\E[d_i] = \theta_i \sum_{j \neq i} \theta_j \bB_{\ell(i), \ell(j)}$. 
Consider node $i$. 
Under Assumption \ref{aspt:dcsbm}, there exists $k$ such that $\bB_{\ell(i), k} > c_B$. Therefore,
\[
\frac{\rho n c_B \tmax^2}{C^2} \leq \theta_i \sum_{j: \ell(j) = k} \bB_{\ell(i), k} \theta_j \leq \E[d_i] \leq 
n \theta_i \tmax \leq n \tmax^2.
\]
Recall the Chernoff bound in \cite{chernoff}, where for $C > 1$,  
\begin{equation}
\label{tmp:chern1}
P(d_i > C\E[d_i]) \leq \exp(-\E[d_i] + C\E[d_i] \log (e/C)),     
\end{equation}
and for $c < 1$, 
\[
P(d_i < c\E[d_i]) \leq \exp(-\E[d_i] + c\E[d_i] \log (e/c)).
\]
If we pick $C = 1 + \delta$, $c = 1 - \delta$ for a constant $\delta > 0$, we have 
\begin{equation}
\label{tmp:chern2}
P\left(\left|{d_i}/{\E[d_i]} - 1\right| \geq \delta\right) \leq 2\exp(-\delta^2 \E[d_i]).
\end{equation}
Recall that $n\theta_0^2 \geq c_{\theta} \log n$, so the probability above will be no larger than $n^{-\delta^2 \rho c_B c_{\theta} / C^2}$. When $c_{\theta}$ is sufficiently large, it is $O(n^{-2})$. Using the union bound gives our claim. 
\end{proof}

\begin{lemma}
\label{lem:diRDPG}[Concentration of degrees for RDPG]
Under Assumptions \ref{aspt:latentrandomnew}, denote the degree conditioned on $\bY$ as $\E[d_i] = \sum_{j \neq i} \rho_n \bY_i^{\top} \bY_j$. 
There exists thresholds $n_0$ and $c_0$ such that if $n > n_0$ and $c_d > c_0$, then with high probability, for a small constant $\delta >0$,
\[
c n \rho_n \leq \E[d_i] \leq n \rho_n, \quad
\left|{d_i}/{\E[d_i]} - 1\right| \leq \delta, \quad i \in [n].
\]
\end{lemma}
\begin{proof}
Since $\bY_i$ are distributed on a unit ball, $\bY_i^{\top} \bY_j \leq 1$ always holds. Hence, 
\[
\E[d_i] \leq \rho_n \sum_{j} 1 = n \rho_n.
\]

Now we consider the lower bound of $\E[d_i]$. For any vector $\ba$, $\ba^{\top} \bY_i$ are i.i.d. distributed with $|\ba^{\top} \bY_i| \leq \|\ba\|$. By Hoeffding's inequality,   
\[
P\left(\left|\frac{1}{n}\sum \ba^{\top} \bY_i - \E[\ba^{\top} \bY_j]\right| \geq \|\ba\|\sqrt{\frac{2\log n}{n}}\right) \leq 2\exp\{-2\log n\} = 2n^{-2}.
\]
Letting $\ba = \bY_i$ where $\|\ba\| \leq 1$, then with probability $1 - O(n^{-2})$, 
\[
\E[d_i] \geq \rho_n \bY_i^{\top}\left(\sum_{j \neq i} \bY_j\right) \geq \rho_n \left((n-1) \bY_i^{\top} \E[\bY_j] - \sqrt{2n\log n}\right) \geq c n \rho_n/2. 
\]
The last inequality follows from the assumption that $\bY_i^{\top} \E[\bY_j] \geq c > 0$. 

With the results about $\E[d_i]$, we apply the analysis in Lemma \ref{lem:di} on $d_i$. It suggests: 
\[
P\left(\left|{d_i}/{\E[d_i]} - 1\right| \geq \delta\right) \leq 2\exp(-\delta^2 \E[d_i]) \leq 2\exp(-\delta^2 c n \rho_n). 
\]

Recall that $n \rho_n \geq c_d \log n$. 
Thus, there exists $n_0$ such that when $n > n_0$, $\delta^2 c n \rho_n > 2\log n$. It then follows that the inequality holds with probability $1 - O(n^{-2})$. Using the union bound, we establish our claim. 
\end{proof}

\begin{lemma}
\label{lem:LA}[Spectrum of adjacency and Laplacian] 
Under Assumptions \ref{aspt:dcsbm}, denote $\bar{\bA} = \bTheta \bY \bB \bY^{\top} \bTheta$. 
The following holds with probability $1 - O(n^{-1})$:
\[
\rho c_B^2 n \tmax^2 / C^2 \leq \lambda_K(\bar{\bA}) \leq \lambda_1(\bar{\bA}) \leq n \tmax^2, 
\]
\[
0.9 \rho c_B^2 / C^2 \leq \lambda_{K}(\bD^{-1/2} \bar{\bA} \bD^{-1/2}) \leq \lambda_1(\bD^{-1/2} \bar{\bA} \bD^{-1/2}) \leq {1.1}/{(\rho c_B)}.
\]
\end{lemma}

\begin{proof}
For the first claim, by Lemma \ref{lem:Anorm} and Lemma \ref{lem:di}, we have  
\[
\rho c_B n \tmax^2 \leq \frac{1}{n} \sum_{i=1}^n \E[d_i] \leq \|\bar{\bA}\| \leq \|\bar{\bA}\|_{\infty} = \max_i \E[d_i] \leq n \tmax^2.
\]

For the lower bound, by Lemma \ref{lem:condprod}, we have 
\[
\frac{\lambda_{K}(\bar{\bA})}{\lambda_1(\bar{\bA})}
\geq \frac{\lambda_{K}^2(\bTheta \bY) \lambda_{K}(\bB)}{\lambda_1^2(\bTheta \bY) \lambda_1(\bB)}.
\]
Under the condition that the condition number of $\bB$ satisfies ${\lambda_{K}(\bB)}/{\lambda_{1}(\bB)} \geq c_B$, and noting that $\bY^{\top} \bTheta^2 \bY$ is diagonal with $(k, k)$-th entries $\sum_{i: \ell(i) = k} \theta_i^2 \in [\rho c_B n \tmax^2 / C^2, n \tmax^2]$ by Lemma \ref{lem:di}, the condition number is constant. Therefore, 
\[
\lambda_{K}(\bar{\bA}) \geq  \frac{\rho c_B^2}{C^2} \lambda_1(\bar{\bA}) \geq \frac{\rho c_B^2}{C^2} \rho c_B n \tmax^2. 
\]

Using Lemma \ref{lem:condprod} with $\delta = 0.1$, we have with probability $1 - O(n^{-1})$, 
\[
0.9 \rho^2 c_B^3 / C^2 \leq \lambda_{K}(\bD^{-1/2} \bar{\bA} \bD^{-1/2}) \leq \lambda_1(\bD^{-1/2} \bar{\bA} \bD^{-1/2}) \leq \frac{1.1}{\rho c_B}.
\]
This proves the second claim. 
\end{proof}

\begin{lemma}
\label{lem:LArdpg}[Spectrum of adjacency and Laplacian under RDPG] 
Under Assumptions \ref{aspt:dcsbm}, denote $\bar{\bA} = \E[\bA|\bY] = \rho_n \bY \bY^{\top}$. 
The following holds with high probability:
\[
c_Y^2 c n \rho_n \leq \lambda_K(\bar{\bA}) \leq \lambda_1(\bar{\bA}) \leq n \rho_n, 
\]
\[
0.9 c_Y^2 c \leq \lambda_{K}(\bD^{-1/2} \bar{\bA} \bD^{-1/2}) \leq \lambda_1(\bD^{-1/2} \bar{\bA} \bD^{-1/2}) \leq {1.1}/{c}.
\]
\end{lemma}

\begin{proof}
For the first claim, by Lemma \ref{lem:Anorm} and Lemma \ref{lem:diRDPG}, we have  
\[
c n \rho_n \leq \frac{1}{n} \sum_{i=1}^n \E[d_i] \leq \|\bar{\bA}\| \leq \|\bar{\bA}\|_{\infty} = \max_i \E[d_i]
 \leq n \rho_n.
\]

For the lower bound, by Lemma \ref{lem:ymspectrum}, we have 
$\lambda_K(\bar{\bA}) = \rho_n \lambda_K(\bY \bY^{\top}) \geq c_Y^2 \rho_n \lambda_1(\bY \bY^{\top}) \geq c_Y^2 c n \rho_n.$

Consider the spectrum of $\bL$. By Lemma \ref{lem:diRDPG} with $\delta = 0.1$, with probability $1 - O(n^{-1})$, 
\[
0.9 c_Y^2 c \leq \lambda_{K}(\bD^{-1/2} \bar{\bA} \bD^{-1/2}) \leq \lambda_1(\bD^{-1/2} \bar{\bA} \bD^{-1/2}) \leq {1.1}/{c}.
\]

This establishes the second claim. 
\end{proof}

\subsection{Proof of Lemma \ref{lem:ymspectrum}}
\begin{proof}
Consider the first bullet about the spectrum of $\bY$. 
Since $\|\by_i\| \leq 1$ almost surely, the spectral norm $\|\bY\| = \lambda_1(\bY)$ can be bounded as follows:
\[
\lambda^2_1(\bY) \leq \|\bY\|_F^2 = \sum_{i=1}^n \|\by_i\|^2 \leq n.
\]

To discuss $\lambda_K(\bY)$, we denote $\bmu = \E[\by_i]$ and $\bw_i = \by_i - \bmu$ as the noise vector and $\bar{\bw} = (\sum_i \bw_i)/n$. The latent factor $\bY = \bW + \bmu \cdot \mathbf{1}_n^{\top}$, where $\mathbf{1}_n$ is the vector of ones. Hence, 
\[
\bY \bY^{\top} = (\bW + \bmu \cdot \mathbf{1}_n^{\top})( \bW + \bmu \cdot \mathbf{1}_n^{\top}) = \bW\bW^{\top} + n\bmu\bar{\bw}^{\top} + n\bar{\bw}\bmu^{\top} + n\bmu\bmu^{\top}.
\]
Therefore,
$\lambda_K(\bY\bY^{\top}) \geq \lambda_K(\bW\bW^{\top} + n\bmu\bmu^{\top}) - 2n\|\bmu\| \|\bar{\bw}\|$.

By Theorem 2 in \cite{denoising}, $\lambda_K(\bW\bW^{\top}) \geq cn$ with probability $1 - O(n^{-2})$. Furthermore, since $n\bmu\bmu'$ is a semi-positive matrix, $\lambda_K(\bW\bW' + n\bmu\bmu') \geq \lambda_K(\bW\bW') \geq cn$. Meanwhile, since $\bw_i$ is bounded with mean 0, by Hoeffding's inequality, $\|\bar{\bw}\| \leq 2\sqrt{K \log n / n}$ with probability $1 - O(n^{-2})$. Thus, $2n\|\bmu\| \|\bar{\bw}\| \leq 4C\sqrt{n \log n}$. When $n$ is large enough and a small $c_Y \leq c$, we have 
$\lambda_K(\bY\bY^{\top}) \geq cn \geq c_Y \|\bY\|$.

The second bullet can be found by the Frobenius norm.  
Since $\|\bM_j\| = 0$ when $j \notin \cS$ and $\|\bM_j\| \geq \kappa$ for $j \in \cS$, we have 
$\|\bM^{\cShat}\| \geq \|\bM^{\cShat}\|_F / \sqrt{K} \geq \kappa \sqrt{s / K}$.

The third statement is a direct result from Lemma \ref{lem:condprod}, where 
\[
\lambda_K(\bY \bM^{\cShat}) \geq \lambda_K(\bY)\lambda_K(\bM^{\cShat}) 
\geq c_Y \|\bY\| c_M \|\bM^{\cShat}\| \geq c_Y c_M \|\bY \bM^{\cShat}\|.
\]

The final statement is about $\bX^{\cShat} = \bY \bM^{\cShat} + \bZ^{\cShat}$, where $\bZ$ has i.i.d. $\mathcal{N}(0,1)$ entries. 
According to the random matrix theory in \cite{vershynin2010introduction}, with probability at least $1 - O(n^{-4})$, there is $\|\bZ^{\cShat}\| \leq 2(\sqrt{\cshat} + \sqrt{n})$. Therefore, 
$\|\bX^{\cShat}\| \leq \|\bY \bM^{\cShat}\| + 2(\sqrt{\cshat} + \sqrt{n})$,
and 
\[
\lambda_K(\bX^{\cShat}) \geq \lambda_K(\bY \bM^{\cShat}) - \|\bW^{\cShat}\| \geq c \|\bY \bM^{\cShat}\| - 2(\sqrt{\cshat} + \sqrt{n}).
\]
The results are proven.
\end{proof}
\end{document}